\documentclass[ reprint, amsmath,amssymb,aps]{revtex4-1}
\pdfoutput=1
\usepackage{epsfig,graphicx,xcolor,amsbsy,amssymb,latexsym,amsfonts,amsmath,setspace}
\usepackage{pstricks}
\usepackage{color}
\usepackage{blindtext}
\usepackage{eurosym}

\usepackage{graphicx}
\usepackage{tikz}
\usepackage{xcolor}
\usepackage{amsmath,amssymb,amsfonts,pstricks,setspace}
\usepackage[enableskew]{youngtab}

\numberwithin{equation}{section}

\newcommand{\bea}{\begin{eqnarray}\displaystyle}
\newcommand{\eea}{\end{eqnarray}}

\newcommand{\figref}[1]{Fig.~\protect\ref{#1}}

\newcommand{\pf}[2]{\mathcal{Z}_{#1,#2}}

\begin{document}
\title{\vspace{-1.8cm}
\bf{Dual Little Strings and their Partition Functions}\\[15pt]}
\author{ {Brice Bastian$^{\S}$,~Stefan~Hohenegger$^{\S}$,~Amer Iqbal$^{\dagger\star}$,~ Soo-Jong Rey$^{\diamond}$}
\\
\hfill\break
}
\date{\today}
\affiliation{${}^{\S}$ Universit\'e de Lyon UMR 5822, CNRS/IN2P3, Institut de Physique Nucl\'eaire de Lyon, 4 rue Enrico Fermi, 69622 Villeurbanne Cedex {\rm FRANCE}\\
${}^{\dagger}$ Abdus Salam School of Mathematical Sciences, G.C. University, Lahore {\rm PAKISTAN} \\
${}^{\star}$ Center for Theoretical Physics, Lahore, {\rm PAKISTAN}
\\
${}^{\diamond}$ School of Physics and Astronomy \& Center for Theoretical Physics, Seoul National University, Seoul 08826  {\rm KOREA}}

\begin{abstract}
We study the topological string partition function of a class of toric, double elliptically fibered 
Calabi-Yau threefolds $X_{N,M}$ at a generic point in the K\"ahler moduli space. These manifolds engineer little string theories in five dimensions or lower and are dual to stacks of M5-branes probing a transverse orbifold singularity. Using the refined topological vertex formalism, we explicitly calculate a generic building block which allows to compute the topological string partition function of $X_{N,M}$ as a series expansion in different K\"ahler parameters. Using this result we give further explicit proof for a duality found previously in the literature, which relates $X_{N,M}\sim X_{N',M'}$ for $NM=N'M'$ and $\text{gcd}(N,M)=\text{gcd}(N',M')$. 
 \end{abstract}

\maketitle


\vskip1cm

\section{Introduction}

Recently, the study of little string theories (LSTs) has received renewed interest: first proposed two decades ago \cite{Berkooz:1997cq,Seiberg:1997zk,Losev:1997hx,Aharony:1998ub} (see \cite{Aharony:1999ks,Kutasov:2001uf} for a review), during the last years, the construction of LSTs from various M-brane constructions and their dual geometric description in F-theory \cite{Bhardwaj:2015oru} has led to a better understanding of supersymmetric quantum theories in six dimensions. Moreover, in the process new and unexpected dualities have been unravelled. The geometric description of these in terms of F-theory compactification on Calabi-Yau threefolds has been particularly useful in understanding various stringy properties such as LST T-duality \cite{Seiberg:1997zk,Intriligator:1997dh,Intriligator:1999cn}.

A particular class of LSTs, dual to $N$  M5-branes transverse to a $\mathbb{Z}_{M}$ orbifold, preserving eight supercharges can be realized in F-theory using toric, non-compact Calabi-Yau manifolds denoted by $X_{N,M}$ \cite{Haghighat:2013gba,Hohenegger:2015btj}. The latter have the structure of a double elliptic fibration, in which one elliptic fibration has a singularity of type $I_{N-1}$ and the other one of $I_{M-1}$. The exchange of the two elliptic fibrations is related to the T-duality of LST \cite{Bhardwaj:2015oru,Hohenegger:2015btj}. Moreover, using for example the refined topological vertex formalism, the topological string partition function $\pf{N}{M}$ on $X_{N,M}$ can be computed explicitly \cite{Hohenegger:2015btj}. By analysing the web diagrams associated with $X_{N,M}$, it was shown in \cite{Hohenegger:2016yuv} that $X_{N,M}\sim X_{N',M'}$ for $MN=M'N'$ and $\text{gcd}(M,N)=k=\text{gcd}(M',N')$, \emph{i.e.} the two Calabi-Yau threefolds lie in the same extended K\"ahler moduli space and can be related by flop transitions. The partition function of the two dual theories should be the same and so it was expected that the topological string partition function of $X_{N,M}$ and $X_{N',M'}$ should be the same \emph{i.e.}
\bea\label{relation}
\pf{N}{M}(\omega,\epsilon_{1,2})=\pf{N'}{M'}(\omega',\epsilon_{1,2})\,
\eea
for $\omega$ and $\omega'$ the K\"ahler forms of the two Calabi-Yau threefolds related by flop transitions. Using a general building block $W^{\{\alpha\}}_{\{\beta\}}$ to compute $\pf{N}{M}(\omega,\epsilon_{1,2})$, we explicitly confirm Eq.(\ref{relation}) for $\text{gcd}(N,M)=1$, thus verifying the duality proposed in \cite{Hohenegger:2016yuv} explicitly at the level of the partition function.

This paper is organised as follows. In section 2, we discuss the $(N,M)=(3,2)$ case in detail and show that it is dual to the $(6,1)$ case, \emph{i.e.} related by a combination of flop  and symmetry transforms. We also discuss the relation between the general $(N,M)$ and $(k,MN/k)$ case. In section 3, we calculate the partition function of the $(3,2)$ case and show that, after a change of K\"ahler parameters, it is equal to the partition function of the $(6,1)$ configuration. In section 4, we discuss the open string amplitudes which make up the building block for generic brane configurations. We also discuss the general ``twisted" $(1,L)$ case, which is related by flop transitions to the standard $(1,L)$ case. In section 5, we present our conclusions and future directions. Some of the detailed calculations and our notation and conventions are relegated to three appendices.


\section{Little Strings and Dual Brane Webs}

At low energies, a stack of $N$ coincident M5-branes engineers a six-dimensional ${\cal N} = (2,0)$ superconformal field theory of $A_{N-1}$-type. Upon replacing the $\mathbb{R}^5$ transverse to the M5-branes by an $\mathbb{R}_{\text{trans}}\times \mathbb{R}^{4}/\mathbb{Z}_{M}$ orbifold geometry, the world-volume theory on the branes becomes an ${\cal N} = (1,0)$ SCFT. One may move away from the conformal point by separating the M5-branes along the transverse $\mathbb{R}_{\text{trans}}$, as massive states appear in the form of M2-branes that end on the separated M5-branes. Note that if $\mathbb{R}_{\text{trans}}$ is compactified to a circle $\mathbb{S}^1_{\rho}$ of circumference $\rho$ so that the M5-branes are points on $\mathbb{S}^1_\rho$, then the theory arising on the M5-branes (whether coincident or not) gives rise to a LST whose defining scale is $\rho$. 

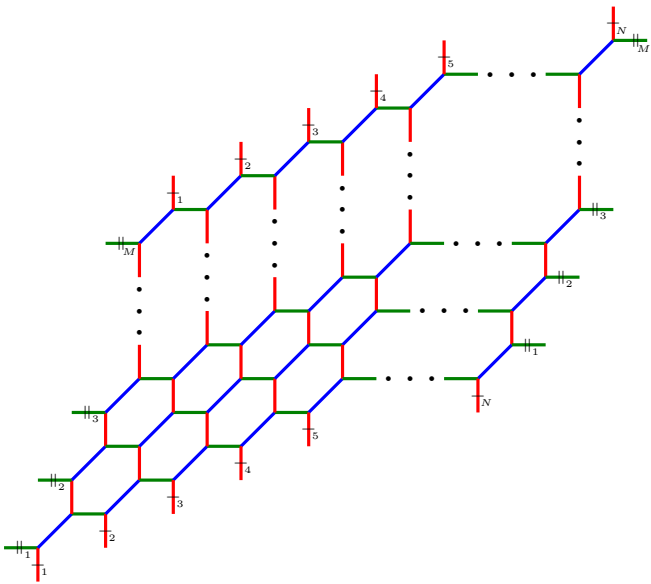
\begin{figure}[htb]
\begin{center}
\scalebox{0.75}{\parbox{10.6cm}{\begin{tikzpicture}[scale = 0.6]
\draw[ultra thick,green!50!black] (-6,0) -- (-5,0);
\draw[ultra thick,red] (-5,-1) -- (-5,0);
\draw[ultra thick,blue] (-5,0) -- (-4,1);
\draw[ultra thick,green!50!black] (-4,1) -- (-3,1);
\draw[ultra thick,red] (-4,1) -- (-4,2);
\draw[ultra thick,red] (-3,1) -- (-3,0);
\draw[ultra thick,blue] (-3,1) -- (-2,2);
\draw[ultra thick,blue] (-4,2) -- (-3,3);
\draw[ultra thick,green!50!black] (-5,2) -- (-4,2);
\draw[ultra thick,red] (-3,3) -- (-3,4);
\draw[ultra thick,green!50!black] (-3,3) -- (-2,3);
\draw[ultra thick,red] (-2,2) -- (-2,3);
\draw[ultra thick,green!50!black] (-2,2) -- (-1,2);
\draw[ultra thick,blue] (-2,3) -- (-1,4);
\draw[ultra thick,blue] (-1,2) -- (0,3);
\draw[ultra thick,red] (-1,2) -- (-1,1);
\draw[ultra thick,green!50!black] (-4,4) -- (-3,4);
\draw[ultra thick,blue] (-3,4) -- (-2,5);
\draw[ultra thick,green!50!black] (-2,5) -- (-1,5);
\draw[ultra thick,green!50!black] (-1,4) -- (0,4);
\draw[ultra thick,green!50!black] (0,3) -- (1,3);
\draw[ultra thick,red] (-2,5) -- (-2,6);
\draw[ultra thick,red] (-1,4) -- (-1,5);
\draw[ultra thick,red] (0,3) -- (0,4);
\draw[ultra thick,blue] (1,3) -- (2,4);
\draw[ultra thick,blue] (0,4) -- (1,5);
\draw[ultra thick,blue] (-1,5) -- (0,6);
%
\draw[ultra thick,green!50!black] (2,4) -- (3,4);
\draw[ultra thick,green!50!black] (1,5) -- (2,5);
\draw[ultra thick,green!50!black] (0,6) -- (1,6);
%
\draw[ultra thick,red] (1,3) -- (1,2);
\draw[ultra thick,red] (2,4) -- (2,5);
\draw[ultra thick,red] (1,5) -- (1,6);
\draw[ultra thick,red] (0,6) -- (0,7);
\draw[ultra thick,green!50!black] (-1,10) -- (0,10);
\node[rotate=90] at (0,8) {{\Huge$\ldots$}};
\node[rotate=90] at (2,9) {{\Huge$\ldots$}};
\node[rotate=90] at (4,10) {{\Huge$\ldots$}};
\node[rotate=90] at (6,11) {{\Huge$\ldots$}};
\node[rotate=90] at (-2,7) {{\Huge$\ldots$}};
\draw[ultra thick,red] (0,9) -- (0,10);
\draw[ultra thick,blue] (0,10) -- (1,11);
\draw[ultra thick,red] (1,11) -- (1,12);
\draw[ultra thick,red] (2,11) -- (2,10);
\draw[ultra thick,green!50!black] (1,11) -- (2,11);
\draw[ultra thick,blue] (2,11) -- (3,12);
\draw[ultra thick,red] (3,12) -- (3,13);
\draw[ultra thick,red] (4,12) -- (4,11);
\draw[ultra thick,green!50!black] (3,12) -- (4,12);
\draw[ultra thick,blue] (4,12) -- (5,13);
\draw[ultra thick,green!50!black] (5,13) -- (6,13);
\draw[ultra thick,red] (5,13) -- (5,14);
\draw[ultra thick,blue] (6,13) -- (7,14);
\draw[ultra thick,red] (6,13) -- (6,12);
\draw[ultra thick,green!50!black] (7,14) -- (8,14);
\draw[ultra thick,red] (7,14) -- (7,15);
\draw[ultra thick,blue] (1,6) -- (2,7);
\draw[ultra thick,red] (2,7) -- (2,8);
\draw[ultra thick,red] (3,7) -- (3,6);
\draw[ultra thick,green!50!black] (2,7) -- (3,7);
\draw[ultra thick,blue] (3,7) -- (4,8);
\draw[ultra thick,red] (4,8) -- (4,9);
\draw[ultra thick,red] (5,8) -- (5,7);
\draw[ultra thick,green!50!black] (4,8) -- (5,8);
\draw[ultra thick,green!50!black] (3,6) -- (4,6);
\draw[ultra thick,green!50!black] (5,7) -- (6,7);
\draw[ultra thick,blue] (5,8) -- (6,9);
\draw[ultra thick,red] (6,9) -- (6,10);
\draw[ultra thick,blue] (2,5) -- (3,6) ;
\draw[ultra thick,blue] (4,6) -- (5,7) ;
\draw[ultra thick,green!50!black] (6,9) -- (7,9);
\node at (8,9) {{\Huge $\ldots$}};
\draw[ultra thick,blue] (3,4) -- (4,5);
\draw[ultra thick,red] (3,4) -- (3,3);
\draw[ultra thick,red] (4,5) -- (4,6);
\draw[ultra thick,green!50!black] (4,5) -- (5,5);
\node at (7,7) {{\Huge $\ldots$}};
\draw[ultra thick,green!50!black] (10,14) -- (11,14);
\draw[ultra thick,red] (11,14) -- (11,13);
\draw[ultra thick,blue] (11,14) -- (12,15);
\node at (9,14) {{\Huge $\ldots$}};
\draw[ultra thick,red] (12,15) -- (12,16);
\draw[ultra thick,green!50!black] (12,15) -- (13,15);

\node[rotate=90] at (11,12) {{\Huge $\ldots$}};
\node at (6,5) {{\Huge $\ldots$}};
\draw[ultra thick,blue] (10,9) -- (11,10);
\draw[ultra thick,green!50!black] (9,9) -- (10,9);
\draw[ultra thick,red] (11,10) -- (11,11);
\draw[ultra thick,green!50!black] (11,10) -- (12,10);
\draw[ultra thick,red] (10,9) -- (10,8);
\draw[ultra thick,blue] (9,7) -- (10,8);
\draw[ultra thick,green!50!black] (8,7) -- (9,7);
\draw[ultra thick,red] (9,6) -- (9,7);
\draw[ultra thick,green!50!black] (10,8) -- (11,8);
\draw[ultra thick,blue] (8,5) -- (9,6);
\draw[ultra thick,green!50!black] (9,6) -- (10,6);
\draw[ultra thick,green!50!black] (7,5) -- (8,5);
\draw[ultra thick,red] (8,4) -- (8,5);
%
%
%
\draw[ultra thick,green!50!black] (-3,9) -- (-2,9);
\draw[ultra thick,blue] (-2,9) -- (-1,10);
\draw[ultra thick,red] (-1,10) -- (-1,11);
\draw[ultra thick,red] (-2,8) -- (-2,9);
\node[rotate=90] at (-5.5,0) {$=$};
\node at (-5.3,-0.2) {{\tiny$1$}};
\node[rotate=90] at (-4.5,2) {$=$};
\node at (-4.3,1.8) {{\tiny$2$}};
\node[rotate=90] at (-3.5,4) {$=$};
\node at (-3.3,3.8) {{\tiny$3$}};
\node[rotate=90] at (-2.5,9) {$=$};
\node at (-2.3,8.75) {{\tiny$M$}};
\node[rotate=90] at (9.5,6) {$=$};
\node at (9.7,5.8) {{\tiny$1$}};
\node[rotate=90] at (10.5,8) {$=$};
\node at (10.7,7.8) {{\tiny$2$}};
\node[rotate=90] at (11.5,10) {$=$};
\node at (11.7,9.8) {{\tiny$3$}};
\node[rotate=90] at (12.7,15) {$=$};
\node at (12.9,14.75) {{\tiny$M$}};
\node at (-5,-0.5) {$-$};
\node at (-4.8,-0.7) {{\tiny $1$}};
\node at (-3,0.5) {$-$};
\node at (-2.8,0.3) {{\tiny $2$}};
\node at (-1,1.5) {$-$};
\node at (-0.8,1.3) {{\tiny $3$}};
\node at (1,2.5) {$-$};
\node at (1.2,2.3) {{\tiny $4$}};
\node at (3,3.5) {$-$};
\node at (3.2,3.3) {{\tiny $5$}};
\node at (8,4.5) {$-$};
\node at (8.25,4.3) {{\tiny $N$}};
\node at (-1,10.5) {$-$};
\node at (-0.8,10.3) {{\tiny $1$}};
\node at (1,11.5) {$-$};
\node at (1.2,11.3) {{\tiny $2$}};
\node at (3,12.5) {$-$};
\node at (3.2,12.3) {{\tiny $3$}};
\node at (5,13.5) {$-$};
\node at (5.2,13.3) {{\tiny $4$}};
\node at (7,14.5) {$-$};
\node at (7.2,14.3) {{\tiny $5$}};
\node at (12,15.5) {$-$};
\node at (12.25,15.3) {{\tiny $N$}};
%
%
\end{tikzpicture}}}

\end{center}
\caption{\sl The 5-brane web in type IIB string theory dual to the Calabi-Yau threefold $X_{N,M}$. The K\"ahler parameters associated with the horizontal, vertical and diagonal lines in the web are collectively denoted by ${\bf h}, {\bf v}$ and ${\bf m}$, respectively. More precise labelings will be given for concrete examples below (see \emph{e.g.}~\figref{Fig:Newton32}(a)).}
\label{Fig:WebToric}
\end{figure}

As a next step, we compactify the worldvolume of the M5-branes on a circle $\mathbb{S}^1_\tau$ of circumference $\tau$. This M-brane configuration is dual \cite{Cecotti:2013mba} to a $(p,q)$ 5-brane web in type IIB string theory given by $N$ coincident NS5-branes wrapped on $\mathbb{S}^1_\tau$ that intersect with $M$ coincident D5-branes wrapped on $\mathbb{S}^1_\rho$. Turning on deformations which separate the intersections of NS5-branes and D5-branes and stretching the NS5-D5 bound-state branes \cite{Aharony:1997bh} yields the brane web shown in \figref{Fig:WebToric}.
Finally, this brane web in type IIB string theory is dual to a toric Calabi-Yau threefold \cite{Leung:1997tw}, which was first studied in \cite{Hohenegger:2015btj} and was denoted by $X_{N,M}$, and provides a more geometric description of the theory via F-theory compactification \cite{Morrison:1996na,Morrison:1996pp}. A general classification of toric and non-toric base manifolds in F-theory models was discussed in \cite{Morrison:2012np,Morrison:2012js} and, more recently, in \cite{Choi:2017vtd}. For general properties of topological strings on elliptic Calabi-Yau threefolds, see \cite{Alim:2012ss,Klemm:2012sx,Huang:2015sta}.

The brane web shown in \figref{Fig:WebToric} has $3NM$ parameters, related to the sizes of various line segments corresponding to the 5-branes. Specifically, we denote the parameters associated with horizontal (green) lines as $\mathbf{h}=\{h_1,\ldots,h_{NM}\}$, with vertical (red) lines as $\mathbf{v}=\{v_1,\ldots,v_{NM}\}$ and with diagonal (blue) lines as $\mathbf{m}=\{m_1,\ldots,m_{NM}\}$. Out of these, only $NM+2$ are independent  (after imposing a number of consistency conditions related to the the periodic identification of the web diagram \cite{Haghighat:2013tka}).
In the dual Calabi-Yau threefold $X_{N,M}$, these $(NM+2)$ parameters are independent K\"ahler moduli. For a discussion of the geometry of $X_{N,M}$ and the corresponding mirror curves, see \cite{Kanazawa:2016tnt}.

The supersymmetric or BPS-state counting partition function of the above mentioned LST can be computed as the topological string partition function $\pf{N}{M}(\mathbf{h},\mathbf{v},\mathbf{m},\epsilon_1,\epsilon_2)$ of $X_{N,M}$. The latter depends on the K\"ahler parameters of $X_{N,M}$ (taking into account that only $NM+2$ of them are independent) as well as two regularisation parameters $\epsilon_{1,2}$ introduced to render the
partition function well-defined. From the perspective of the field theory that
describes the low-energy limit of the LST engineered by $X_{N,M}$, these parameters can be interpreted geometrically as the $\Omega$-background \cite{Nekrasov:2002qd, Losev:2003py}. The quantization of the K\"ahler parameters in units of unrefined $\epsilon$, which correspond to Coulomb branch parameters \cite{Nekrasov:2003rj}, leads to a description of the partition function in terms of an irreducible representation of an affine group. For details, see \cite{Bastian:2017jje}.

Given the web diagram, $\pf{N}{M}$ can be computed efficiently using the (refined) topological vertex formalism: to this end, a preferred direction in the web needs to be chosen, which from the perspective of LSTs determines the parameters that play the role of coupling constants. Concretely, different choices of the preferred direction lead to different (but equivalent) power series expansions of $\pf{N}{M}$ in terms of either $\mathbf{h}$, $\mathbf{v}$ or $\mathbf{m}$. In this paper, we shall use this fact to prove a duality (that was first proposed in \cite{Hohenegger:2016yuv}) at the level of the partition function for generic values of the K\"ahler parameters: indeed, it was argued in \cite{Hohenegger:2016yuv} that $X_{N,M}$ and $X_{N',M'}$ are related to each other through a series of $SL(2, \mathbb{Z})$ symmetry and flop transformations if $NM=N'M'$ and $\text{gcd}(N,M)=k=\text{gcd}(N',M')$. 
This suggests that
\begin{align}
\pf{N}{M}(\mathbf{h},\mathbf{v},\mathbf{m},\epsilon_1,\epsilon_2)=\pf{N'}{M'}(\mathbf{h}',\mathbf{v}',\mathbf{m}',\epsilon_1,\epsilon_2)\,,\label{DualityIdentPartFct}
\end{align}
where the duality map $(\mathbf{h},\mathbf{v},\mathbf{m})\longmapsto (\mathbf{h}',\mathbf{v}',\mathbf{m}')$ was proposed in \cite{Hohenegger:2016yuv} (see also appendix~\ref{StripFlop} for an example), taking into account the consistency conditions on both sides. The relation (\ref{DualityIdentPartFct}) was tested in \cite{Hohenegger:2016yuv} (based on a conjecture put forward in~\cite{Hohenegger:2016eqy}) at a particular region in the K\"ahler moduli space of $X_{N,M}$ (with $h_i=h_j$, $v_i=v_j$ and $m_i=m_j$ for $i,j=1,\ldots,NM$) and the Nekrasov-Shatashvili limit $\epsilon_2\to 0$ \cite{Nekrasov:2009rc, Mironov:2009uv}. In the following, by computing $\pf{N}{M}$ at a generic point in the K\"ahler moduli space, we give explicit evidence of (\ref{DualityIdentPartFct}) in general. However, for simplicity, we shall limit ourselves to $\text{gcd}(N,M)=1$.

\section{Partition Function of $(3,2)$ Case}\label{Sect:DirectComputation32}
As checking (\ref{DualityIdentPartFct}) is rather complicated for generic $(N,M)$, before presenting the generic case in the following section, we first discuss as a (nontrivial) example the case $(N,M)=(3,2)$. Specifically, we shall demonstrate in the following that
\begin{align}
\pf{3}{2}(\omega,\epsilon_1,\epsilon_2)=\pf{6}{1}(\omega',\epsilon_1,\epsilon_2)\,,\label{PartitionFc61}
\end{align}
where $\omega$ and $\omega'$ denote the (independent) K\"ahler parameters of $X_{N,M}$ and $X_{N',M'}$, respectively. We shall verify (\ref{PartitionFc61}) explicitly at a generic point in the moduli space of K\"ahler parameters by performing a novel expansion of the left hand side of this relation.
\subsection{Consistency Conditions and Parametrisation}
\begin{figure*}
\scalebox{0.6}{\parbox{13cm}{\begin{tikzpicture}[scale = 1.50]
\draw[ultra thick] (-6,0) -- (-5,0);
\draw[ultra thick] (-5,-1) -- (-5,0);
\draw[ultra thick] (-5,0) -- (-4,1);
\draw[ultra thick] (-4,1) -- (-3,1);
\draw[ultra thick] (-4,1) -- (-4,2);
\draw[ultra thick] (-3,1) -- (-3,0);
\draw[ultra thick] (-3,1) -- (-2,2);
\draw[ultra thick] (-4,2) -- (-3,3);
\draw[ultra thick] (-5,2) -- (-4,2);
\draw[ultra thick] (-3,3) -- (-3,4);
\draw[ultra thick] (-3,3) -- (-2,3);
\draw[ultra thick] (-2,2) -- (-2,3);
\draw[ultra thick] (-2,2) -- (-1,2);
\draw[ultra thick] (-2,3) -- (-1,4);
\draw[ultra thick] (-1,2) -- (0,3);
\draw[ultra thick] (-1,2) -- (-1,1);
\draw[ultra thick] (-1,4) -- (0,4);
\draw[ultra thick] (0,3) -- (1,3);
\draw[ultra thick] (0,3) -- (0,4);
\draw[ultra thick] (0,4) -- (1,5);
\draw[ultra thick] (1,5) -- (2,5);
\draw[ultra thick] (1,5) -- (1,6);
\draw[ultra thick] (-1,4) -- (-1,5);
\node[rotate=90] at (-5.7,0) {$=$};
\node at (-5.6,-0.15) {{\tiny$1$}};
\node[rotate=90] at (-4.7,2) {$=$};
\node at (-4.6,1.85) {{\tiny$2$}};
\node[rotate=90] at (0.8,3) {$=$};
\node at (0.9,2.8) {{\tiny$1$}};
\node[rotate=90] at (1.8,5) {$=$};
\node at (1.9,4.8) {{\tiny$2$}};
\node at (-5,-0.75) {$-$};
\node at (-4.85,-0.75) {{\tiny $1$}};
\node at (-3,0.25) {$-$};
\node at (-2.85,0.25) {{\tiny $2$}};
\node at (-1,1.25) {$-$};
\node at (-0.85,1.25) {{\tiny $3$}};
\node at (-3,3.7) {$-$};
\node at (-2.85,3.65) {{\tiny $1$}};
\node at (-1,4.75) {$-$};
\node at (-0.85,4.65) {{\tiny $2$}};
\node at (1,5.75) {$-$};
\node at (1.15,5.65) {{\tiny $3$}};
\node at (-4.3,0.3) {{\small $m_1$}};
\node at (-2.3,1.3) {{\small $m_2$}};
\node at (-0.3,2.3) {{\small $m_3$}};
\node at (-3.3,2.3) {{\small $m_4$}};
\node at (-1.3,3.3) {{\small $m_5$}};
\node at (0.7,4.3) {{\small $m_6$}};
\node at (-3.5,1.2) {{\small $h_5$}};
\node at (-1.5,2.2) {{\small $h_3$}};
\node at (0.5,3.3) {{\small $h_1$}};
\node at (-2.5,3.2) {{\small $h_2$}};
\node at (-0.5,4.2) {{\small $h_6$}};
\node at (1.5,5.3) {{\small $h_4$}};
\node at (-5.3,-0.5) {{\small $v_1$}};
\node at (-3.3,0.5) {{\small $v_5$}};
\node at (-1.3,1.5) {{\small $v_3$}};
\node at (-4.2,1.5) {{\small $v_4$}};
\node at (-2.2,2.5) {{\small $v_2$}};
\node at (-0.2,3.5) {{\small $v_6$}};
\node at (-5,1) {\bf IV};
\node at (-3,2) {\bf II};
\node at (-1,3) {\bf VI};
\node at (-4,3) {\bf I};
\node at (-2,4) {\bf V};
\node at (-2,1) {\bf III};
\draw[fill=black] (-5,0) circle (0.05cm);
\node at (-4.85,-0.15) {{\small $1$}};
\draw[fill=black] (-4,1) circle (0.05cm);
\node at (-3.9,0.8) {{\small $4$}};
\draw[fill=black] (-3,1) circle (0.05cm);
\node at (-2.85,0.85) {{\small $2$}};
\draw[fill=black] (-2,2) circle (0.05cm);
\node at (-1.9,1.8) {{\small $5$}};
\draw[fill=black] (-1,2) circle (0.05cm);
\node at (-0.85,1.85) {{\small $3$}};
\draw[fill=black] (0,3) circle (0.05cm);
\node at (0.1,2.8) {{\small $6$}};
\draw[fill=black] (-4,2) circle (0.05cm);
\node at (-3.85,1.85) {{\small $7$}};
\draw[fill=black] (-3,3) circle (0.05cm);
\node at (-2.9,2.8) {{\small $10$}};
\draw[fill=black] (-2,3) circle (0.05cm);
\node at (-1.85,2.85) {{\small $8$}};
\draw[fill=black] (-1,4) circle (0.05cm);
\node at (-0.9,3.8) {{\small $11$}};
\draw[fill=black] (0,4) circle (0.05cm);
\node at (0.15,3.85) {{\small $9$}};
\draw[fill=black] (1,5) circle (0.05cm);
\node at (1.1,4.8) {{\small $12$}};
%
\node at (-2,-1.5) {\Large (a)};
\end{tikzpicture}}}
\hspace{1cm}
\scalebox{0.6}{\parbox{13cm}{\begin{tikzpicture}[scale = 1.0]
\draw[fill=black!20!white] (0,0) -- (3,0) -- (3,2) -- (0,2) -- (0,0);
\draw[-] (-1,0) -- (10,0);
\draw[-] (0,-1) -- (0,7);
\draw[ultra thick] (0,0) -- (9,0);
\draw[-] (0,1) -- (9,1);
\draw[ultra thick] (0,2) -- (9,2);
\draw[-] (0,3) -- (9,3);
\draw[ultra thick] (0,4) -- (9,4);
\draw[-] (0,5) -- (9,5);
\draw[ultra thick] (0,6) -- (9,6);
\draw[ultra thick] (0,0) -- (0,6);
\draw[-] (1,0) -- (1,6);
\draw[-] (2,0) -- (2,6);
\draw[ultra thick] (3,0) -- (3,6);
\draw[-] (4,0) -- (4,6);
\draw[-] (5,0) -- (5,6);
\draw[ultra thick] (6,0) -- (6,6);
\draw[-] (7,0) -- (7,6);
\draw[-] (8,0) -- (8,6);
\draw[ultra thick] (9,0) -- (9,6);
\draw[-] (0,1) -- (1,0);
\draw[-] (0,2) -- (2,0);
\draw[-] (0,3) -- (3,0);
\draw[-] (0,4) -- (4,0);
\draw[-] (0,5) -- (5,0);
\draw[-] (0,6) -- (6,0);
\draw[-] (1,6) -- (7,0);
\draw[-] (2,6) -- (8,0);
\draw[-] (3,6) -- (9,0);
\draw[-] (4,6) -- (9,1);
\draw[-] (5,6) -- (9,2);
\draw[-] (6,6) -- (9,3);
\draw[-] (7,6) -- (9,4);
\draw[-] (8,6) -- (9,5);
\node[rotate=90] at (4.5,-1.5) {\Large $\cdots$};
\node[rotate=90] at (4.5,7.5) {\Large $\cdots$};
\node at (-1.5,3) {\Large $\cdots$};
\node at (10.5,3) {\Large $\cdots$};
\node at (0.3,0.3) {\footnotesize \bf 1};
\node at (1.3,0.3) {\footnotesize \bf 2};
\node at (2.3,0.3) {\footnotesize \bf 3};
\node at (3.3,0.3) {\footnotesize \bf 1};
\node at (4.3,0.3) {\footnotesize \bf 2};
\node at (5.3,0.3) {\footnotesize \bf 3};
\node at (6.3,0.3) {\footnotesize \bf 1};
\node at (7.3,0.3) {\footnotesize \bf 2};
\node at (8.3,0.3) {\footnotesize \bf 3};
\node at (0.7,0.7) {\footnotesize \bf 4};
\node at (1.7,0.7) {\footnotesize \bf 5};
\node at (2.7,0.7) {\footnotesize \bf 6};
\node at (3.7,0.7) {\footnotesize \bf 4};
\node at (4.7,0.7) {\footnotesize \bf 5};
\node at (5.7,0.7) {\footnotesize \bf 6};
\node at (6.7,0.7) {\footnotesize \bf 4};
\node at (7.7,0.7) {\footnotesize \bf 5};
\node at (8.7,0.7) {\footnotesize \bf 6};
\node at (0.3,1.3) {\footnotesize \bf 7};
\node at (1.3,1.3) {\footnotesize \bf 8};
\node at (2.3,1.3) {\footnotesize \bf 9};
\node at (3.3,1.3) {\footnotesize \bf 7};
\node at (4.3,1.3) {\footnotesize \bf 8};
\node at (5.3,1.3) {\footnotesize \bf 9};
\node at (6.3,1.3) {\footnotesize \bf 7};
\node at (7.3,1.3) {\footnotesize \bf 8};
\node at (8.3,1.3) {\footnotesize \bf 9};
\node at (0.7,1.7) {\footnotesize \bf 10};
\node at (1.7,1.7) {\footnotesize \bf 11};
\node at (2.7,1.7) {\footnotesize \bf 12};
\node at (3.7,1.7) {\footnotesize \bf 10};
\node at (4.7,1.7) {\footnotesize \bf 11};
\node at (5.7,1.7) {\footnotesize \bf 12};
\node at (6.7,1.7) {\footnotesize \bf 10};
\node at (7.7,1.7) {\footnotesize \bf 11};
\node at (8.7,1.7) {\footnotesize \bf 12};
\node at (0.3,2.3) {\footnotesize \bf 1};
\node at (1.3,2.3) {\footnotesize \bf 2};
\node at (2.3,2.3) {\footnotesize \bf 3};
\node at (3.3,2.3) {\footnotesize \bf 1};
\node at (4.3,2.3) {\footnotesize \bf 2};
\node at (5.3,2.3) {\footnotesize \bf 3};
\node at (6.3,2.3) {\footnotesize \bf 1};
\node at (7.3,2.3) {\footnotesize \bf 2};
\node at (8.3,2.3) {\footnotesize \bf 3};
\node at (0.7,2.7) {\footnotesize \bf 4};
\node at (1.7,2.7) {\footnotesize \bf 5};
\node at (2.7,2.7) {\footnotesize \bf 6};
\node at (3.7,2.7) {\footnotesize \bf 4};
\node at (4.7,2.7) {\footnotesize \bf 5};
\node at (5.7,2.7) {\footnotesize \bf 6};
\node at (6.7,2.7) {\footnotesize \bf 4};
\node at (7.7,2.7) {\footnotesize \bf 5};
\node at (8.7,2.7) {\footnotesize \bf 6};
\node at (0.3,3.3) {\footnotesize \bf 7};
\node at (1.3,3.3) {\footnotesize \bf 8};
\node at (2.3,3.3) {\footnotesize \bf 9};
\node at (3.3,3.3) {\footnotesize \bf 7};
\node at (4.3,3.3) {\footnotesize \bf 8};
\node at (5.3,3.3) {\footnotesize \bf 9};
\node at (6.3,3.3) {\footnotesize \bf 7};
\node at (7.3,3.3) {\footnotesize \bf 8};
\node at (8.3,3.3) {\footnotesize \bf 9};
\node at (0.7,3.7) {\footnotesize \bf 10};
\node at (1.7,3.7) {\footnotesize \bf 11};
\node at (2.7,3.7) {\footnotesize \bf 12};
\node at (3.7,3.7) {\footnotesize \bf 10};
\node at (4.7,3.7) {\footnotesize \bf 11};
\node at (5.7,3.7) {\footnotesize \bf 12};
\node at (6.7,3.7) {\footnotesize \bf 10};
\node at (7.7,3.7) {\footnotesize \bf 11};
\node at (8.7,3.7) {\footnotesize \bf 12};
\node at (0.3,4.3) {\footnotesize \bf 1};
\node at (1.3,4.3) {\footnotesize \bf 2};
\node at (2.3,4.3) {\footnotesize \bf 3};
\node at (3.3,4.3) {\footnotesize \bf 1};
\node at (4.3,4.3) {\footnotesize \bf 2};
\node at (5.3,4.3) {\footnotesize \bf 3};
\node at (6.3,4.3) {\footnotesize \bf 1};
\node at (7.3,4.3) {\footnotesize \bf 2};
\node at (8.3,4.3) {\footnotesize \bf 3};
\node at (0.7,4.7) {\footnotesize \bf 4};
\node at (1.7,4.7) {\footnotesize \bf 5};
\node at (2.7,4.7) {\footnotesize \bf 6};
\node at (3.7,4.7) {\footnotesize \bf 4};
\node at (4.7,4.7) {\footnotesize \bf 5};
\node at (5.7,4.7) {\footnotesize \bf 6};
\node at (6.7,4.7) {\footnotesize \bf 4};
\node at (7.7,4.7) {\footnotesize \bf 5};
\node at (8.7,4.7) {\footnotesize \bf 6};
\node at (0.3,5.3) {\footnotesize \bf 7};
\node at (1.3,5.3) {\footnotesize \bf 8};
\node at (2.3,5.3) {\footnotesize \bf 9};
\node at (3.3,5.3) {\footnotesize \bf 7};
\node at (4.3,5.3) {\footnotesize \bf 8};
\node at (5.3,5.3) {\footnotesize \bf 9};
\node at (6.3,5.3) {\footnotesize \bf 7};
\node at (7.3,5.3) {\footnotesize \bf 8};
\node at (8.3,5.3) {\footnotesize \bf 9};
\node at (0.7,5.7) {\footnotesize \bf 10};
\node at (1.7,5.7) {\footnotesize \bf 11};
\node at (2.7,5.7) {\footnotesize \bf 12};
\node at (3.7,5.7) {\footnotesize \bf 10};
\node at (4.7,5.7) {\footnotesize \bf 11};
\node at (5.7,5.7) {\footnotesize \bf 12};
\node at (6.7,5.7) {\footnotesize \bf 10};
\node at (7.7,5.7) {\footnotesize \bf 11};
\node at (8.7,5.7) {\footnotesize \bf 12};
\node at (4.7,-3.25) {\Large (b)};
\end{tikzpicture}}}
\caption{\sl (a) Web diagram of $X_{3,2}$ with explicit labelling of all line segments. The Roman numerals I,$\ldots$,VI denote the different hexagons for which the consistency conditions (\ref{InitC1}) -- (\ref{InitC6}) are imposed. (b) Newton polygon and tiling of the plane: The grey area is dual to the web diagram of $X_{N,M}$. Due to the periodic identification of the web, the Newton polygon has to be periodically continued, leading to a tiling of the plane. For better readability, we have labelled the vertices of the web-diagram, as well as all of the equivalent dual faces of the Newton polygon.}
\label{Fig:Newton32}
\end{figure*}
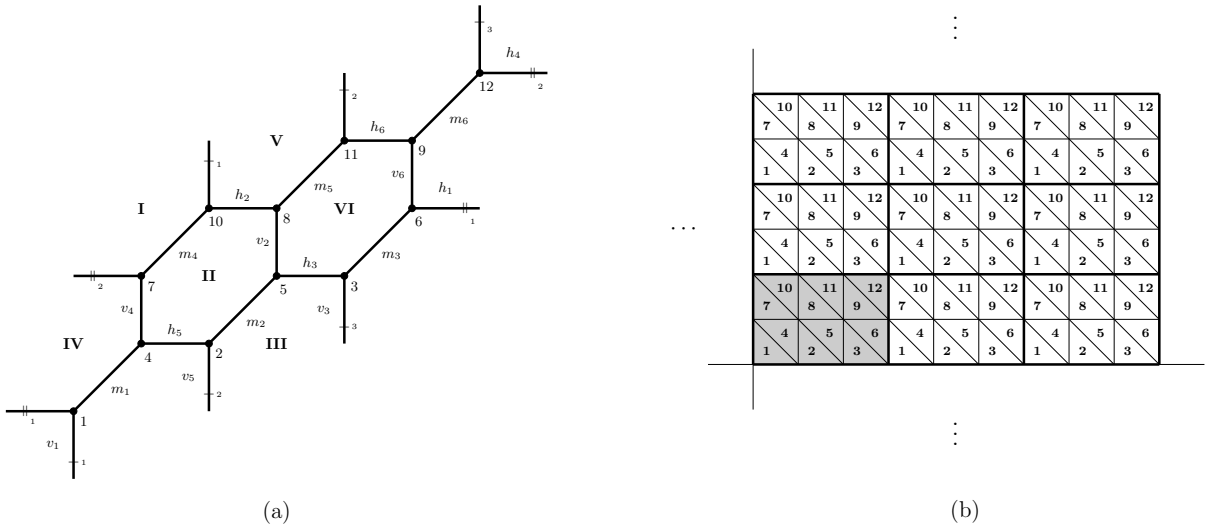

The web diagram of $X_{3,2}$ is shown in \figref{Fig:Newton32}(a) and the K\"ahler parameters labelling various rational curves are given by
\begin{align}
&h_{1,\ldots,6}\,,&&v_{1,\ldots,6}\,,&&m_{1,\ldots,6}\,.
\end{align}

\noindent
Of these 18 parameters, however, only 8 are independent: indeed, for each of the six compact divisors of $X_{3,2}$ (each of which is a $\mathbb{P}^{1}\times \mathbb{P}^{1}$ blown up at two points), we have two consistency conditions. These conditions arise because each compact divisor (represented as a hexagon in the web) has six rational curves which are toric divisors of $\mathbb{P}^{1}\times \mathbb{P}^{1}$ blown up at two points but only four of these are independent. Explicitly, we therefore obtain the following conditions:
{\allowdisplaybreaks
\begin{align}
&\bullet\, \text{hexagon I}:\nonumber\\*
&~~~~~h_4+m_4=m_3+h_1\,,v_3+m_3=m_4+v_1\,,\label{InitC1}\\
&\bullet\, \text{hexagon II}:\nonumber\\*
&~~~~~h_5+m_2=m_4+h_2\,,v_4+m_4=m_2+v_2\,,\label{InitC2}\\
&\bullet\, \text{hexagon III}:\nonumber\\*
&~~~~~h_6+m_6=m_2+h_3\,,v_5+m_2=m_6+v_3\,,\label{InitC3}\\
&\bullet\, \text{hexagon IV}:\nonumber\\*
&~~~~~h_1+m_1=m_6+h_4\,,v_6+m_6=m_1+v_4\,,\label{InitC4}\\
&\bullet\, \text{hexagon V}:\nonumber\\*
&~~~~~h_2+m_5=m_1+h_5\,,v_1+m_1=m_5+v_5\,,\label{InitC5}\\
&\bullet\, \text{hexagon VI}:\nonumber\\*
&~~~~~h_3+m_3=m_5+h_6\,,v_2+m_5=m_3+v_6\,.\label{InitC6}
\end{align}}
The above conditions indeed leave $8$ independent parameters.  However, there is a priori an ambiguity in choosing the latter. A suitable parametrisation can be found by comparison with the web diagram of $X_{6,1}$ in \figref{Fig:Flop61}, which has been shown in \cite{Hohenegger:2016yuv} to be dual to $X_{3,2}$ by flop and symmetry transforms. The $(6,1)$ diagram is parametrised in terms of the maximal set of variables $(M,V,H_{1,\ldots,6})$, which satisfies all consistency conditions. Through the explicit duality map found in \cite{Hohenegger:2016yuv}, we can also express the $(\mathbf{h},\mathbf{v},\mathbf{m})$ of the $(3,2)$ web diagram in terms of $(M,V,H_{1,\ldots,6})$
{\allowdisplaybreaks\begin{align}
&h_1=-M - H_5  - H_6 \,,\hspace{0.3cm}h_2= -M - H_1  - H_6\nonumber\\
&h_3= -M - H_1  - H_2 \,,\hspace{0.3cm}h_4= -M - H_2  - H_3\nonumber\\
&h_5= -M - H_3  - H_4 \,,\hspace{0.3cm}h_6= -M - H_4  - H_5 \nonumber\\[4pt]
&v_1= 2 M + H_1  + H_5  + H_6 \,,\hspace{0.1cm}v_2= 2 M + H_1  + H_2  + H_6 \,,\nonumber\\
&v_3= 2 M + H_1  + H_2  + H_3\,,\hspace{0.1cm}v_4= 2 M + H_2  + H_3  + H_4\,, \nonumber\\
&v_5= 2 M + H_3  + H_4  + H_5 \,,\hspace{0.1cm}v_6=2 M + H_4  + H_5  + H_6 \,,\nonumber\\[4pt]
&m_1= 2 M + H_3  + 2H_4  + 2 H_5  + H_6  + V\,,\nonumber\\
&m_2=2 M + H_1  + 2 H_2  + 2 H_3  + H_4  + V\,,\nonumber\\
&m_3= 2 M + 2 H_1  + H_2  + H_5  + 2H_6  + V\,,\nonumber\\
&m_4=2 M + 2H_1  + 2 H_2  + H_3  + H_6  + V\,,\nonumber\\
&m_5= 2 M + H_1  + H_4  + 2H_5  + 2 H_6  + V\,,\nonumber\\
&m_6=2 M + H_2  + 2 H_3  + 2 H_4  + H_5  + V\,.\label{Para323}
\end{align}}
One can indeed check that (\ref{Para323}) identically satisfies (\ref{InitC1}) -- (\ref{InitC6}). Moreover, as we shall discuss in the following, the parametrisation (\ref{Para323}) is crucial for showing the identity (\ref{PartitionFc61}). To this end, we first review the topological string partition function of $X_{6,1}$ in the following.
\subsection{The Partition Function $\pf{6}{1}$}
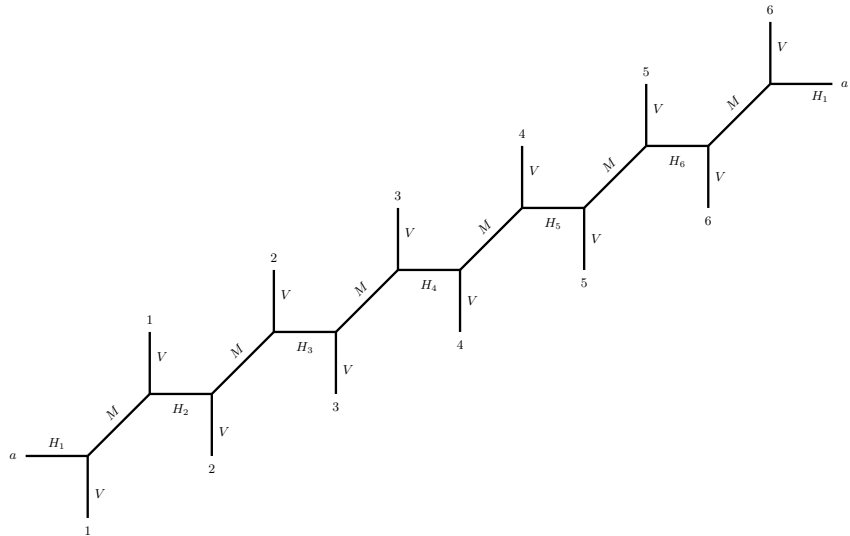
\begin{figure*}
\begin{center}
\scalebox{0.55}{\parbox{22.7cm}{\begin{tikzpicture}[scale = 1.5]
\draw[ultra thick] (5,2) -- (5,1);
\draw[ultra thick] (6,3) -- (6,4);
\draw[ultra thick] (7,3) -- (7,2);
\draw[ultra thick] (8,4) -- (8,5);
\draw[ultra thick] (9,4) -- (9,3);
\draw[ultra thick] (10,5) -- (10,6);
\draw[ultra thick] (11,5) -- (11,4);
\draw[ultra thick] (12,6) -- (12,7);
\draw[ultra thick] (13,6) -- (13,5);
\draw[ultra thick] (14,7) -- (14,8);
\draw[ultra thick] (15,7) -- (15,6);
\draw[ultra thick] (16,8) -- (16,9);
\draw[ultra thick] (4,2) -- (5,2);
\draw[ultra thick] (6,3) -- (7,3);
\draw[ultra thick] (8,4) -- (9,4);
\draw[ultra thick] (10,5) -- (11,5);
\draw[ultra thick] (12,6) -- (13,6);
\draw[ultra thick] (14,7) -- (15,7);
\draw[ultra thick] (16,8) -- (17,8);
\draw[ultra thick] (5,2) -- (6,3);
\draw[ultra thick] (7,3) -- (8,4);
\draw[ultra thick] (9,4) -- (10,5);
\draw[ultra thick] (11,5) -- (12,6);
\draw[ultra thick] (13,6) -- (14,7);
\draw[ultra thick] (15,7) -- (16,8);
\node[rotate=45] at (5.4,2.7) {{\small $M$}};
\node[rotate=45] at (7.4,3.7) {{\small $M$}};
\node[rotate=45] at (9.4,4.7) {{\small $M$}};
\node[rotate=45] at (11.4,5.7) {{\small $M$}};
\node[rotate=45] at (13.4,6.7) {{\small $M$}};
\node[rotate=45] at (15.4,7.7) {{\small $M$}};
\node at (4.5,2.2) {{\small $H_1$}};
\node at (6.5,2.75) {{\small $H_2$}};
\node at (8.5,3.75) {{\small $H_3$}};
\node at (10.5,4.75) {{\small $H_4$}};
\node at (12.5,5.75) {{\small $H_5$}};
\node at (14.5,6.75) {{\small $H_6$}};
\node at (16.8,7.8) {{\small $H_1$}};
\node at (5.2,1.4) {{\small $V$}};
\node at (7.2,2.4) {{\small $V$}};
\node at (9.2,3.4) {{\small $V$}};
\node at (11.2,4.5) {{\small $V$}};
\node at (13.2,5.5) {{\small $V$}};
\node at (15.2,6.5) {{\small $V$}};
\node at (6.2,3.6) {{\small $V$}};
\node at (8.2,4.6) {{\small $V$}};
\node at (10.2,5.6) {{\small $V$}};
\node at (12.2,6.6) {{\small $V$}};
\node at (14.2,7.6) {{\small $V$}};
\node at (16.2,8.6) {{\small $V$}};
\node at (3.8,2) {{\small \bf $a$}};
\node at (17.2,8) {{\small \bf $a$}};
\node at (5,0.8) {{\small \bf $1$}};
\node at (7,1.8) {{\small \bf $2$}};
\node at (9,2.8) {{\small \bf $3$}};
\node at (11,3.8) {{\small \bf $4$}};
\node at (13,4.8) {{\small \bf $5$}};
\node at (15,5.8) {{\small \bf $6$}};
\node at (6,4.2) {{\small \bf $1$}};
\node at (8,5.2) {{\small \bf $2$}};
\node at (10,6.2) {{\small \bf $3$}};
\node at (12,7.2) {{\small \bf $4$}};
\node at (14,8.2) {{\small \bf $5$}};
\node at (16,9.2) {{\small \bf $6$}};
%
%
%
%
%
%
%
%
%
\end{tikzpicture}}
}
\end{center}
\caption{\sl Toric diagram dual to the configuration $(3,2)$. This configuration is obtained \cite{Hohenegger:2016yuv} by performing a flop transformation on the line segments $h_i$ and $v_i$ (for $i=1,\ldots,6$) (and a combination of additional symmetry transforms) in the original diagram in the left portion of \figref{Fig:Newton32}. Notice that the parameters $(M,V,H_{1,\ldots,6})$ are the same as in  \figref{Fig:Newton32}.}
\label{Fig:Flop61}
\end{figure*}

The $(6,1)$ web is shown in \figref{Fig:Flop61} in which the K\"ahler parameters are also labelled.
Here, the parameters $(M,V,H_{1,\ldots,6})$ are the same that appear in (\ref{Para323}) (after applying the duality map proposed in \cite{Hohenegger:2016yuv}). Furthermore, the partition function $\pf{6}{1}$ was explicitly calculated in \cite{Haghighat:2013gba,Haghighat:2013tka,Hohenegger:2013ala} using the (refined) topological vertex \cite{Aganagic:2003db, Iqbal:2007ii}. The latter has a preferred direction which can be chosen in different ways to obtain different series expansions of $\pf{6}{1}$. For example, choosing the preferred direction to be vertical (\emph{i.e.} along the $(0,1)$-direction in the web diagram in \figref{Fig:Flop61}), the topological string partition function takes the form
{\allowdisplaybreaks
\begin{align}
&\pf{6}{1}(V,M,H_i,\epsilon_{1,2})=W_6(\emptyset)\label{Zflop32Part}\\
&\times \sum_{\alpha_{1},\ldots,\alpha_{6}}(-Q_V Q_M)^{|\alpha_1|+\ldots+|\alpha_6|}\left(\prod_{a=1}^6\frac{\vartheta_{\alpha_a\alpha_a}(Q_M;\rho)}{\vartheta_{\alpha_a\alpha_a}(\sqrt{t/q};\rho)}\right)\nonumber\\
&\times\left(\prod_{1\leq a<b\leq 6}\frac{\vartheta_{\alpha_a\alpha_b}(Q_{ab}Q_M^{-1};\rho)\,\vartheta_{\alpha_a\alpha_b}(Q_{ab}Q_M;\rho)}{\vartheta_{\alpha_a\alpha_b}(Q_{ab}\sqrt{t/q};\rho)\,\vartheta_{\alpha_a\alpha_b}(Q_{ab}\sqrt{q/t};\rho)}\right), \nonumber
\end{align}}
where we adopted the following definitions
{ \allowdisplaybreaks
\begin{align}
&Q_V=e^{-V}\,, &&Q_M=e^{-M}\,, &&Q_{H_i}=e^{-H_i}\,,\nonumber\\
&q=e^{2\pi i\epsilon_1}\,,&&t=e^{-2\pi i\epsilon_2}\,.\label{Notation1}
\end{align}}
and
{\allowdisplaybreaks
\begin{align}
&Q_{\rho}=e^{2\pi i \rho}=e^{-6M-\sum_{i=1}^6 H_i}\,,\nonumber
\\
&Q_{ab}=Q_M^{b-a}Q_{H_a} \dots Q_{H_{b-1}}\,,&&\forall\, 1\leq a< b\,.\label{DefQab}
\end{align}}
The summation is over integer partitions $\alpha_{1,\ldots,6}$ and $\vartheta_{\mu\nu}$ is defined in equation~(\ref{DefTheta1}) in appendix~\ref{App:Identities}.
\subsection{Diagonal Expansion of $\pf{3}{2}$}
From the viewpoint of independent variables $(M,V,H_{1,\ldots,6})$, the (vertical) partition function (\ref{Zflop32Part}) is written as a power series expansion in $Q_V$, since neither of the $\vartheta$-functions depends on $V$. Thus, in order to match (\ref{PartitionFc61}) order by order in $Q_V$, we have to write $\pf{3}{2}$ in a similar fashion. Upon inspection of (\ref{Para323}), this can indeed be achieved by choosing the preferred direction of the refined topological vertex to be along $(1,1)$ (\emph{i.e.} diagonally in the web diagram in \figref{Fig:Newton32}(a)). Indeed, in this manner, the partition function is seen to take the form
\begin{widetext}
\begin{align}
\pf{3}{2}(\mathbf{h},\mathbf{v},\mathbf{m},\epsilon_{1,2})=\sum_{\alpha_1,\ldots,\alpha_6}Q_{m_1}^{|\alpha_4|}\,Q_{m_2}^{|\alpha_2|}\,Q_{m_3}^{|\alpha_6|}\,Q_{m_4}^{|\alpha_1|}\,Q_{m_5}^{|\alpha_5|}\,Q_{m_6}^{|\alpha_3|}\,W_{\alpha_1\,\alpha_2\,\alpha_3\,\alpha_4\,\alpha_5\,\alpha_6} ^{\alpha_4\,\alpha_5\,\alpha_6\,\alpha_1\,\alpha_2\,\alpha_3}(Q_{h_i},Q_{v_i},\epsilon_{1,2})\,,\label{PartFunc}
\end{align}
\end{widetext}
where $\alpha_{1,\ldots,6}$ are integer partitions. Furthermore, we used the notation (for $i=1,\ldots,6$)
\begin{align}
&Q_{h_i}=e^{-h_i}\,,&&Q_{v_i}=e^{-v_i}\,,&&Q_{m_i}=e^{-m_i}\,,\label{Qnotation}
\end{align}
together with (\ref{Notation1}). The coefficient $W_{\alpha_1\,\alpha_2\,\alpha_3\,\alpha_4\,\alpha_5\,\alpha_6} ^{\alpha_4\,\alpha_5\,\alpha_6\,\alpha_1\,\alpha_2\,\alpha_3}$ shall be computed in the following. As neither $\mathbf{h}$ nor $\mathbf{v}$ depend on $V$, (\ref{PartFunc}) is also a series expansion in $e^{-V}$, which can be compared order by order to (\ref{Zflop32Part}) thereby allowing to verify (\ref{PartitionFc61}).

\begin{figure*}
\begin{center}
\scalebox{0.65}{\parbox{11cm}{\begin{tikzpicture}[scale = 1.0]
\draw[fill=green!50!white] (0,5) -- (6,-1) -- (6,-2) -- (0,4) -- (0,5);
\draw[-] (-1,6) -- (7,6);
\draw[-] (0,-3) -- (0,7);
\draw[ultra thick] (0,-2) -- (6,-2);
\draw[-] (0,-1) -- (6,-1);
\draw[ultra thick] (0,0) -- (6,0);
\draw[-] (0,1) -- (6,1);
\draw[ultra thick] (0,2) -- (6,2);
\draw[-] (0,3) -- (6,3);
\draw[ultra thick] (0,4) -- (6,4);
\draw[-] (0,5) -- (6,5);
\draw[ultra thick] (0,6) -- (6,6);
\draw[ultra thick] (0,-2) -- (0,6);
\draw[-] (1,-2) -- (1,6);
\draw[-] (2,-2) -- (2,6);
\draw[ultra thick] (3,-2) -- (3,6);
\draw[-] (4,-2) -- (4,6);
\draw[-] (5,-2) -- (5,6);
\draw[ultra thick] (6,-2) -- (6,6);
\draw[-] (0,-1) -- (1,-2);
\draw[-] (0,0) -- (2,-2);
\draw[-] (0,1) -- (3,-2);
\draw[-] (0,2) -- (4,-2);
\draw[-] (0,3) -- (5,-2);
\draw[-] (0,4) -- (6,-2);
\draw[-] (0,5) -- (6,-1);
\draw[-] (0,6) -- (6,0);
\draw[-] (1,6) -- (6,1);
\draw[-] (2,6) -- (6,2);
\draw[-] (3,6) -- (6,3);
\draw[-] (4,6) -- (6,4);
\draw[-] (5,6) -- (6,5);
\node[rotate=90] at (3,-3) {\Large $\cdots$};
\node[rotate=90] at (3,7) {\Large $\cdots$};
\node at (-1,2) {\Large $\cdots$};
\node at (8,2) {\Large $\cdots$};
\node at (0.3,-1.7) {\footnotesize \bf 1};
\node at (1.3,-1.7) {\footnotesize \bf 2};
\node at (2.3,-1.7) {\footnotesize \bf 3};
\node at (3.3,-1.7) {\footnotesize \bf 1};
\node at (4.3,-1.7) {\footnotesize \bf 2};
\node at (5.3,-1.7) {\footnotesize \bf 3};
\node at (0.7,-1.3) {\footnotesize \bf 4};
\node at (1.7,-1.3) {\footnotesize \bf 5};
\node at (2.7,-1.3) {\footnotesize \bf 6};
\node at (3.7,-1.3) {\footnotesize \bf 4};
\node at (4.7,-1.3) {\footnotesize \bf 5};
\node at (5.7,-1.3) {\footnotesize \bf 6};
\node at (0.3,-0.7) {\footnotesize \bf 7};
\node at (1.3,-0.7) {\footnotesize \bf 8};
\node at (2.3,-0.7) {\footnotesize \bf 9};
\node at (3.3,-0.7) {\footnotesize \bf 7};
\node at (4.3,-0.7) {\footnotesize \bf 8};
\node at (5.3,-0.7) {\footnotesize \bf 9};
\node at (0.7,-0.3) {\footnotesize \bf 10};
\node at (1.7,-0.3) {\footnotesize \bf 11};
\node at (2.7,-0.3) {\footnotesize \bf 12};
\node at (3.7,-0.3) {\footnotesize \bf 10};
\node at (4.7,-0.3) {\footnotesize \bf 11};
\node at (5.7,-0.3) {\footnotesize \bf 12};
\node at (0.3,0.3) {\footnotesize \bf 1};
\node at (1.3,0.3) {\footnotesize \bf 2};
\node at (2.3,0.3) {\footnotesize \bf 3};
\node at (3.3,0.3) {\footnotesize \bf 1};
\node at (4.3,0.3) {\footnotesize \bf 2};
\node at (5.3,0.3) {\footnotesize \bf 3};
\node at (0.7,0.7) {\footnotesize \bf 4};
\node at (1.7,0.7) {\footnotesize \bf 5};
\node at (2.7,0.7) {\footnotesize \bf 6};
\node at (3.7,0.7) {\footnotesize \bf 4};
\node at (4.7,0.7) {\footnotesize \bf 5};
\node at (5.7,0.7) {\footnotesize \bf 6};
\node at (0.3,1.3) {\footnotesize \bf 7};
\node at (1.3,1.3) {\footnotesize \bf 8};
\node at (2.3,1.3) {\footnotesize \bf 9};
\node at (3.3,1.3) {\footnotesize \bf 7};
\node at (4.3,1.3) {\footnotesize \bf 8};
\node at (5.3,1.3) {\footnotesize \bf 9};
\node at (0.7,1.7) {\footnotesize \bf 10};
\node at (1.7,1.7) {\footnotesize \bf 11};
\node at (2.7,1.7) {\footnotesize \bf 12};
\node at (3.7,1.7) {\footnotesize \bf 10};
\node at (4.7,1.7) {\footnotesize \bf 11};
\node at (5.7,1.7) {\footnotesize \bf 12};
\node at (0.3,2.3) {\footnotesize \bf 1};
\node at (1.3,2.3) {\footnotesize \bf 2};
\node at (2.3,2.3) {\footnotesize \bf 3};
\node at (3.3,2.3) {\footnotesize \bf 1};
\node at (4.3,2.3) {\footnotesize \bf 2};
\node at (5.3,2.3) {\footnotesize \bf 3};
\node at (0.7,2.7) {\footnotesize \bf 4};
\node at (1.7,2.7) {\footnotesize \bf 5};
\node at (2.7,2.7) {\footnotesize \bf 6};
\node at (3.7,2.7) {\footnotesize \bf 4};
\node at (4.7,2.7) {\footnotesize \bf 5};
\node at (5.7,2.7) {\footnotesize \bf 6};
\node at (0.3,3.3) {\footnotesize \bf 7};
\node at (1.3,3.3) {\footnotesize \bf 8};
\node at (2.3,3.3) {\footnotesize \bf 9};
\node at (3.3,3.3) {\footnotesize \bf 7};
\node at (4.3,3.3) {\footnotesize \bf 8};
\node at (5.3,3.3) {\footnotesize \bf 9};
\node at (0.7,3.7) {\footnotesize \bf 10};
\node at (1.7,3.7) {\footnotesize \bf 11};
\node at (2.7,3.7) {\footnotesize \bf 12};
\node at (3.7,3.7) {\footnotesize \bf 10};
\node at (4.7,3.7) {\footnotesize \bf 11};
\node at (5.7,3.7) {\footnotesize \bf 12};
\node at (0.3,4.3) {\footnotesize \bf 1};
\node at (1.3,4.3) {\footnotesize \bf 2};
\node at (2.3,4.3) {\footnotesize \bf 3};
\node at (3.3,4.3) {\footnotesize \bf 1};
\node at (4.3,4.3) {\footnotesize \bf 2};
\node at (5.3,4.3) {\footnotesize \bf 3};
\node at (0.7,4.7) {\footnotesize \bf 4};
\node at (1.7,4.7) {\footnotesize \bf 5};
\node at (2.7,4.7) {\footnotesize \bf 6};
\node at (3.7,4.7) {\footnotesize \bf 4};
\node at (4.7,4.7) {\footnotesize \bf 5};
\node at (5.7,4.7) {\footnotesize \bf 6};
\node at (0.3,5.3) {\footnotesize \bf 7};
\node at (1.3,5.3) {\footnotesize \bf 8};
\node at (2.3,5.3) {\footnotesize \bf 9};
\node at (3.3,5.3) {\footnotesize \bf 7};
\node at (4.3,5.3) {\footnotesize \bf 8};
\node at (5.3,5.3) {\footnotesize \bf 9};
\node at (0.7,5.7) {\footnotesize \bf 10};
\node at (1.7,5.7) {\footnotesize \bf 11};
\node at (2.7,5.7) {\footnotesize \bf 12};
\node at (3.7,5.7) {\footnotesize \bf 10};
\node at (4.7,5.7) {\footnotesize \bf 11};
\node at (5.7,5.7) {\footnotesize \bf 12};
\node at (3,-4.3) {\Large (a)};
\end{tikzpicture}}}
\hspace{1.25cm}
\scalebox{0.6}{\parbox{12cm}{\begin{tikzpicture}[scale = 1.50]
\draw[ultra thick] (0,0) -- (1,0) -- (1,-1) -- (2,-1) -- (2,-2) -- (3,-2) -- (3,-3) -- (4,-3) -- (4,-4) -- (5,-4) -- (5,-5) -- (6,-5) -- (6,-6) -- (7,-6);
\draw[ultra thick] (1,0) -- (1.5,0.5);
\draw[ultra thick] (2,-1) -- (2.5,-0.5);
\draw[ultra thick] (3,-2) -- (3.5,-1.5);
\draw[ultra thick] (4,-3) -- (4.5,-2.5);
\draw[ultra thick] (5,-4) -- (5.5,-3.5);
\draw[ultra thick] (6,-5) -- (6.5,-4.5);
\draw[ultra thick] (1,-1) -- (0.5,-1.5);
\draw[ultra thick] (2,-2) -- (1.5,-2.5);
\draw[ultra thick] (3,-3) -- (2.5,-3.5);
\draw[ultra thick] (4,-4) -- (3.5,-4.5);
\draw[ultra thick] (5,-5) -- (4.5,-5.5);
\draw[ultra thick] (6,-6) -- (5.5,-6.5);
\node at (1.7,0.6) {{\small  $4$}};
\node at (2.7,-0.4) {{\small $5$}};
\node at (3.7,-1.4) {{\small $6$}};
\node at (4.7,-2.4) {{\small $1$}};
\node at (5.7,-3.4) {{\small $2$}};
\node at (6.7,-4.4) {{\small $3$}};
\node at (0.3,-1.7) {{\small $1$}};
\node at (1.3,-2.7) {{\small $2$}};
\node at (2.3,-3.7) {{\small $3$}};
\node at (3.3,-4.7) {{\small $4$}};
\node at (4.3,-5.7) {{\small $5$}};
\node at (5.3,-6.7) {{\small $6$}};
\node at (-0.2,0) {{\small \bf $a$}};
\node at (7.2,-6) {{\small \bf $a$}};
\node at (0.5,0.2) {{\small $h_1$}};
\node at (1.5,-0.8) {{\small $h_2$}};
\node at (2.5,-1.8) {{\small $h_3$}};
\node at (3.5,-2.8) {{\small $h_4$}};
\node at (4.5,-3.8) {{\small $h_5$}};
\node at (5.5,-4.8) {{\small $h_6$}};
\node at (6.5,-5.8) {{\small $h_1$}};
\node at (0.8,-0.5) {{\small $v_1$}};
\node at (1.8,-1.5) {{\small $v_2$}};
\node at (2.8,-2.5) {{\small $v_3$}};
\node at (3.8,-3.5) {{\small $v_4$}};
\node at (4.8,-4.5) {{\small $v_5$}};
\node at (5.8,-5.5) {{\small $v_6$}};
\node at  (1.5,0.1) {{\small $m_1$}};
\node at  (2.5,-0.9) {{\small $m_5$}};
\node at  (3.5,-1.9) {{\small $m_3$}};
\node at  (4.5,-2.9) {{\small $m_4$}};
\node at  (5.5,-3.9) {{\small $m_2$}};
\node at  (6.5,-4.9) {{\small $m_6$}};
\node at (0.6,-1) {{\small $m_4$}};
\node at (1.6,-2) {{\small $m_2$}};
\node at (2.6,-3) {{\small $m_6$}};
\node at (3.6,-4) {{\small $m_1$}};
\node at (4.6,-5) {{\small $m_5$}};
\node at (5.6,-6) {{\small $m_3$}};
\node at (3,-7.8) {\Large (b)};
\end{tikzpicture}}}
\end{center}
\caption{\sl (a) Different choice of fundamental polygon in the tiling of the plane for the configuration $(N,M)=(3,2)$. (b) The twisted web diagram as the dual of the green-colored fundamental domain.}
\label{Fig:DiagNewtonPoly}
\end{figure*}
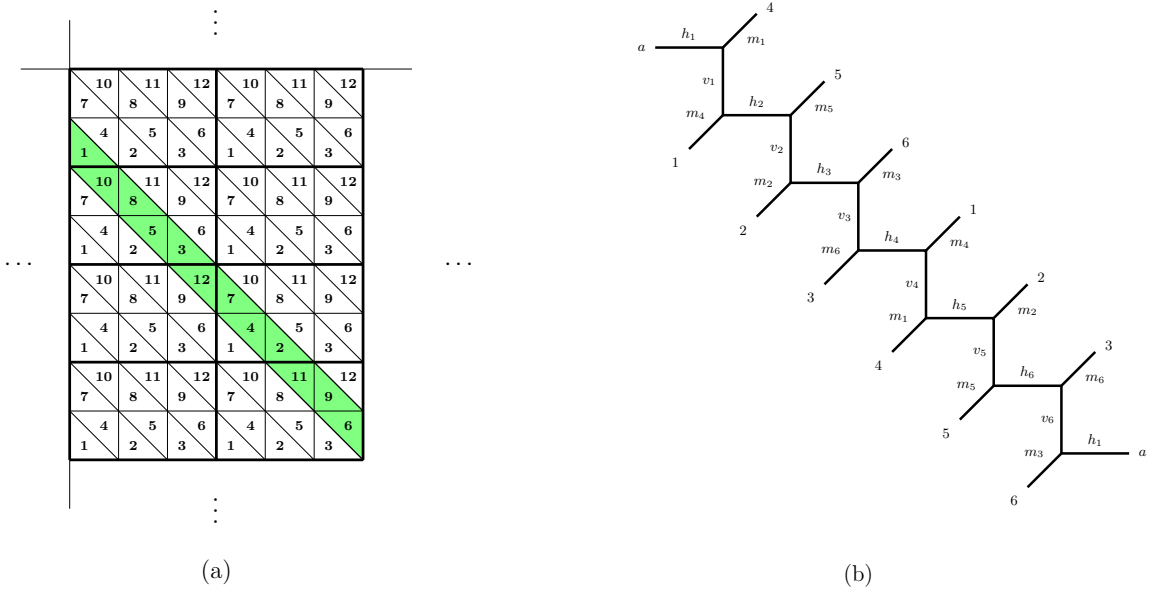

To our knowledge, while the topological string partition function has been studied in recent literature~\cite{Haghighat:2013gba,Haghighat:2013tka,Hohenegger:2015btj,Hohenegger:2016yuv,Hohenegger:2016eqy,Hohenegger:2013ala,Bastian:2017jje}, diagonal expansions of the form (\ref{PartFunc}) for  $\pf{N}{M}$ have not been studied so far. From geometric perspective, this expansion has a very natural interpretation: as the vertical and horizontal legs of the $(3,2)$ web diagram are pairwise glued together, the corresponding Newton polygon should be considered on a torus, equivalently,  a tiling of the plane should be considered with the fundamental domain (the tile) given in \figref{Fig:Newton32}(b). The choice of fundamental domain, however, is not unique and different choices lead to different (but equivalent) presentations of the web. An alternative such choice is shown in \figref{Fig:DiagNewtonPoly}(a). Indeed, the green region contains all 12 faces exactly once. This new fundamental domain, however, gives rise to an equivalent representations of the web as given in the \figref{Fig:DiagNewtonPoly}(b), which we refer to as the ``twisted" $(6,1)$ web. After an $SL(2,\mathbb{Z})$ transformation, this twisted $(6,1)$ web becomes very similar to the $(6,1)$ web in \figref{Fig:Flop61} except that the upper diagonal and lower diagonal legs that are glued together are not aligned (Indeed, it was shown in \cite{Hohenegger:2016yuv} that, by a combination of symmetry and flop transforms, the twisted $(6,1)$ web can be converted into the standard $(6,1)$ web which is shown in \figref{Fig:Flop61}). 


\subsection{Diagonal Partition Function}
Using the presentation of $(3,2)$ diagram shown in \figref{Fig:DiagNewtonPoly}(b), it remains to compute the coefficient $W_{\alpha_1\,\alpha_2\,\alpha_3\,\alpha_4\,\alpha_5\,\alpha_6} ^{\alpha_4\,\alpha_5\,\alpha_6\,\alpha_1\,\alpha_2\,\alpha_3}$ in (\ref{PartFunc}) to obtain $\pf{3}{2}$. To facilitate this computation (and also to explain the structure of the $\alpha_i$ that appear in this expression), we redraw \figref{Fig:DiagNewtonPoly}(b) in \figref{Twisted61} to include the integer partitions associated with each interval. Here, we have also indicated the parameter $\rho$, which shall play an important role in the explicit computation below.
\begin{figure*}
\begin{center}
\scalebox{0.65}{\parbox{13cm}{\begin{tikzpicture}[scale = 1.50]
\draw[ultra thick] (0,0) -- (1,0) -- (1,-1) -- (2,-1) -- (2,-2) -- (3,-2) -- (3,-3) -- (4,-3) -- (4,-4) -- (5,-4) -- (5,-5) -- (6,-5) -- (6,-6) -- (7,-6);
\draw[ultra thick] (1,0) -- (1.5,0.5);
\draw[ultra thick] (2,-1) -- (2.5,-0.5);
\draw[ultra thick] (3,-2) -- (3.5,-1.5);
\draw[ultra thick] (4,-3) -- (4.5,-2.5);
\draw[ultra thick] (5,-4) -- (5.5,-3.5);
\draw[ultra thick] (6,-5) -- (6.5,-4.5);
\draw[ultra thick] (1,-1) -- (0.5,-1.5);
\draw[ultra thick] (2,-2) -- (1.5,-2.5);
\draw[ultra thick] (3,-3) -- (2.5,-3.5);
\draw[ultra thick] (4,-4) -- (3.5,-4.5);
\draw[ultra thick] (5,-5) -- (4.5,-5.5);
\draw[ultra thick] (6,-6) -- (5.5,-6.5);
\node at (1.7,0.6) {{\small  $\alpha_4$}};
\node at (2.7,-0.4) {{\small $\alpha_5$}};
\node at (3.7,-1.4) {{\small $\alpha_6$}};
\node at (4.7,-2.4) {{\small $\alpha_1$}};
\node at (5.7,-3.4) {{\small $\alpha_2$}};
\node at (6.7,-4.4) {{\small $\alpha_3$}};
\node at (0.3,-1.7) {{\small $\alpha^t_1$}};
\node at (1.3,-2.7) {{\small $\alpha^t_2$}};
\node at (2.3,-3.7) {{\small $\alpha^t_3$}};
\node at (3.3,-4.7) {{\small $\alpha^t_4$}};
\node at (4.3,-5.7) {{\small $\alpha^t_5$}};
\node at (5.3,-6.7) {{\small $\alpha^t_6$}};
\node at (-0.2,0) {{\small \bf $a$}};
\node at (7.2,-6) {{\small \bf $a$}};
\node at (0.5,0.2) {{\small $h_1,\mu_1$}};
\node at (1.5,-0.8) {{\small $h_2,\mu_2$}};
\node at (2.5,-1.8) {{\small $h_3,\mu_3$}};
\node at (3.5,-2.8) {{\small $h_4,\mu_4$}};
\node at (4.5,-3.8) {{\small $h_5,\mu_5$}};
\node at (5.5,-4.8) {{\small $h_6,\mu_6$}};
\node at (6.5,-5.8) {{\small $h_1,\mu_1$}};
\node at (0.65,-0.5) {{\small $v_1,\nu_1$}};
\node at (1.65,-1.5) {{\small $v_2,\nu_2$}};
\node at (2.65,-2.5) {{\small $v_3,\nu_3$}};
\node at (3.65,-3.5) {{\small $v_4,\nu_4$}};
\node at (4.65,-4.5) {{\small $v_5,\nu_5$}};
\node at (5.65,-5.5) {{\small $v_6,\nu_6$}};
\node at  (1.5,0.1) {{\small $m_1$}};
\node at  (2.5,-0.9) {{\small $m_5$}};
\node at  (3.5,-1.9) {{\small $m_3$}};
\node at  (4.5,-2.9) {{\small $m_4$}};
\node at  (5.5,-3.9) {{\small $m_2$}};
\node at  (6.5,-4.9) {{\small $m_6$}};
\node at (0.6,-1) {{\small $m_4$}};
\node at (1.6,-2) {{\small $m_2$}};
\node at (3,-3.5) {{\small $m_6$}};
\node at (3.6,-4) {{\small $m_1$}};
\node at (4.6,-5) {{\small $m_5$}};
\node at (5.6,-6) {{\small $m_3$}};
\draw[thick,red,<->] (2.05,0.95) -- (2.95,0.05);
\node[red,rotate=315] at (2.7,0.55) {{\small $H_5+M$}};
\draw[thick,red,<->] (3.05,-0.05) -- (3.95,-0.95);
\node[red,rotate=315] at (3.7,-0.45) {{\small $H_6+M$}};
\draw[thick,red,<->] (4.05,-1.05) -- (4.95,-1.95);
\node[red,rotate=315] at (4.7,-1.45) {{\small $H_1+M$}};
\draw[thick,red,<->] (5.05,-2.05) -- (5.95,-2.95);
\node[red,rotate=315] at (5.7,-2.45) {{\small $H_2+M$}};
\draw[thick,red,<->] (6.05,-3.05) -- (6.95,-3.95);
\node[red,rotate=315] at (6.7,-3.45) {{\small $H_3+M$}};
\draw[thick,red,<->] (7.05,-4.05) -- (7.95,-4.95);
\node[red,rotate=315] at (7.7,-4.45) {{\small $H_4+M$}};
\draw[thick,red,<->] (0,-2) -- (2.5,-4.5);
\node[red] at (1.1,-3.5) {{\small $M$}};
\draw[thick,red,<->] (-0.5,-1.55) -- (-0.5,-2.95);
\node[red,rotate=90] at (-0.3,-2.3) {{\small $V-2M$}};
\draw[dashed] (-0.5,-1.5) -- (0.5,-1.5);
\draw[dashed] (-0.5,-3) -- (4,-3);
\end{tikzpicture}}}
\end{center}
\caption{\sl Detailed parametrisation of the twisted web diagram associated with $X_{3,2}$: The choice of K\"ahler parameters $(M,V,H_{1,\ldots,6})$ is inspired by comparison to the web diagram of $X_{6,1}$ shown in \figref{Fig:Newton32}. Furthermore, the diagonal intervals are labelled by integer partitions $\alpha_{1,\ldots,6}$ (and their transposed versions $\alpha^t_{1,\ldots,6}$), which indicate the gluing of the web.}
\label{Twisted61}
\end{figure*}
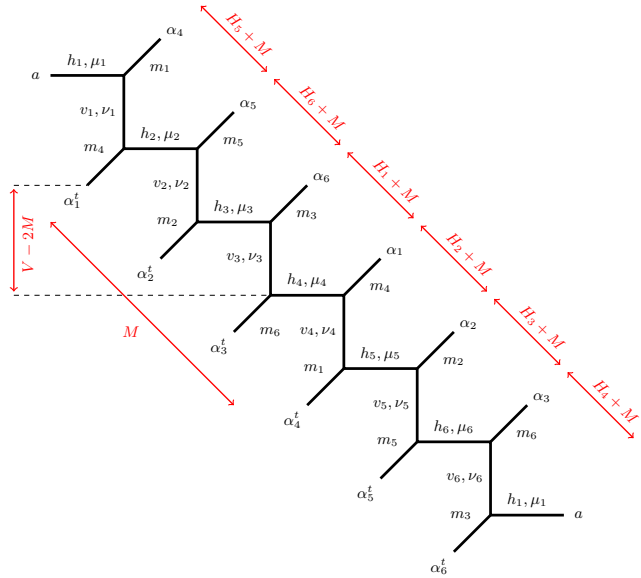
\begin{widetext}
With this notation, we have the following expression for $W$ in terms of the refined topological vertex
{\allowdisplaybreaks
\begin{align}
&W_{\alpha_1\,\alpha_2\,\alpha_3\,\alpha_4\,\alpha_5\,\alpha_6}^{\alpha_4\,\alpha_5\,\alpha_6\,\alpha_1\,\alpha_2\,\alpha_3}
(Q_{h_i},Q_{v_i},\epsilon_1,\epsilon_2)=\sum_{{\mu_1,\ldots,\mu_6}\atop{\nu_1,\ldots,\nu_6}}(-Q_{h_1})^{|\mu_1|}\,(-Q_{h_2})^{|\mu_2|}\,(-Q_{h_3})^{|\mu_3|}\,(-Q_{h_4})^{|\mu_4|}\,\nonumber\\
&\hspace{0.5cm}\times (-Q_{h_5})^{|\mu_5|}\, (-Q_{h_6})^{|\mu_6|}\,(-Q_{v_1})^{|\nu_1|}\,(-Q_{v_2})^{|\nu_2|}\,(-Q_{v_3})^{|\nu_3|}\,(-Q_{v_4})^{|\nu_4|}\,(-Q_{v_5})^{|\nu_5|}\,(-Q_{v_6})^{|\nu_6|}\,\nonumber\\
&\hspace{0.5cm}\times C_{\mu^t_1\nu_1\alpha_4}(q,t) \,C_{\mu_2\nu^t_1,\alpha^t_1}(t,q)\, C_{\mu^t_2\nu_2\alpha_5}(q,t)\,C_{\mu_3\nu_2^t\alpha^t_2}(t,q)\,C_{\mu^t_6\nu_3\alpha_3}(q,t)\,C_{\mu_4\nu^t_3\alpha^t_3}(t,q)\,\nonumber\\
&\hspace{0.5cm}\times C_{\mu^t_4\nu_4\alpha_1}(q,t)\,C_{\mu_5\nu^t_4\alpha_4^t}(t,q)\, C_{\mu_5^t\nu_5\alpha_2}(q,t)\, C_{\mu_6\nu_5^t\alpha_5^t}(t,q)\,C_{\mu^t_6\nu_6\alpha_3}(q,t)\,C_{\mu_1\nu_6^t\alpha_6^t}(t,q)\,. \label{Wblock}
\end{align}}
\end{widetext}

\noindent
Information of our conventions regarding integer partitions (and functional identities which shall be important in the following) is compiled in appendix~\ref{App:Identities},
and the refined topological vertex $C_{\mu\nu\rho}(q,t)$ is defined in (\ref{DefRefinedVertex}). Using the expression for the latter, we can write (\ref{Wblock}) as
\begin{widetext}
\vskip-0.3cm
\begin{align}
&W_{\alpha_1\,\alpha_2\,\alpha_3\,\alpha_4\,\alpha_5\,\alpha_6}^{\alpha_4\,\alpha_5\,\alpha_6\,\alpha_1\,\alpha_2\,\alpha_3}(Q_{h_i},Q_{v_i},\epsilon_1,\epsilon_2)= \Big( \prod_{k=1}^6 t^{\frac{||\alpha_k||^2}{2}} q^{\frac{||\alpha_k^t||^2}{2}} \tilde{Z}_{\alpha_k}(q,t) \tilde{Z}_{\alpha_k^t}(t,q)  \Big) \nonumber \\
&\hspace{1cm}\times \sum_{\substack{\{\mu\} \{\nu\} \\ \{\eta\} \{\tilde{\eta}\}}} \prod_{i=1}^6 (-Q_{h_i})^{|\mu_i|} (-Q_{v_i})^{|\nu_i|} s_{\mu_i/ \tilde{\eta}_{i+3}}(x_{i+3}) s_{\mu_i^t/ \eta_{i+5}}(y_{i+5}) s_{\nu_i^t/ \eta_i}(w_i) s_{\nu_i/ \tilde{\eta}_{i+3}}(z_{i+3})\,,  \label{W}
\end{align}
\end{widetext}
where $s_{\mu/\nu}$ are (skew) Schur functions (see \cite{Macdo} for the definition). The partitions in (\ref{W}) are cyclically identified (\emph{e.g.} $\eta_{i+6}=\eta_i$) and variables are defined as
\begin{align}
&x_i= q^{-\rho+\frac{1}{2}}t^{-\alpha_{i}-\frac{1}{2}}\,, &&y_i=t^{-\rho+\frac{1}{2}}q^{-\alpha_{i}^t-\frac{1}{2}}\,,\nonumber\\
&w_i= q^{-\rho}t^{-\alpha_i }\,, &&z_i=t^{-\rho}q^{-\alpha_{i}^t }\,.
\end{align}
Using the identities given in Appendix~\ref{App:Identities}, we can write Eq.(\ref{W}) in the form (see Appendix~\ref{App:32ExpComputation} for details)
\begin{widetext}
\begin{align}
&W_{\alpha_1\,\alpha_2\,\alpha_3\,\alpha_4\,\alpha_5\,\alpha_6} ^{\alpha_4\,\alpha_5\,\alpha_6\,\alpha_1\,\alpha_2\,\alpha_3}= W_{2,3}(\emptyset)\, \left[ \left( \frac{t}{q} \right)^{\frac{5}{2}}\prod_{l=1}^6 Q_{h_l} \right]^{\sum_{i=1}^6|\alpha_i|}  \prod_{i,j=1}^6  \prod_{(r,s)\in \alpha_i} \frac{\theta_1(\rho,\hat{z}^{(i,j)}_{r,s})\theta_1(\rho,\hat{u}^{(i,j)}_{r,s})}{\theta_1(\rho,\hat{w}^{(i,j)}_{r,s})\theta_1(\rho,\hat{v}^{(i,j)}_{r,s})}\,, \label{blockW}
\end{align}
\end{widetext}
where $\theta_1$ is the Jacobi-theta function
\begin{widetext}
{\allowdisplaybreaks
\begin{align}
\theta_1&(\rho;z)=-ie^{\frac{i\pi \tau}{4}} e^{i \pi z} \prod_{k=1}^{\infty} (1-e^{2 \pi i k \rho})\, (1-e^{2 \pi i k \rho}e^{2\pi i z})(1-e^{2 \pi i (k-1)\rho}e^{-2\pi i z})\,, \nonumber
\end{align}}
\end{widetext}
and $W_{2,3}(\emptyset)=W_6(\emptyset)=W^{\emptyset\emptyset\emptyset\emptyset\emptyset\emptyset}_{\emptyset\emptyset\emptyset\emptyset\emptyset\emptyset}$ is a normalisation factor (see \cite{Haghighat:2013gba,Hohenegger:2013ala} and Eq.(\ref{WemptyDef}) for $L=6$ below). Furthermore, the arguments of theta functions are given by
{\allowdisplaybreaks
\begin{align}
&\hat{z}^{(i,j)}_{r,s}= \frac{\beta_{i+3,j}}{2 \pi i} + \epsilon_1 (\alpha_{i+3-j,s}^t-r+\tfrac{1}{2}) - \epsilon_2(\alpha_{i,r}-s +\tfrac{1}{2})\,, \nonumber \\
&\hat{u}^{(i,j)}_{r,s}=  \frac{\beta_{i+j,j}}{2 \pi i} - \epsilon_1 (\alpha_{i+j-3,s}^t-r+\tfrac{1}{2}) + \epsilon_2(\alpha_{i,r}-s + \tfrac{1}{2})\,,   \nonumber \\
&\hat{w}^{(i,j)}_{r,s}=  \frac{\lambda_{i+3,j}}{2 \pi i} + \epsilon_1 (\alpha_{i-j,s}^t-r) - \epsilon_2(\alpha_{i,r}-s +1)\,,  \nonumber \\
&\hat{v}^{(i,j)}_{r,s}=   \frac{\lambda_{i+j-3,j}}{2 \pi i}   - \epsilon_1 (\alpha_{i+j,s}^t-r+1) + \epsilon_2(\alpha_{i,r}-s )\,
\nonumber
\end{align}}
\noindent with the shorthand notation 
{\allowdisplaybreaks
\begin{align}
&\beta_{i,j}=h_i+\sum_{k=1}^{j-1}(h_{i-k}+v_{i-k})\,,
\nonumber \\
&\lambda_{i,j}= \begin{cases}
							0 & \quad \text{if } j=6 \\
							h_i+\sum_{k=1}^{j-1}(h_{i-k}+v_{i-k})+v_{i-j} & \quad \text{else}  \\
						\end{cases}\nonumber
\end{align}}

\subsection{Comparison of $\pf{6}{1}$ with $\pf{3}{2}$}
In order to verify the relation (\ref{PartitionFc61}), we have to compare (\ref{PartFunc}) with (\ref{Zflop32Part}). To this end, we first express the arguments of $\theta_{1}$-functions in (\ref{blockW}) in terms of the parameters $(V,M,H_i)$ and using Eq.~(\ref{Para323}):
{\allowdisplaybreaks
\begin{align}
&\beta_{i,1}=-M-H_{i-1}-H_i\,, && \beta_{i,4}=2M+H_{i-2}\,,\nonumber\\
&\beta_{i,2}=-H_{i-1}\,,&& \beta_{i,5}=3M+H_{i-3}+H_{i-2}\,,\nonumber\\
&\beta_{i,3}=M\,, && \beta_{i,6}=4M+\sum_{r=1}^{3}H_{i-1-r}\,,\nonumber
\end{align}}
as well as
\begin{equation}
\lambda_{i,j}=\begin{cases}
						0 & \quad \text{if}~~~~ j=6 \\
                      jM+H_{i-2}+\dots+H_{i-1-j} & \quad \text{else.}
                      \end{cases}
                      \nonumber
\end{equation}
The prefactor in Eq.~(\ref{blockW}) then becomes
\begin{align}
\Big[ \Big( \frac{t}{q} \Big)^{\frac{5}{2}}\prod_{l=1}^6 Q_{h_l} \Big]^{\sum_{i=1}^{6}|\alpha_i|} = \Big[ \Big(\frac{t}{q} \Big)^{\frac{5}{2}} Q_{\rho}^{-1} \prod_{i=1}^6 Q_{H_i}^{-1} \Big]^{\sum_{i=1}^{6}|\alpha_i|}\,.
\nonumber
\end{align}
Furthermore, we can change the arguments of the $\theta_{1}$-functions in Eq.~(\ref{Wblock}) using the following identity of Jacobi theta functions under the shift by an integer $n$:
\begin{align}
\theta_1\left(\rho;z+n\,\rho\right)=Q_{\rho}^{-\frac{n^2}{2}}\,(-e^{-2\pi i z})^n\,\theta_1(\rho;z)\, . 
\label{ThetaFctIdentityShift}
\end{align}
In this manner, Eq.~(\ref{Wblock}) can be written as
\begin{widetext}
{\allowdisplaybreaks\begin{align}
&W_{\alpha_1\,\alpha_2\,\alpha_3\,\alpha_4\,\alpha_5\,\alpha_6} ^{\alpha_4\,\alpha_5\,\alpha_6\,\alpha_1\,\alpha_2\,\alpha_3}(Q_{H_i},Q_M,\epsilon_1,\epsilon_2)= \left[ \left(\frac{t}{q} \right)^{5/2} Q_{\rho}^{-1} \prod_{i=1}^6 Q_{H_i}^{-1} \right]^{\sum_{i=1}^6|\alpha_i|} \, \Big[ -\Big( \frac{q}{t} \Big)^{5/2} \Big]^{\sum_{i=1}^6|\alpha_i|}  \nonumber \\
& \hspace{1.5cm}\times (Q^2_\rho Q_M^{-7}Q_{H_1}^{-1}Q_{H_2}^{-2}Q_{H_3}^{-2}Q_{H_4}^{-1})^{|\alpha_1|} \,(Q^2_\rho Q_M^{-7}Q_{H_2}^{-1}Q_{H_3}^{-2}Q_{H_4}^{-2}Q_{H_5}^{-1})^{|\alpha_2|} (Q_\rho Q_M^{-1}Q_{H_1}Q_{H_2}Q_{H_4}^{-1}Q_{H_5}^{-1})^{|\alpha_3|} \nonumber \\
&  \hspace{1.5cm}\times (Q_M^5Q_{H_1}Q_{H_2}^2Q_{H_3}^2Q_{H_4})^{|\alpha_4|}(Q_M^5Q_{H_2}Q_{H_3}^2Q_{H_4}^2Q_{H_5})^{|\alpha_5|} (Q_\rho Q_M^{-1}Q_{H_1}^{-1}Q_{H_2}^{-1}Q_{H_4}Q_{H_5})^{|\alpha_6|} \nonumber \\
& \hspace{1.5cm} \times  \left( \prod_{a=1}^6 \frac{\vartheta_{\alpha_a \alpha_a}(Q_M;\rho)}{\vartheta_{\alpha_a \alpha_a}(\sqrt{q/t};\rho)}\right)\,\left( \prod_{1 \leq a < b \leq 6} \frac{\vartheta_{\alpha_a \alpha_b}(Q_{ab}Q_M;\rho)\vartheta_{\alpha_a \alpha_b}(Q_{ab}Q_M^{-1};\rho)}{\vartheta_{\alpha_a \alpha_b}(Q_{ab}\sqrt{t/q} ;\rho)\vartheta_{\alpha_a \alpha_b}(Q_{ab}\sqrt{q/t} ;\rho)} \right)\,,
\end{align}}
\end{widetext}
where we combined the $\theta_1$-functions into $\vartheta$-functions which are defined in Appendix A (using also Eq.~(\ref{TranSum})). Thus, the whole partition function $\pf{3}{2}(V,M,H_{1,\ldots,6},\epsilon_{1,2})$ in Eq.~(\ref{PartFunc}) becomes
\begin{widetext}
{\allowdisplaybreaks\begin{align}
{\cal Z}_{X_{3,2}}(V,M,H_{1,\ldots,6},\epsilon_1,\epsilon_2)&=W_{2,3}  (\emptyset) \times \sum_{\alpha_1,\ldots,\alpha_6} (-Q_VQ_M)^{|\alpha_1|+\dots+|\alpha_6|}  \left( \prod_{a=1}^6 \frac{\vartheta_{\alpha_a \alpha_a}(Q_M;\rho)}{\vartheta_{\alpha_a \alpha_a}(\sqrt{q/t};\rho)}\right) \nonumber \\
&\times\left( \prod_{1 \leq a < b \leq 6} \frac{\vartheta_{\alpha_a \alpha_b}(Q_{ab}Q_M;\rho)\vartheta_{\alpha_a \alpha_b}(Q_{ab}Q_M^{-1};\rho)}{\vartheta_{\alpha_a \alpha_b}(Q_{ab}\sqrt{t/q} ;\rho)\vartheta_{\alpha_a \alpha_b}(Q_{ab}\sqrt{q/t} ;\rho)} \right) \, . 
\end{align}}
\end{widetext}
We see that this agrees with the partition function $\pf{6}{1}(V,M,H_i,\epsilon_{1,2})$ in Eq.~(\ref{Zflop32Part}). This proves (\ref{PartitionFc61}) at a generic point in the K\"ahler moduli space and for generic values of $\epsilon_{1,2}$. This result provides a very strong support to the duality proposed in \cite{Hohenegger:2016yuv}.

\section{Building Block for General $(N,M)$ Web}
\label{sec:strip}
In the last section, we computed the partition function $\pf{3}{2}$ at a generic point in the moduli space by choosing the preferred direction of the (refined) topological vertex along the diagonal. As we have seen, this leads to a series representation of $\pf{3}{2}$ that was instrumental in proving the duality (\ref{PartitionFc61}). In this section, we generalise this result to generic configurations of the type $(N,M)$. To this end, we first discuss the 'diagonal expansion' for generic $(N,M)$ and, in a second step, derive a building block (the generalisation of $W$ in (\ref{PartFunc})) which allows the computation of $\pf{N}{M}$ in full generality.

\subsection{Diagonal Expansion of a Generic $(N,M)$ Web}
\begin{figure*}
\begin{center}
\scalebox{0.4}{\parbox{14cm}{\begin{tikzpicture}[scale = 1.0]
\draw[ultra thick,fill=gray!50!white] (0,10) -- (6,10) -- (6,6) -- (0,6) -- (0,10);
\draw[ultra thick,fill=green!50!white] (0,10) -- (12,-2) -- (12,0) -- (0,12) -- (0,10);
\draw[-] (-1,10) -- (13,10);
\draw[-] (0,-3) -- (0,13);
\draw[-,line width=1.25mm] (0,-2) -- (12,-2);
\draw[-] (0,-1) -- (12,-1);
\draw[-] (0,0) -- (12,0);
\draw[-] (0,1) -- (12,1);
\draw[-,line width=1.25mm] (0,2) -- (12,2);
\draw[-] (0,3) -- (12,3);
\draw[-] (0,4) -- (12,4);
\draw[-] (0,5) -- (12,5);
\draw[-,line width=1.25mm] (0,6) -- (12,6);
\draw[-] (0,7) -- (12,7);
\draw[-] (0,8) -- (12,8);
\draw[-] (0,9) -- (12,9);
\draw[-,line width=1.25mm] (0,10) -- (12,10);
\draw[-] (0,11) -- (12,11);
\draw[-] (0,12) -- (12,12);
\draw[-,line width=1.25mm] (0,-2) -- (0,12);
\draw[-] (1,-2) -- (1,12);
\draw[-] (2,-2) -- (2,12);
\draw[-] (3,-2) -- (3,12);
\draw[-] (4,-2) -- (4,12);
\draw[-] (5,-2) -- (5,12);
\draw[-,line width=1.25mm] (6,-2) -- (6,12);
\draw[-] (7,-2) -- (7,12);
\draw[-] (8,-2) -- (8,12);
\draw[-] (9,-2) -- (9,12);
\draw[-] (10,-2) -- (10,12);
\draw[-] (11,-2) -- (11,12);
\draw[-,line width=1.25mm] (12,-2) -- (12,12);
\draw[-] (0,-1) -- (1,-2);
\draw[-] (0,0) -- (2,-2);
\draw[-] (0,1) -- (3,-2);
\draw[-] (0,2) -- (4,-2);
\draw[-] (0,3) -- (5,-2);
\draw[-] (0,4) -- (6,-2);
\draw[-] (0,5) -- (7,-2);
\draw[-] (0,6) -- (8,-2);
\draw[-] (0,7) -- (9,-2);
\draw[-] (0,8) -- (10,-2);
\draw[-] (0,9) -- (11,-2);
\draw[-] (0,10) -- (12,-2);
\draw[-] (0,11) -- (12,-1);
\draw[-] (0,12) -- (12,0);
\draw[-] (1,12) -- (12,1);
\draw[-] (2,12) -- (12,2);
\draw[-] (3,12) -- (12,3);
\draw[-] (4,12) -- (12,4);
\draw[-] (5,12) -- (12,5);
\draw[-] (6,12) -- (12,6);
\draw[-] (7,12) -- (12,7);
\draw[-] (8,12) -- (12,8);
\draw[-] (9,12) -- (12,9);
\draw[-] (10,12) -- (12,10);
\draw[-] (11,12) -- (12,11);
\draw[ultra thick, red, line width=1.25mm] (0,12) -- (1,11);
\draw[ultra thick, red, line width=1.25mm] (6,4) -- (7,3);
%
%
\draw[ultra thick,<->] (-0.5,10) -- (-0.5,12);
\draw[ultra thick,<->] (0,-2.5) -- (12,-2.5);
\draw[ultra thick,<->] (0,12.5) -- (6,12.5);
\node at (-0.8,11) {\Large $k$};
\node at (6,-3.1) {\huge $\frac{NM}{k}$};
\node at (3,13) {\Large $\delta$};
\draw[ultra thick,<->] (-0.5,6) -- (-0.5,2);
\node at (-0.85,4) {\Large $M$};
\draw[ultra thick,<->] (0,-3.8) -- (6,-3.8);
\node at (3,-4.1) {\Large $N$};
\node at (6,16) {\Huge \phantom{(a)}};
\node at (6,-6.5) {\Huge (a)};
\end{tikzpicture}}}
\hspace{1cm}
\scalebox{0.45}{\parbox{21cm}{\begin{tikzpicture}[scale = 1.0]
\draw[red!30!yellow!25!white,fill=red!30!yellow!25!white] (0.15,3.25) -- (-1.4,1.6) -- (13.1,-7.3) -- (14.55,-5.55) -- (0.15,3.25);

\draw[ultra thick] (-1,-2) -- (0,-1);
\draw[ultra thick] (0,-3) -- (1,-2);
\draw[ultra thick] (-1,0) -- (0,0) -- (0,-1) -- (1,-1) -- (1,-2) -- (2,-2);
\draw[ultra thick] (0,0) -- (1,1);
\draw[ultra thick] (1,-1) -- (2,0);
\draw[ultra thick] (0,2) -- (1,2) -- (1,1) -- (2,1) -- (2,0) -- (3,0);
\draw[ultra thick] (1,2) -- (2,3);
\draw[ultra thick] (2,1) -- (3,2);
\node[rotate=45] at (3,4){\Huge $\cdots$};
\node[rotate=45] at (4,3){\Huge $\cdots$};
\draw[ultra thick] (4,5) -- (5,6);
\draw[ultra thick] (5,4) -- (6,5);
\draw[ultra thick] (4,7) -- (5,7) -- (5,6) -- (6,6) -- (6,5) -- (7,5);
\draw[ultra thick] (5,7) -- (6,8);
\draw[ultra thick] (6,6) -- (7,7);
\draw[ultra thick] (5,9) -- (6,9) -- (6,8) -- (7,8) -- (7,7) -- (8,7);
\draw[ultra thick] (6,9) -- (7,10);
\draw[ultra thick] (7,8) -- (8,9);
\node[rotate=-22.5] at (3,-2.5) {\Huge $\cdots$};
\node[rotate=-22.5] at (4,-0.5) {\Huge $\cdots$};
\node[rotate=-22.5] at (8,4.5) {\Huge $\cdots$};
\node[rotate=-22.5] at (9,6.5) {\Huge $\cdots$};
\draw[ultra thick,xshift=5cm,yshift=-3cm] (-1,-2) -- (0,-1);
\draw[ultra thick,xshift=5cm,yshift=-3cm] (0,-3) -- (1,-2);
\draw[ultra thick,xshift=5cm,yshift=-3cm] (-1,0) -- (0,0) -- (0,-1) -- (1,-1) -- (1,-2) -- (2,-2);
\draw[ultra thick,xshift=5cm,yshift=-3cm] (0,0) -- (1,1);
\draw[ultra thick,xshift=5cm,yshift=-3cm] (1,-1) -- (2,0);
\draw[ultra thick,xshift=5cm,yshift=-3cm] (0,2) -- (1,2) -- (1,1) -- (2,1) -- (2,0) -- (3,0);
\draw[ultra thick,xshift=5cm,yshift=-3cm] (1,2) -- (2,3);
\draw[ultra thick,xshift=5cm,yshift=-3cm] (2,1) -- (3,2);
\node[,xshift=5cm,yshift=-3cm,rotate=45] at (3,4){\Huge $\cdots$};
\node[xshift=5cm,yshift=-3cm,rotate=45] at (4,3){\Huge $\cdots$};
\draw[ultra thick,xshift=5cm,yshift=-3cm] (4,5) -- (5,6);
\draw[ultra thick,xshift=5cm,yshift=-3cm] (5,4) -- (6,5);
\draw[ultra thick,xshift=5cm,yshift=-3cm] (4,7) -- (5,7) -- (5,6) -- (6,6) -- (6,5) -- (7,5);
\draw[ultra thick,xshift=5cm,yshift=-3cm] (5,7) -- (6,8);
\draw[ultra thick,xshift=5cm,yshift=-3cm] (6,6) -- (7,7);
\draw[ultra thick,xshift=5cm,yshift=-3cm] (5,9) -- (6,9) -- (6,8) -- (7,8) -- (7,7) -- (8,7);
\draw[ultra thick,xshift=5cm,yshift=-3cm] (6,9) -- (7,10);
\draw[ultra thick,xshift=5cm,yshift=-3cm] (7,8) -- (8,9);
\node[xshift=5cm,yshift=-3cm,rotate=-22.5] at (3,-2.5) {\Huge $\cdots$};
\node[xshift=5cm,yshift=-3cm,rotate=-22.5] at (4,-0.5) {\Huge $\cdots$};
\node[xshift=5cm,yshift=-3cm,rotate=-22.5] at (8,4.5) {\Huge $\cdots$};
\node[xshift=5cm,yshift=-3cm,rotate=-22.5] at (9,6.5) {\Huge $\cdots$};
\draw[ultra thick,xshift=10cm,yshift=-6cm] (-1,-2) -- (0,-1);
\draw[ultra thick,xshift=10cm,yshift=-6cm] (0,-3) -- (1,-2);
\draw[ultra thick,xshift=10cm,yshift=-6cm] (-1,0) -- (0,0) -- (0,-1) -- (1,-1) -- (1,-2) -- (2,-2);
\draw[ultra thick,xshift=10cm,yshift=-6cm] (0,0) -- (1,1);
\draw[ultra thick,xshift=10cm,yshift=-6cm] (1,-1) -- (2,0);
\draw[ultra thick,xshift=10cm,yshift=-6cm] (0,2) -- (1,2) -- (1,1) -- (2,1) -- (2,0) -- (3,0);
\draw[ultra thick,xshift=10cm,yshift=-6cm] (1,2) -- (2,3);
\draw[ultra thick,xshift=10cm,yshift=-6cm] (2,1) -- (3,2);
\node[,xshift=10cm,yshift=-6cm,rotate=45] at (3,4){\Huge $\cdots$};
\node[xshift=10cm,yshift=-6cm,rotate=45] at (4,3){\Huge $\cdots$};
\draw[ultra thick,xshift=10cm,yshift=-6cm] (4,5) -- (5,6);
\draw[ultra thick,xshift=10cm,yshift=-6cm] (5,4) -- (6,5);
\draw[ultra thick,xshift=10cm,yshift=-6cm] (4,7) -- (5,7) -- (5,6) -- (6,6) -- (6,5) -- (7,5);
\draw[ultra thick,xshift=10cm,yshift=-6cm] (5,7) -- (6,8);
\draw[ultra thick,xshift=10cm,yshift=-6cm] (6,6) -- (7,7);
\draw[ultra thick,xshift=10cm,yshift=-6cm] (5,9) -- (6,9) -- (6,8) -- (7,8) -- (7,7) -- (8,7);
\draw[ultra thick,xshift=10cm,yshift=-6cm] (6,9) -- (7,10);
\draw[ultra thick,xshift=10cm,yshift=-6cm] (7,8) -- (8,9);
\node at (-1.4,0) {\large $a_1$};
\node at (-0.4,2) {\large $a_2$};
\node at (3.5,7) {\large $a_{k-1}$};
\node at (4.6,9) {\large $a_{k}$};
\node at (12.5,-8) {\large $a_1$};
\node at (13.5,-6) {\large $a_2$};
\node at (17.6,-1) {\large $a_{k-1}$};
\node at (18.5,1) {\large $a_{k}$};
\node at (7.2,10.2) {\large $1$};
\node at (8.2,9.2) {\large $2$};
\node at (12.2,7.2) {\large $\Delta$};
\node at (13.2,6.3) {\large $\Delta+1$};
\node at (17.2,4.4) {\large $\tfrac{NM}{k}-1$};
\node at (18.2,3.3) {\large $\tfrac{NM}{k}$};
\node at (-1.4,-2.3) {\large $\tfrac{NM}{k}-\Delta$};
\node at (0.3,-3.4) {\large $\tfrac{NM}{k}+1-\Delta$};
\node at (3.8,-5.4) {\large $\tfrac{NM}{k}$};
\node at (4.8,-6.3) {\large $1$};
\node at (8.8,-8.3) {\large $\Delta-1$};
\node at (9.8,-9.3) {\large $\Delta$};
\node at (6,-10) {{\fontsize{22}{50}{$\text{(b)}$}}};
\end{tikzpicture}}}

\end{center}
\caption{\sl (a) Newton polygon of $X_{N,M}$ (for better readability we have chosen $(N,M)=(6,4)$): The grey and green region highlight two different (but equivalent) fundamental domains that are used to tile the plane. Indeed, the grey region is the dual polygon of the web shown in \figref{Fig:WebToric}, while the dual of the green region is the twisted web diagram. The distance $\delta$ between the two equivalent intervals colored in red determines the shift $\Delta$ (for the particular case $(6,4)$ we have $\delta=6=N$ and $\Delta=1$). (b) Twisted web diagram for generic $(N,M)$. The orange region highlights one of the building blocks which can be used to compute the partition function $\pf{N}{M}$ and which is shown in more detail in \figref{ArbitraryStrip} along with a labelling of the relevant parameters and integer partitions in preparation of the topological vertex computation.}
\label{NewtonPolygonNM}
\end{figure*}
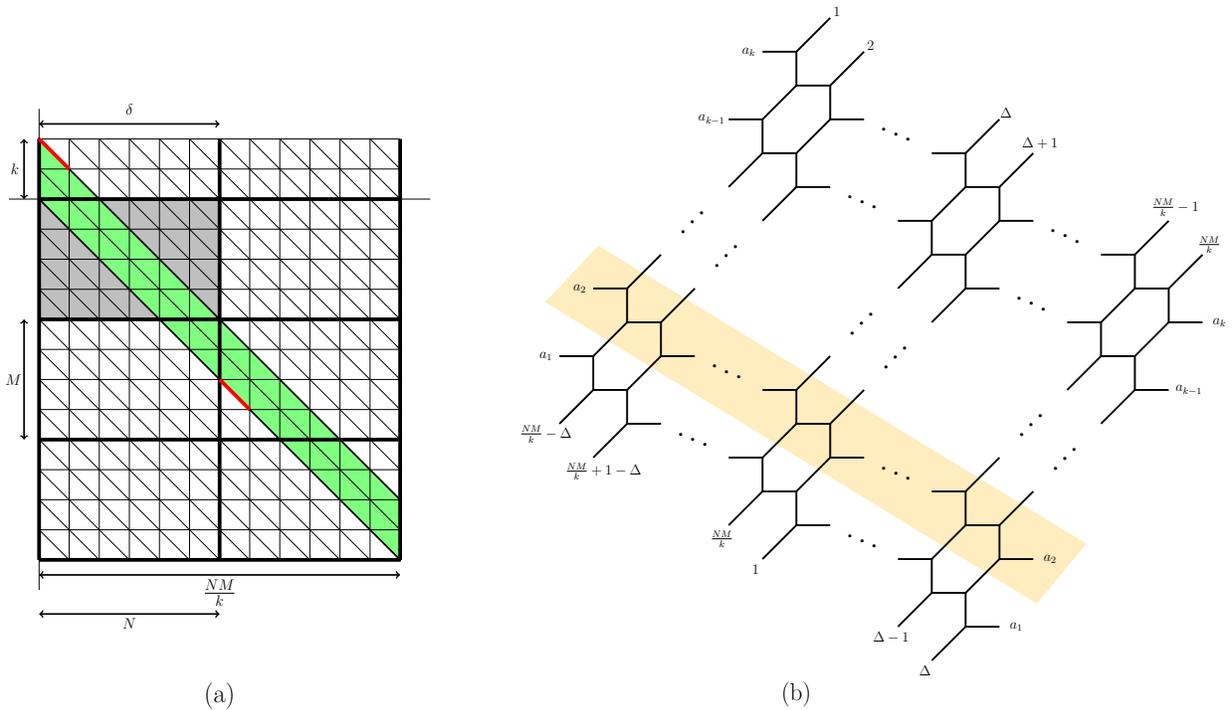

The most direct way to understand generalisation of the diagonal expansion for a generic $(N,M)$ web is from the perspective of dual Newton polygon: if the external legs of the web in \figref{Fig:WebToric} were not glued together (\emph{i.e.} mutually identified), the corresponding Newton polygon would be an $N\times M$ rectangle in which each $1\times 1$ square is triangulated in the same way (as all the diagonal intervals are parallel). The gluing of external lines creates a periodic web, whose Newton polygon can be obtained by drawing the $N\times M$ rectangle on the torus (\emph{i.e.} gluing the parallel edges of $N\times M$ polygon). We can, however, also consider the periodic Newton polygon as a tiling of the plane in terms of basic $N\times M$ rectangle. This is shown for the case $(3,2)$ in \figref{Fig:Newton32}(b) and more generically in \figref{NewtonPolygonNM}(a). In both cases, the grey region is the $N\times M$ rectangle that is dual to the web diagram of $X_{N,M}$.

As in the $(3,2)$ case, however, the fundamental domain of a given tiling is not unique. Indeed, the green region in \figref{NewtonPolygonNM}(a) also contains all distinct faces of the Newton polygon, and is therefore a possible choice of the fundamental domain. However, the web diagram dual to this green region (drawn in \figref{NewtonPolygonNM}(b)) shows that $X_{N,M}$ can also be drawn as a $(\frac{NM}{k},k)$ web (with $k=\text{gcd}(N,M)$) but not with `adjacent' external legs on opposite sides glued together. Indeed, the `off-set' $\Delta=\delta/N$ can be read off by comparing the positions of the two red intervals in \figref{NewtonPolygonNM}(a) and is given by the minimal (integer) solution to the diophantine equation
\bea\label{Off-SetEq}
\delta=\Delta N=nM-k\,,~~\text{for} ~~n\in \mathbb{Z}\,.
\eea
We will call generic web diagrams with $\delta\neq 0$ {\sl twisted}, the twisted $(6,1)$ diagram being an example already encountered in the last section. Note, however, that twisted webs of the type \figref{NewtonPolygonNM}(b) can also be brought to the form of an untwisted $(\frac{NM}{k},k)$ web by a series of flop transitions \cite{Hohenegger:2016yuv}. It turns out, however, that the presentation of the $(N,M)$ web diagram as a twisted $(\frac{NM}{k},k)$ web is more suited for the computation of the partition function.

\subsection{Generic Building Block}
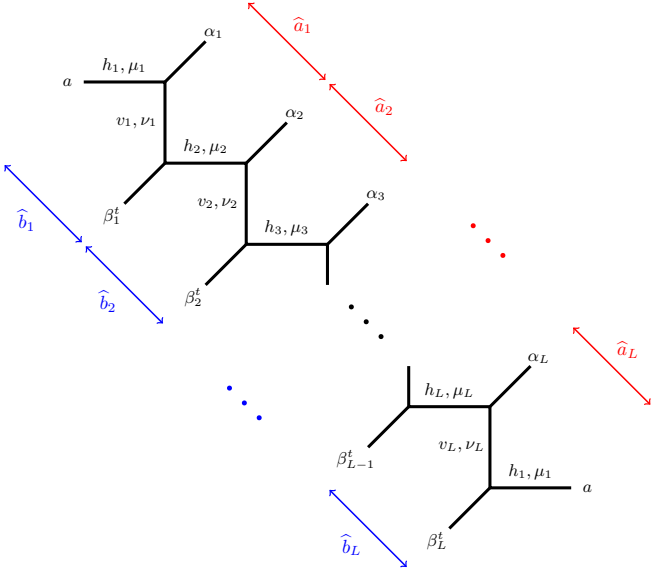
\begin{figure}[h]
\begin{center}
\scalebox{0.72}{\parbox{12cm}{\begin{tikzpicture}[scale = 1.50]
\draw[ultra thick] (0,0) -- (1,0) -- (1,-1) -- (2,-1) -- (2,-2) -- (3,-2) -- (3,-2.5) ;
\node[rotate=315] at (3.5,-3) {\Huge $\cdots$};
\draw[ultra thick] (1,0) -- (1.5,0.5);
\draw[ultra thick] (2,-1) -- (2.5,-0.5);
\draw[ultra thick] (3,-2) -- (3.5,-1.5);
\draw[ultra thick] (1,-1) -- (0.5,-1.5);
\draw[ultra thick] (2,-2) -- (1.5,-2.5);
\node at (1.6,0.6) {{\small \bf $\alpha_1$}};
\node at (2.6,-0.4) {{\small \bf $\alpha_2$}};
\node at (3.6,-1.4) {{\small \bf $\alpha_3$}};
\node at (0.35,-1.65) {{\small \bf $\beta^t_1$}};
\node at (1.35,-2.65) {{\small \bf $\beta^t_2$}};
\node at (-0.2,0) {{\small \bf $a$}};
\node at (0.5,0.2) {{\small $h_1,\mu_1$}};
\node at (1.5,-0.8) {{\small $h_2,\mu_2$}};
\node at (2.5,-1.8) {{\small $h_3,\mu_3$}};
\node at (0.65,-0.5) {{\small $v_1,\nu_1$}};
\node at (1.65,-1.5) {{\small $v_2,\nu_2$}};
\begin{scope}[xshift=-1cm,yshift=1cm]
\node at (5.65,-5.5) {{\small $v_L,\nu_L$}};
\draw[ultra thick] (5,-4.5) -- (5,-5) -- (6,-5) -- (6,-6) -- (7,-6);
\node at (5.5,-4.8) {{\small $h_L,\mu_L$}};
\node at (6.5,-5.8) {{\small $h_1,\mu_1$}};
\node at (4.35,-5.65) {{\small \bf $\beta^t_{L-1}$}};
\node at (5.35,-6.65) {{\small \bf $\beta^t_L$}};
\node at (6.6,-4.4) {{\small \bf $\alpha_L$}};
\node at (7.2,-6) {{\small \bf $a$}};
\draw[ultra thick] (6,-5) -- (6.5,-4.5);
\draw[ultra thick] (5,-5) -- (4.5,-5.5);
\draw[ultra thick] (6,-6) -- (5.5,-6.5);
\end{scope}
\draw[thick,<->,red] (2.025,0.975) -- (2.975,0.025);
\node[red] at (2.7,0.7) {$\widehat{a}_1$};
\draw[thick,<->,red] (3.025,-0.025) -- (3.975,-0.975);
\node[red] at (3.7,-0.3) {$\widehat{a}_2$};
\node[red,rotate=315] at (5,-2) {\Huge$\cdots$};
\draw[thick,<->,red] (6.025,-3.025) -- (6.975,-3.975);
\node[red] at (6.7,-3.3) {$\widehat{a}_L$};
\draw[thick,<->,blue] (-0.975,-1.025) -- (-0.025,-1.975);
\node[blue] at (-0.7,-1.7) {$\widehat{b}_1$};
\draw[thick,<->,blue] (0.025,-2.025) -- (0.975,-2.975);
\node[blue] at (0.3,-2.7) {$\widehat{b}_2$};
\node[blue,rotate=315] at (2,-4) {\Huge$\cdots$};
\draw[thick,<->,blue] (3.025,-5.025) -- (3.975,-5.975);
\node[blue] at (3.3,-5.7) {$\widehat{b}_L$};
\end{tikzpicture}}}
\end{center}
\caption{\sl `Staircase' strip of length $L$ with a labelling of the K\"ahler parameters and integer partitions.}
\label{ArbitraryStrip}
\end{figure}

The twisted web diagram is decomposable into $k=\text{gcd}(N,M)$ basic strips (one of them is highlighted in orange color in \figref{NewtonPolygonNM}) which are glued together along the diagonal intervals. We refer to these strips sometimes as `staircase' diagrams and a generic such strip of length $L$ (along with a suitable labelling of K\"ahler parameters associated with the various intervals, as well as integer partitions associated with the external diagonal legs) is shown in \figref{ArbitraryStrip}. The open string amplitudes for such strips can be glued together to form the topological string partition function of $X_{N,M}$ (see \cite{Haghighat:2013gba} for the case $X_{N,1}$, \cite{Hohenegger:2013ala} for $X_{N,M}$ with a non-maximal set of K\"ahler parameters and \cite{, Haghighat:2013tka} for the limit in which one of the elliptic fibrations of $X_{N,M}$ degenerates). Below, we compute the generic building block associated with the strip in \figref{ArbitraryStrip} at a generic point in moduli space, which allows to evaluate $\pf{N}{M}$ at an arbitrary point in the K\"ahler moduli space.
Using the refined topological vertex, we can express the generic building block for the strip in \figref{ArbitraryStrip} as
\begin{widetext}
\begin{align}
W^{\alpha_1 \dots \alpha_L}_{\beta_1 \dots \beta_L}(Q_{h_i},Q_{v_i},\epsilon_1,\epsilon_2)= \hat{Z} \times \sum_{\substack{\{\mu\}\{\nu\} \\ \{\eta\} \{\tilde{\eta}\}}} \prod_{i=1}^L Q_{h_i}^{|\mu_i|}Q_{v_i}^{|\nu_i|} s_{\mu_i/\eta_i}(x_i) s_{\mu_i^t/\tilde{\eta}_{i-1}}(y_{i-1})s_{\nu_i^t/\tilde{\eta}_i}(w_i) s_{\nu_i/ \eta_i}(z_i) \label{WBlock}
\end{align}
\end{widetext}
where the prefactor is given by
\begin{equation}
\hat{Z}= \prod_{i=1}^L t^{\frac{||\alpha_k||^2}{2}} q^{\frac{||\alpha_k^t||^2}{2}} \tilde{Z}_{\alpha_k}(q,t) \tilde{Z}_{\alpha_k^t}(t,q)\,,
\end{equation}
and the arguments of skew Schur functions are defined as 
\begin{align}
&x_i= q^{-\xi+\frac{1}{2}}t^{-\alpha_{i}-\frac{1}{2}}\,, && y_i=t^{-\xi+\frac{1}{2}}q^{-\beta_{i}^t-\frac{1}{2}}\,,\nonumber\\
&w_i= q^{-\xi}t^{-\beta_i } \,, && z_i=t^{-\xi}q^{-\alpha_{i}^t }\,,\label{DefThetaArguments}
\end{align}
where $\xi=\left\{-\tfrac{1}{2}\,,-\tfrac{3}{2}\,,-\tfrac{5}{2}\,,\ldots\right\}$. It is important to recall that we associate arbitrary (and independent) partitions $\alpha_{1,\ldots,L}$ and $\beta_{1,\ldots,L}$ with the upper and lower diagonal legs of the staircase strip (as shown in \figref{ArbitraryStrip}). Put differently, the diagonal line segments of this staircase are not glued together and therefore no consistency conditions are imposed on the K\"ahler parameters. Thus, the $2L$ variables
\begin{align}
&h_{1,\ldots,L}\,,&&\text{and} &&v_{1,\ldots,L}\,,
\end{align}
are independent of each other (in the following, we will use the same notation as in (\ref{Qnotation})). Note, however, that this will be changed in section~\ref{Sect:TwistedStrip}, when we consider a specific gluing of the external lines of the staircase, as shown in \figref{Fig:StripLdelta}. For later convenience, we also introduce the parameters
\begin{align}
&\widehat{a}_i= h_{i+1} + v_{i}    &&\text{and} &&\widehat{b}_i= h_i + v_i\,,&&\label{DefRootsStrip}
\end{align}
for all $i=1,\ldots,L$. These are not all independent of each other, but satisfy the consistency condition $\sum_{i=1}^L\widehat{a}_i=\sum_{i=1}^L\widehat{b}_i$.

Comparing with the case $(N,M)=(3,2)$ as discussed in the previous section, we note that the structure of (\ref{WBlock}) is very similar to (\ref{W}), except that the summation involves a larger number of integer partitions. We remark that physically, the sizes of the integer partitions labelling the fundamental building block $W^{\alpha_1 \dots \alpha_L}_{\beta_1 \dots \beta_L}$ give the instanton charges of individual $U(1)$'s  in the dual gauge theory. The dual gauge theory and the role of these partitions labelling the building block will be discussed in detail in \cite{paper2}. However, as any given summand in Eq.(\ref{WBlock}) still contains products of skew Schur functions which can be manipulated using the relations (\ref{Schur1}), we can follow the same steps as in appendix~\ref{App:32ExpComputation} to work out (\ref{WBlock}). Specifically, due to the identification of the horizontal ends of the strip (denoted by $a$ in \figref{ArbitraryStrip}), we can generalise the recursion relation (\ref{Recursion6}) for generic $L$, which allows us to write a product representation of (\ref{WBlock}). As the computation is lengthy and very tedious, we refrain from giving the details of various steps and simply state the result in the following form
\begin{widetext}
\vskip-0.5cm
\begin{align}
W^{\alpha_1 \dots \alpha_L}_{\beta_1 \dots \beta_L} \sim \hat{Z} \cdot \prod_{i,j=1}^L \prod_{k,r,s=1}^{\infty} \frac{(1-\widehat{Q}_{i,j}\,Q_\rho^{k-1} x_{i,r}y_{i-j,s})(1-\widehat{Q}^v_{i,j}Q_\rho^{k-1}z_{i,s}w_{i+j-1,r})}{(1-\overline{Q}_{i,j}Q_\rho^{k-1} x_{i,r}z_{i-j,s})(1-\dot{Q}_{i,j}Q_\rho^{k-1} y_{i,s}w_{i+j,r})}\,, \label{ArbStrip}
\end{align}
\end{widetext}
where for brevity we omitted an overall prefactor independent of the external legs, which will be fixed by the normalisation in the end. The modular parameter is given by
\begin{align}
&Q_\rho=\widetilde{Q}_1 \widetilde{Q}_2 \dots \widetilde{Q}_L && \text{with} && \widetilde{Q}_i=Q_{h_i}Q_{v_i}\,.
\end{align}
In (\ref{ArbStrip}), the arguments $x_i$, $y_i$, $w_i$ and $z_i$ are given in Eq.(\ref{DefThetaArguments}) and we have introduced the following shorthand notation 
{\allowdisplaybreaks
\begin{align}
&\widehat{Q}_{i,j} = Q_{h_i} \prod_{k=1}^{j-1}\widetilde{Q}_{i-k}\,,\hspace{0.5cm} \widehat{Q}_{i,j}^v=Q_{v_i}\prod_{k=1}^{j-1}\widetilde{Q}_{i+k}\nonumber\\
&\overline{Q}_{i,j} = \begin{cases}
							1 & \quad \text{if } j=L \\
							Q_{h_i}Q_{v_{i-j}}\prod_{k=1}^{j-1}\widetilde{Q}_{i-k}  & \quad \text{if }j\neq L \\
						\end{cases} \nonumber\\		
&\dot{Q}_{i,j} = \widetilde{Q}_{i+1} \dots \widetilde{Q}_{i+j}		\,.		 \label{QPara}
\end{align}}
Finally, upon reinstating the appropriate normalisation factor $W_L(\emptyset)$ (which is interpreted as the closed string amplitude and is defined below), Eq. (\ref{ArbStrip}) can be written as follows (see Eq.~(\ref{JfncPartBuilding}) for the definition of functions $\mathcal{J}_{\alpha\beta}$),
\begin{widetext}
\vskip-0.5cm
\begin{align}
W^{\alpha_1 \dots \alpha_L}_{\beta_1 \dots \beta_L} = W_L(\emptyset) \cdot \hat{Z}\cdot \prod_{i,j=1}^L \frac{\mathcal{J}_{\alpha_i \beta_{j}}(\widehat{Q}_{i,i-j};q,t)\mathcal{J}_{\beta_{j}\alpha_i}((\widehat{Q}_{i,i-j})^{-1}Q_\rho;q,t)}{\mathcal{J}_{\alpha_i \alpha_{j}}(\overline{Q}_{i,i-j}\sqrt{q/t};q,t)\mathcal{J}_{\beta_{j}\beta_i}(\dot{Q}_{i,j-i}\sqrt{t/q};q,t)}\,, \label{BBlock}
\end{align}
where \begin{align}
W_L(\emptyset)=\prod_{i,j=1}^L \prod_{k,r,s=1}^{\infty} \frac{(1-\widehat{Q}_{i,j}\,Q_\rho^{k-1} q^{r-\frac{1}{2}}t^{s-\frac{1}{2}})(1-\widehat{Q}^{-1}_{i,j}Q_\rho^{k}q^{s-\frac{1}{2}}t^{r-\frac{1}{2}})}{(1-\overline{Q}_{i,j}Q_\rho^{k-1} q^{r}t^{s-1})(1-\dot{Q}_{i,j}Q_\rho^{k-1} q^{s-1}t^r)}\,.\label{WemptyDef}
\end{align}
\end{widetext}
While the numerator of (\ref{BBlock}) can in principle be further simplified by combining the $\mathcal{J}_{\alpha\beta}\mathcal{J}_{\beta\alpha}$ into $\vartheta_{\alpha\beta}$ following (\ref{App:CurlJ1}), a similar simplification for the denominator requires a gluing of the external legs. We therefore postpone these steps to the next section, where the latter is considered in detail.

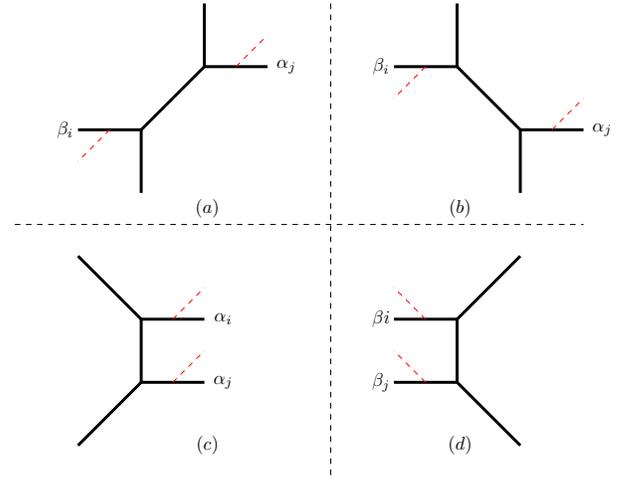
\begin{figure}
\begin{center}
\scalebox{0.7}{\parbox{12cm}{\begin{tikzpicture}[scale = 1.2]
\draw[ultra thick, black] (0,0) -- (1,0)-- (2,1)--(3,1);
\draw[ultra thick, black] (1,0) -- (1,-1);
\draw[ultra thick, black] (2,1) -- (2,2) ;
\node at (-0.25,0) {{ $\beta_{i}$}};
\node at (3.25,1) {{ $\alpha_{j}$}};
\draw[dashed, red] (0.5,0)--(0,-0.5);
\draw[dashed, red] (2.5,1) -- (3,1.5);

\draw[ultra thick, black] (5,1) -- (6,1)-- (6,2);
\draw[ultra thick, black] (6,1) -- (7,0) -- (8,0);
\draw[ultra thick, black] (7,0) -- (7,-1) ;
\node at (4.75,1) {{ $\beta_{i}$}};
\node at (8.25,0) {{ $\alpha_{j}$}};
\draw[dashed, red] (7.5,0)--(8,0.5);
\draw[dashed, red] (5.5,1) -- (5,0.5);

\draw[ultra thick, black] (0,-2) -- (1,-3)-- (2,-3);
\draw[ultra thick, black] (1,-3) -- (1,-4)--(2,-4);
\draw[ultra thick, black] (1,-4) -- (0,-5) ;
\node at (2.25,-3) {{ $\alpha_{i}$}};
\node at (2.25,-4) {{ $\alpha_{j}$}};
\draw[dashed, red] (1.5,-3)--(2,-2.5);
\draw[dashed, red] (1.5,-4) -- (2,-3.5);

\draw[ultra thick, black] (5,-3) -- (6,-3)-- (7,-2);
\draw[ultra thick, black] (6,-3) -- (6,-4) -- (5,-4);
\draw[ultra thick, black] (6,-4) -- (7,-5) ;
\node at (4.75,-3) {{ $\beta{i}$}};
\node at (4.75,-4) {{ $\beta_{j}$}};
\draw[dashed, red] (5.5,-3)--(5,-2.5);
\draw[dashed, red] (5.5,-4) -- (5,-3.5);

\draw[ dashed, black] (-1,-1.5) -- (8,-1.5);
\draw[ dashed, black] (4,2) -- (4,-5.5);
\node at (2,-1.25) {{ $(a)$}};
\node at (6,-1.25) {{ $(b)$}};
\node at (2,-5) {{ $(c)$}};
\node at (6,-5) {{ $(d)$}};

\end{tikzpicture}}}
\end{center}
\caption{\sl (a) Lagrangian branes (respresented as dashed red lines) on the resolved conifold and (b) after flop transition. (c) Lagrangian branes on ${\cal O}(-2)\oplus {\cal O}(0)\mapsto \mathbb{P}^{1}$ and (d) Lagrangian branes on ${\cal O}(0)\oplus {\cal O}(-2)\mapsto \mathbb{P}^{1}$. }
\label{OpenAmplitude}
\end{figure}

There is a more intuitive way of understanding the structure of (\ref{BBlock}). The open string amplitude in Eq.~(\ref{ArbStrip}) can be understood in terms of counting holomorphic curves in the presence of Lagrangian branes \cite{Iqbal:2004ne}. Recall that, if the external legs of the strip diagram are not glued (as in \figref{ArbitraryStrip}), there are only two types of curves:
\begin{enumerate}
\item[\emph{(i)}] curves with local geometry ${\cal O}(-1,-1)\mapsto \mathbb{P}^1$ 
\item[\emph{(ii)}] curves with local geometry ${\cal O}(-2,\ \ 0)\mapsto \mathbb{P}^1$
\end{enumerate}
If we place Lagrangian branes \cite{Aganagic:2000gs} on (pairs of) external legs of the web diagram, the curves contributing to the open string amplitude~(\ref{ArbStrip}) are of type \emph{(i)} if the external legs are on different sides of the diagram and of type \emph{(ii)} if they are on the same side, respectively. In the former case, for two external legs labelled by $\alpha_{i}$ and $\beta_{j}$, these curves are shown in \figref{OpenAmplitude}(a) and \figref{OpenAmplitude}(b), respectively, and their contributions to the open string amplitude $W^{\alpha_1 \dots \alpha_L}_{\beta_1 \dots \beta_L}$ in~(\ref{ArbStrip}) are
{\allowdisplaybreaks 
\bea
Z^{(a)}_{\alpha_{i}\beta_{j}}=\prod_{a,b=1}^{\infty}(1-Q\,q^{a-\alpha^{t}_{i,b}-\frac{1}{2}}t^{b-\beta_{j,a}-\frac{1}{2}})\,,\nonumber\\
Z^{(b)}_{\beta_{i}\alpha_{j}}=\prod_{a,b=1}^{\infty}(1-Q\,q^{a-\beta^{t}_{i,b}-\frac{1}{2}}t^{b-\alpha_{j,a}-\frac{1}{2}})\,.\label{Open1}
\eea}
The curves of type \emph{(ii)} with branes on external legs on the same side labelled by $\alpha_{i},\alpha_{j}$ or $\beta_{i},\beta_{j}$ are shown in \figref{OpenAmplitude}(b) and contribute to the open string amplitude,
{\allowdisplaybreaks\bea
Z^{(c)}_{\alpha_{i}\alpha_{j}}=\prod_{a,b=1}^{\infty}(1-Q\,q^{a-\alpha^{t}_{j,b}}t^{b-\alpha_{i,a}-1})\,,\nonumber\\
Z^{(d)}_{\beta_{i}\beta_{j}}=\prod_{a,b=1}^{\infty}(1-Q\,q^{a-\beta^{t}_{j,b}-1}t^{b-\beta_{i,a}})\,.\label{Open2}
\eea}
Here, for all four cases $-\ln(Q)$ is the K\"ahler parameter associated with the only holomorphic curve in the geometry. In order to describe the area of this curve in terms of the K\"ahler parameters of the strip diagram, the parameters $\widehat{a}_{1,\ldots,L}$ and $\widehat{b}_{1,\ldots,L}$ in (\ref{DefRootsStrip}) are very useful (see \figref{ArbitraryStrip}). In terms of the latter, we have for (\ref{QPara})
{\allowdisplaybreaks\begin{align}
&\widehat{Q}_{i,j} =\text{exp}(-h_i-\sum_{k=1}^{j-1}\widehat{b}_{i-k})\,,\hspace{0.3cm}
 \dot{Q}_{i,j} = \exp(-\sum_{k=1}^j\widehat{b}_{i+k})\,,		\nonumber \\
&	\overline{Q}_{i,j} = \begin{cases}
							1 & \quad \text{if } j=L\,, \\
							\exp(-\sum_{k=1}^{j}\widehat{a}_{i-k} )  & \quad \text{if }j\neq L\,. \\
						\end{cases} 	
						\label{Qroots}
\end{align}}
Note  that gluing the external legs (as \emph{e.g.} in \figref{Fig:StripLdelta}) of the strip allows infinitely many holomorphic curve between any two external lines by going around the strip a number of times before ending. As there are no other holomorphic curves, the open string amplitude associated with the glued strip is an infinite product over the ``winding" in addition to a product over distinct pairs of external legs. However, each pair of legs still gives rise to factors of the type shown in Eq.~(\ref{Open1}) and Eq.~(\ref{Open2}). 

\subsection{Twisted Strip and Duality for $\text{gcd}(N,M)=1$}\label{Sect:TwistedStrip}
Having computed the general building block in (\ref{BBlock}), arbitrary partition functions of the type $\pf{N}{M}$ can be computed by gluing several of the $W$ together along the external lines. Depending on the choice of gluing parameters (and the orientation of the fundamental building block), we can obtain various different series expansions of $\pf{N}{M}$. Indeed, for the diagonal expansion, the gluing of several strips is indicated in \figref{NewtonPolygonNM}(b).

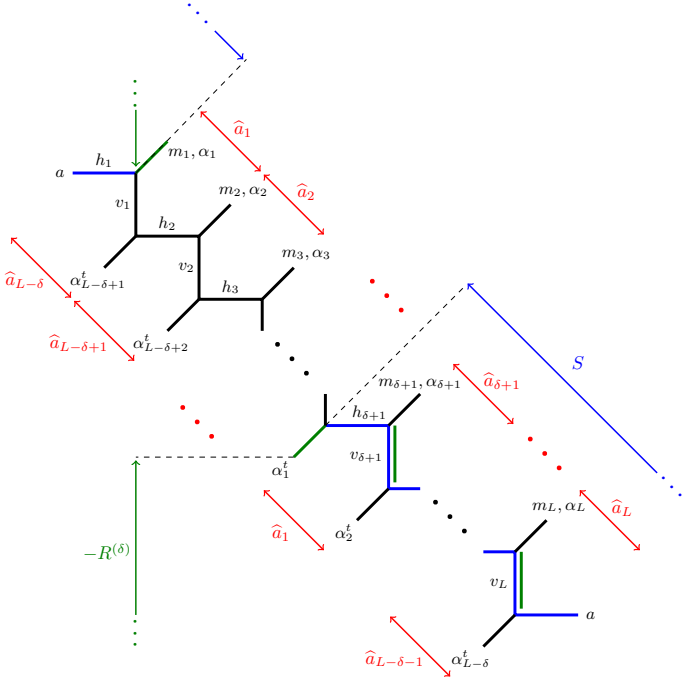
\begin{figure}
\begin{center}
\scalebox{0.7}{\parbox{13cm}{\begin{tikzpicture}[scale = 1.2]
\draw[ultra thick, blue] (0,0) -- (1,0);
\draw[ultra thick] (1,0) -- (1,-1) -- (2,-1) -- (2,-2) -- (3,-2) -- (3,-2.5) ;
\draw[ultra thick,green!50!black] (1,0) -- (1.5,0.5);
\draw[ultra thick] (2,-1) -- (2.5,-0.5);
\draw[ultra thick] (3,-2) -- (3.5,-1.5);
\draw[ultra thick] (1,-1) -- (0.5,-1.5);
\draw[ultra thick] (2,-2) -- (1.5,-2.5);
\node at (1.9,0.3) {{\small \bf $m_1,\alpha_1$}};
\node at (2.7,-0.3) {{\small \bf $m_2,\alpha_2$}};
\node at (3.7,-1.3) {{\small \bf $m_3,\alpha_3$}};
\node at (0.4,-1.7) {{\small \bf $\alpha^t_{L-\delta+1}$}};
\node at (1.4,-2.7) {{\small \bf $\alpha^t_{L-\delta+2}$}};
\node at (-0.2,0) {{\small \bf $a$}};
\node at (0.5,0.2) {{\small $h_1$}};
\node at (1.5,-0.8) {{\small $h_2$}};
\node at (2.5,-1.8) {{\small $h_3$}};
\node at (0.8,-0.5) {{\small $v_1$}};
\node at (1.8,-1.5) {{\small $v_2$}};
\node[rotate=315] at (3.5,-3) {\Huge $\cdots$};
\draw[ultra thick, blue] (4,-4) -- (5,-4) -- (5,-5) -- (5.5,-5);
\draw[ultra thick,green!50!black] (4,-4) -- (3.5,-4.5);
\draw[ultra thick, green!50!black] (5.1,-4) -- (5.1,-4.9);
\draw[ultra thick] (5,-4) -- (5.5,-3.5);
\draw[ultra thick] (5,-5) -- (4.5,-5.5);
\draw[ultra thick] (4,-3.5) -- (4,-4);
\node at (3.3,-4.7) {{\small \bf $\alpha^t_{1}$}};
\node at (4.3,-5.7) {{\small \bf $\alpha^t_{2}$}};
\node at (4.65,-4.5) {{\small $v_{\delta+1}$}};
\node at (4.7,-3.8) {{\small $h_{\delta+1}$}};
\node at (5.5,-3.35) {{\small \bf $m_{\delta+1},\alpha_{\delta+1}$}};
\node[rotate=315] at (6,-5.5) {\Huge $\cdots$};
\draw[ultra thick, blue] (6.5,-6) -- (7,-6) -- (7,-7) -- (8,-7);
\draw[ultra thick, green!50!black] (7.1,-6) -- (7.1,-6.9);
\draw[ultra thick] (7,-7) -- (6.5,-7.5);
\draw[ultra thick] (7,-6) -- (7.5,-5.5);
\node at (7.7,-5.3) {{\small \bf $m_L,\alpha_L$}};
\node at (6.3,-7.7) {{\small \bf $\alpha^t_{L-\delta}$}};
\node at (6.75,-6.5) {{\small $v_{L}$}};
\node at (8.2,-7) {{\small \bf $a$}};
\draw[thick,<->,red] (2.025,0.975) -- (2.975,0.025);
\node[red] at (2.7,0.7) {$\widehat{a}_1$};
\draw[thick,<->,red] (3.025,-0.025) -- (3.975,-0.975);
\node[red] at (3.7,-0.3) {$\widehat{a}_2$};
\node[red,rotate=315] at (5,-2) {\Huge$\cdots$};
\draw[thick,<->,red] (6.025,-3.025) -- (6.975,-3.975);
\node[red] at (6.8,-3.3) {$\widehat{a}_{\delta+1}$};
\node[red,rotate=315] at (7.5,-4.5) {\Huge$\cdots$};
\draw[thick,<->,red] (8.025,-5.025) -- (8.975,-5.975);
\node[red] at (8.7,-5.3) {$\widehat{a}_{L}$};
\draw[thick,<->,red] (-0.975,-1.025) -- (-0.025,-1.975);
\node[red] at (-0.75,-1.7) {$\widehat{a}_{L-\delta}$};
\draw[thick,<->,red] (0.025,-2.025) -- (0.975,-2.975);
\node[red] at (0.1,-2.7) {$\widehat{a}_{L-\delta+1}$};
\node[red,rotate=315] at (2,-4) {\Huge$\cdots$};
\draw[thick,<->,red] (3.025,-5.025) -- (3.975,-5.975);
\node[red] at (3.3,-5.7) {$\widehat{a}_1$};
\draw[thick,<->,red] (5.025,-7.025) -- (5.975,-7.975);
\node[red] at (5.1,-7.7) {$\widehat{a}_{L-\delta-1}$};
\draw[dashed] (4,-4) -- (6.2,-1.8);
\draw[thick,<-,blue] (6.25,-1.75) -- (9.25,-4.75);
\node[blue,rotate=315] at (9.5,-5) {\Large$\cdots$};
\draw[thick,<-,blue] (2.7,1.8) -- (2.25,2.25);
\node[blue,rotate=315] at (2,2.5) {\Large$\cdots$};
\draw[dashed] (1,0) -- (2.75,1.8);
\node at (8,-3) {{\blue $S$}};
\draw[dashed] (3.5,-4.5) -- (1,-4.5);
\draw[thick,<-,green!50!black] (1,-4.55) -- (1,-7);
\node[rotate=90,green!50!black] at (1,-7.25) {\Large$\cdots$};
\draw[thick,<-,green!50!black] (1,0.1) -- (1,1);
\node[rotate=90,green!50!black] at (1,1.3) {\Large$\cdots$};
\node[green!50!black] at (0.5,-6) {{ $-R^{(\delta)}$}};
\end{tikzpicture}}}
\end{center}
\caption{Strip of length $L$ and shift $\delta$ with parametrisation suitable for the topological vertex computation.}
\label{Fig:StripLdelta}
\end{figure}

As a next step, we can use (\ref{BBlock}) to verify explicitly (\ref{DualityIdentPartFct}), thus generalising our checks of (\ref{PartitionFc61}) to more generic configurations. As the explicit computations are rather involved, we limit ourselves to cases with $\text{gcd}(N,M)=1$, in which the shifted web (shown in \figref{NewtonPolygonNM}(b)) takes the form of a single `staircase' strip of length $L=NM$, whose external lines are glued with a shift $\delta$ given by eq.~(\ref{Off-SetEq}). This configuration is schematically shown in \figref{Fig:StripLdelta}. As was explained in \cite{Hohenegger:2016yuv}, the duality $X_{N,M}\sim X_{NM,1}$ (assuming $\text{gcd}(N,M)=1$) relies on iteratively using flop and symmetry transformations to change the twisted strip with shift $\delta>0$ into a twisted strip with shift $\delta=0$. While this procedure changes the individual intervals in the (twisted) web-diagram, it is expected that the partition function $\pf{N}{M}$ remains invariant. In the following, we will show explicitly that the partition function is invariant under the transformations that changes $\delta\longrightarrow\delta+1$, which (by induction) explicitly proves (\ref{DualityIdentPartFct}) for $\text{gcd}(N,M)=1$.
\subsubsection{Parametrisation}
In this section, we consider a twisted web diagram of length $L=NM$ in which the external (diagonal) legs are glued together (albeit with a shift $\delta$), so not all line segments in \figref{Fig:StripLdelta} are independent one another. Instead, they have to satisfy the following consistency conditions 
\begin{align}
v_i+m_i&=v_{i+\delta+1}+m_{i+1}\,,\nonumber\\
h_{i+1}+m_{i+1}&=h_{i+\delta +1} + m_i\,, \label{ConsistCond}
\end{align}
which leave a total of $L+2$ independent parameters. Eliminating the $m_i$ from (\ref{ConsistCond}), we obtain 
\begin{equation}
\widehat{a}_i=v_i + h_{i+1} = v_{i+\delta+1} + h_{i+\delta+1}\,, \label{RootsId}
\end{equation}
which in terms of the variables (\ref{DefRootsStrip}) corresponds to $\widehat{a}_{i}=\widehat{b}_{i+\delta+1} $, as has already been taken into account in \figref{Fig:StripLdelta}. There, we have also introduced the (diagonal) distance $S$ between identified external legs, which is represented by the blue curve. In \figref{Fig:StripLdelta}, we chose to measure this distance between the diagonal intervals carrying the integer partition function $\alpha_1$ and $\alpha_1^t$, respectively. In fact, $S$ is the same for any pair of identified legs (we adopt a notation where $h_i=h_{i+L}$ and similarly $\widehat{a}_i=\widehat{a}_{i+L}$)
\begin{align}
S=h_{i-(L-\delta-1)}+\sum_{r=1}^{L-\delta-1}\widehat{a}_{i-r}\,, \,\,\,\,\,1\leq i\leq L\,,
\label{SPara}
\end{align}
due to the consistency conditions (\ref{ConsistCond}) and (\ref{RootsId}). 
We can thus express the horizontal distances as $h_i=S-\sum_{k=\delta+2}^{L}\widehat{a}_{i-k}\,$
which, for the shorthand notation in (\ref{Qroots}), implies that 
\begin{align}
\widehat{Q}_{i,j}= \begin{cases}
					\exp(-S + \sum_{r=j+\delta+1}^{L}\widehat{a}_{i-r}) & \text{if } j + \delta\leq L\\[4pt]
					\exp(-S -\sum_{r=1}^{j+\delta}\widehat{a}_{i-r} & \text{if } j+\delta>L \\
					\end{cases}\nonumber
\end{align}
With this notation, the partition function associated with the twisted diagram in \figref{Fig:StripLdelta} can be written as
\begin{align}
\pf{L=NM}{1}^{(\delta)}=\sum_{\{\alpha\}}\left(\prod_{i=1}^{L}Q_{m_i}^{|\alpha_i|}\right)\,W^{\alpha_1 \dots \alpha_L}_{\alpha_{L-\delta+1} \dots \alpha_{L-\delta}}(\widehat{a}_i,S,\epsilon_1,\epsilon_2)\,.\label{PartitionFunctionTwistDiagL}
\end{align}
Here, we have added the explicit superscript $(\delta)$ to indicate the partition function associated with the twisted web diagram with shift $\delta$. We will show in the next section that $\delta$ in (\ref{PartitionFunctionTwistDiagL}) (for $\text{gcd}(N,M)=1$) can be arbitrarily shifted, provided that we apply the flop and symmetry transformation reviewed in appendix~\ref{StripFlop} to the K\"ahler parameters. Therefore $\pf{L=NM}{1}^{(\delta)}$ is in fact identical to $\pf{N}{M}(\omega,\epsilon_{1,2})$ up to an appropriate transformation of $\omega$. It was proposed in \cite{Hohenegger:2016yuv} (and checked at a particular region in the moduli space) that this also holds for $\text{gcd}(N,M)>1$.

Furthermore, $W$ in (\ref{PartitionFunctionTwistDiagL}) follows from the generic expression (\ref{BBlock}) by identifying the partitions 
\begin{align}
&\beta_i=\alpha_{i+L-\delta-1}\,,&&\forall i=1,\ldots,L
\end{align} 
where we recall that the latter are cyclically identified, \emph{i.e.} $\alpha_i=\alpha_{i+L}$). Concretely, we find
\begin{widetext}
\vskip-0.5cm
\begin{align}
W^{\alpha_1 \dots \alpha_L}_{\alpha_{L-\delta+1} \dots \alpha_{L-\delta}}(\widehat{a}_i,S,\epsilon_1,\epsilon_2)  = W_L(\emptyset) \times \Big[ \Big( \frac{t}{q} \Big)^{\frac{L-1}{2}} \frac{Q_S^L }{Q_{\rho}^{L-\delta-1}}    \Big]^{|\alpha_1|+\dots+|\alpha_L|}  \times\prod_{i,j=1}^L \frac{\vartheta_{\alpha_i \alpha_j}(\widehat{Q}_{i,i-j-\delta};\rho)}{\vartheta_{\alpha_i \alpha_j}(\overline{Q}_{i,i-j} \sqrt{q/t};\rho)} \label{WstripTheta}
\end{align}
\end{widetext}
Here, we introduced
\begin{align}
&Q_S=e^{-S}\,,&&\text{and} &&Q_{\rho}=e^{2\pi i\rho}\,,
\end{align}
with $\rho=\frac{i}{2 \pi} \sum_{i=1}^L\widehat{a}_i$. Furthermore, we have used the identities (\ref{App:CurlJ1}) and (\ref{App:CurlJ2}) to combine the $\mathcal{J}_{\alpha\beta}$ in (\ref{BBlock}) into $\vartheta$-functions (cancelling the factor $\hat{Z}$ in the process). The full partition function (\ref{PartitionFunctionTwistDiagL}) also depends on the parameters $m_i$, which are not all independent. Indeed, by rewriting (\ref{ConsistCond}) in terms of (independent) parameters $S$ and $\widehat{a}_i$, we get the following recursive relation for $m_i$
\begin{align}
m_{i+1}=m_i+\sum_{k=\delta+1}^{L-1}\widehat{\alpha}_{i-k}-\sum_{k=1}^{L-\delta-1}\widehat{\alpha}_{i-k} \,. \label{Recurm}
\end{align}
This implies that only one of the $L$ many parameters $m_i$ can be chosen to be independent. This freedom is parametrised by $R^{(\delta)}$: in \figref{Fig:StripLdelta}, it is shown as the vertical distance between the external legs labelled by the partition $\alpha_1$ and $\alpha_1^t$. Similar to the parameter $S$ due to the consistency conditions (\ref{ConsistCond}) (and equivalently (\ref{Recurm})), this length is the same for any pair of partitions $(\alpha_i,\alpha_i^t)$
\begin{align}
&R^{(\delta)}=m_i - \sum_{k=1}^{L-\delta-1}v_{i-k}\,,&&\forall\,i=1,\ldots,L=NM\,.\nonumber
\end{align}
We remark that $(R^{(\delta)},S,\widehat{a}_{1,\ldots,L})$ are $L+2=NM+2$ independent variables and therefore describe a maximally independent set of K\"ahler parameters for $X_{N,M}$.
\subsubsection{Flop Transformations}
\label{FlopTransfo}
Our strategy for proving (\ref{DualityIdentPartFct}) for $\text{gcd}(N,M)=1$ is to show that 
\begin{align}
\pf{L}{1}^{(\delta)}(\omega,\epsilon_1,\epsilon_2)=\pf{L}{1}^{(\delta+1)}(\omega',\epsilon_1,\epsilon_2)\,\label{ShiftDualityRel}
\end{align} 
for the partition function defined in (\ref{PartitionFunctionTwistDiagL}): indeed, $\pf{L}{1}^{(\delta+1)}$  is associated with a twisted web diagram (with shift $\delta+1$) that can be related to \figref{Fig:StripLdelta} through a series of flop and symmetry transformations, that also relate the K\"ahler parameters $\omega$ and $\omega'$. This duality was first discussed in detail in \cite{Hohenegger:2016yuv} and is reviewed in appendix~\ref{StripFlop}. 

The parameters (\ref{RootsId}) and (\ref{SPara}) in the transformed diagram are the same as those defined in the original diagram
\allowdisplaybreaks{
\begin{align}
\widehat{a}_i'&=v_i'+h_{i+1}'= -h_i + h_i + v_i + h_{i+1}= \widehat{a}_i \nonumber \\
S'&=h_{i-(L-\delta-1)}+ \widehat{a}_{i-(L-\delta-1)} + \sum_{r=1}^{L-\delta-2}\widehat{a}_{i-r} = S \, . \nonumber
\end{align}}
Hence, the $\widehat{a}_i$ always parametrise the distance between two adjacent legs and $S$ measures the vertical distance between two identified legs. 

In order to show that (\ref{ShiftDualityRel}) holds, we first rewrite the building block (\ref{WstripTheta}) in a way that makes the $\delta$ dependence explicit
\begin{widetext}
\vskip-0.5cm
\begin{align}
&W^{\alpha_1 \dots \alpha_L}_{\alpha_{L-\delta+1} \dots \alpha_{L-\delta}}  =  W_L(\emptyset) \times \Big[ \Big( \frac{t}{q} \Big)^{\frac{L-1}{2}} Q_S^L \, \overbrace{Q_{\rho}^{1-K}}^{=C}\,\Big]^{|\alpha_1|+\dots+|\alpha_L|}  \cdot \left( \prod_{i,j=1}^L \frac{1}{\vartheta_{\alpha_i \alpha_j}(\overline{Q}_{i,i-j} \sqrt{q/t};\rho)} \right)  \nonumber \\
&\hspace{0.5cm}\times \overbrace{\left( \prod_{{i \leq j}\atop{j-i<K}} \vartheta_{\alpha_i \alpha_j}(Q_{i,j}^{-1}Q_S) \right) \left( \prod_{{i \leq j}\atop{j-i \geq K}} \vartheta_{\alpha_i \alpha_j}(Q_{i,j}^{-1}Q_SQ_{\rho}) \right)}^{=A} \cdot \overbrace{\left( \prod_{{i>j}\atop{i-j \leq \delta}} \vartheta_{\alpha_i \alpha_j}(Q_{j,i}Q_S) \right) \left( \prod_{{i>j}\atop{i-j > \delta}}  \vartheta_{\alpha_i \alpha_j}(Q_{j,i}Q_SQ_{\rho}^{-1}) \right)}^{=B}\,, \label{SepBlock}
\end{align}
\end{widetext}
where $K=L-\delta$ and the $Q_{i,j}$ are defined as follows
\begin{align}
Q_{i,j}=\prod_{k=0}^{j-i-1} \exp(-\widehat{\alpha}_{i+k})
\end{align}
The difference between $W^{\alpha_1 \dots \alpha_L}_{\alpha_{L-\delta+1} \dots \alpha_{L-\delta}}$ for the original diagram and $W^{\alpha_1 \dots \alpha_L}_{\alpha_{L-\delta} \dots \alpha_{L-\delta-1}}$ for the diagram obtained by flop and symmetry transformations rests in the three terms $A,B$ and $C$. Their respective counterparts in the shifted diagram, denoted by $A',B'$ and $C'$, are given by 
\allowdisplaybreaks{
\begin{align}
A'&=\left( \prod_{{i \leq j}\atop{j-i< K'}} \hspace{-0.30cm}\vartheta_{\alpha_i \alpha_j}\left(\tfrac{ Q_S}{Q_{i,j}}\right) \right) \left( \prod_{{i \leq j}\atop{j-i \geq K'}} \hspace{-0.30cm} \vartheta_{\alpha_i \alpha_j}\left(\tfrac{ Q_SQ_{\rho}}{Q_{i,j}}\right) \right) \,,     \nonumber \\
B'&=\left( \prod_{{i>j}\atop{i-j \leq \delta'}}\hspace{-0.30cm}  \vartheta_{\alpha_i \alpha_j}(Q_{j,i} Q_S) \right) \left( \prod_{{i>j}\atop{i-j > \delta'}}\hspace{-0.30cm}  \vartheta_{\alpha_i \alpha_j}\left(\tfrac{Q_{j,i} Q_S}{Q_{\rho}}\right) \right)  \,,  \nonumber \\
C'&= Q_{\rho}^{1-K'}, \label{CFac}
\end{align}}
where $K'=K-1$ and $\delta'=\delta+1$.  The difference between $A$ and $A'$ (respectively, $B$ and $B'$) lies in the arguments of those $\vartheta $-functions for which $j-i=K'$ (resp. $i-j=\delta+1$): they differ by a factor of $Q_{\rho}$. The difference between $C$ and $C'$ is also a factor of $ Q_{\rho}$.

Finally, we also need to take account of the factors of $Q_{m_i}$ that appear in the full partition function (\ref{PartitionFunctionTwistDiagL}). In the flopped diagram, these are given by (see (\ref{DualityMapAbstract}))
\begin{equation}
Q_{m'_i} = \left\{\begin{array}{ll} Q_{m_i}Q_S^2 & \text{if } \delta'=L  \\ \tfrac{Q_{m_i} Q_S^2}{ \prod_{r=\delta+2}^{L}Q_{a_{i-r}}\prod_{r=1}^{L-\delta-1}Q_{a_{i-r}}}& \text{else} \end{array}	\right. \label{Contri2}
\end{equation}
where we defined $Q_{a_i}=\exp(-\widehat{a}_i)$. In order to show that the equality (\ref{ShiftDualityRel}) holds, we now show that the difference between $\pf{L}{1}^{(\delta)}(\omega,\epsilon_1,\epsilon_2)$ and $\pf{L}{1}^{(\delta+1)}(\omega',\epsilon_1,\epsilon_2)$ can be canceled by merely applying the shift identity (\ref{ThetaFctIdentityShift}) to the $\vartheta$-functions mentioned above in (\ref{CFac}). 

First, we consider the case when the twist in the external legs is $\delta'=L$. Shifting the required $\vartheta$-functions to regain the $\vartheta$-structure of the $\delta=L-1$ strip gives
\begin{align}
\prod_{i=1}^{L} \vartheta_{\alpha_i \alpha_i}(Q_S Q_{\rho})= \prod_{i=1}^L (Q_S^{-2}Q_{\rho}^{-1})^{|\alpha_i|} \vartheta_{\alpha_i \alpha_i}(Q_S) \label{SimpleShift}
\end{align}
The prefactors in (\ref{SimpleShift}) resulting from the shift identity combine with (\ref{CFac}) and (\ref{Contri2}) to reproduce the expression for $\pf{L}{1}^{(\delta)}(\omega,\epsilon_1,\epsilon_2)$, thus proving that (\ref{ShiftDualityRel}) holds for $\delta=L-1$. 

The computations when $\delta' \neq L$ are more involved, so we will simply sketch them. Below we present the $\vartheta$-functions from $W^{\alpha_1 \dots \alpha_L}_{\alpha_{L-\delta} \dots \alpha_{L-\delta-1}}$ that need to be shifted in order to regain $W^{\alpha_1 \dots \alpha_L}_{\alpha_{L-\delta+1} \dots \alpha_{L-\delta}}$. We need to distinguish different cases depending on the partition $\alpha_i$ in question. For the sake of clarity, we focus only on terms resulting from shifts that come to the power $|\alpha_i|$ in each separate case.
\allowdisplaybreaks{\begin{enumerate}
\item For $i \leq \min(K',\delta')$, we shift the following $\vartheta$-functions 
\begin{align}
&\vartheta_{\alpha_{i+\delta+1}\alpha_i}(Q_{i,i+\delta+1}Q_S)\vartheta_{\alpha_i \alpha_{i+K'}} \left( \frac{Q_SQ_{\rho}}{Q_{i,i+K'}} \right) \nonumber \\
&\sim  \, (Q_S^{-2} Q_{i,i+\delta+1}^{-1}Q_{i,i+K'})^{|\alpha_i|} \, \vartheta_{\alpha_{i+\delta+1}\alpha_i} \left(\frac{Q_{i,i+\delta+1}Q_S}{Q_{\rho}} \right)  \nonumber \\
&\hspace{0.2cm} \times\vartheta_{\alpha_i \alpha_{i+K'}} \left(\frac{Q_S}{Q_{i,i+K'}} \right)
\end{align}
\item For $i > \max(K',\delta')$, we shift the following $\vartheta$-functions
\begin{align}
&\vartheta_{\alpha_i\alpha_{i-\delta-1}}(Q_{i-\delta-1,i}Q_S)\vartheta_{\alpha_{i-K'}\alpha_i } \left(\frac{Q_SQ_{\rho}}{Q_{i-K',i}} \right) \nonumber \\
&\sim  \, (Q_S^{-2} Q_{i-\delta-1,i}^{-1}Q_{i-K',i})^{|\alpha_i|} \, \vartheta_{\alpha_i \alpha_{i-\delta-1}} \left(\frac{Q_{i-\delta-1,i}Q_S}{Q_{\rho}} \right)  \nonumber \\
&\hspace{0.2cm} \times \vartheta_{\alpha_{i-K'} \alpha_i } \left(\frac{Q_S}{Q_{i-K',i}} \right)
\end{align}
\item For $\min(K',\delta') < i \leq \max(K',\delta')$, we need to distinguish between two cases: 
\begin{enumerate}
\item When $K'>\delta'$ we shift
\begin{align}
&\vartheta_{\alpha_{i+\delta+1}\alpha_i}(Q_{i,i+\delta+1}Q_S)\vartheta_{\alpha_i \alpha_{i-\delta-1}} \left(Q_SQ_{i-\delta-1,i} \right) \nonumber \\
&\sim  \, (Q_S^{-2} Q_{\rho} Q_{i-\delta-1,i}^{-1} Q_{i,i+\delta+1}^{-1})^{|\alpha_i|} \, \vartheta_{\alpha_{i+\delta+1} \alpha_i } \left(\frac{Q_{i,i+\delta+1}Q_S}{Q_{\rho}} \right) \nonumber \\
&\hspace{0.2cm} \times  \vartheta_{\alpha_i \alpha_{i-\delta-1} } \left(\frac{Q_SQ_{i-\delta-1,i}}{Q_{\rho}} \right)
\end{align}
\item For $K'< \delta'$, we shift 
\begin{align}
&\vartheta_{\alpha_i \alpha_{i+K'}} \left(\frac{Q_S Q_{\rho}}{Q_{i,i+K'}} \right)\vartheta_{ \alpha_{i-K'}\alpha_i} \left(\frac{Q_S Q_{\rho}}{Q_{i-K',i}}  \right) \nonumber \\
&\sim \, (Q_S^{-2} Q_{\rho}^{-1} Q_{i-K',i} Q_{i,i+K'})^{|\alpha_i|} \, \vartheta_{\alpha_i  \alpha_{i+K'} } \left(\frac{Q_S }{Q_{i,i+K'}}\right) \nonumber \\
&\hspace{0.2cm} \times  \vartheta_{ \alpha_{i-K'} \alpha_i} \left(\frac{Q_S }{Q_{i-K',i}}   \right)
\end{align}
\end{enumerate}
\end{enumerate}}
\noindent 
In each case, the factors resulting from shifting the $\vartheta$-functions combine with (\ref{CFac}) and (\ref{Contri2}) to reproduce the expression for $\pf{L}{1}^{(\delta)}(\omega,\epsilon_1,\epsilon_2)$, thus showing that (\ref{ShiftDualityRel}) holds for a generic off-set $\delta$ with $\omega$ and $\omega'$ related trough the duality transformations (\ref{DualityMapAbstract}). As the relation (\ref{ShiftDualityRel}) can be applied iteratively to the point $\delta = L$ (for which $\pf{L}{1}^{(\delta=L)}=\pf{L}{1}$), this further implies (for $\gcd(N,M)=1$)
\begin{align}
&\pf{N}{M}(\mathbf{h},\mathbf{v},\mathbf{m},\epsilon_{1,2})=\pf{NM}{1}(\mathbf{h}',\mathbf{v}',\mathbf{m}',\epsilon_{1,2})\,,
\end{align}
where the K\"ahler parameters $(\mathbf{h},\mathbf{v},\mathbf{m})$ and $(\mathbf{h}',\mathbf{v}',\mathbf{m}')$ are related by the duality map implied by (\ref{DualityMapAbstract}). This proves (\ref{DualityIdentPartFct}) for $\text{gcd}(N,M)=1$. As was already argued in \cite{Hohenegger:2016yuv}, this behavior is expected to hold also for $\text{gcd}(N,M)>1$, however, extending the above computations to these cases is technically more involved and will be left for forthcoming study.
\section{Conclusions}
In this paper, we studied the topological string partition functions of double elliptically fibered Calabi-Yau threefolds $X_{N,M}$ that give rise to a class of LSTs with 8 supercharges via F-theory compactification. We have shown by an explicit example that $X_{3,2}$ and $X_{6,1}$, which are related by flop and symmetry transforms, have the same topological string partition function, hence providing an explicit proof of the duality proposed in \cite{Hohenegger:2016yuv} between $X_{N,M}$ and $X_{\frac{NM}{k},k}$ ($k=\mbox{gcd}(N,M)$). Indeed, while the case discussed here were characterised by $\text{gcd}(N,M)=1$ (in order to simplify the computations), we expect the duality to straightforwardly extend also to $\text{gcd}(N,M)>1$ on general grounds \cite{Li:1998hba, Liu:2005fz, Konishi:2006ev, Taki:2008hb}. 

A logically natural question is whether the diagonal expansion of $\mathcal{Z}_{N,M}$, performed in this paper, can also be interpreted as an instanton expansion of a new gauge theory engineered from the $X_{N,M}$ web. As we will discuss in \cite{paper2}, this turns out indeed the case. The parameters $m_{i}$ can be expressed in terms of coupling constants of a quiver gauge theory related by U-dualities to the usual quiver theory on the compactified M5-branes dual to $X_{N,M}$.  The dual gauge theories coming from the horizontal and vertical description of the $X_{N,M}$ brane web, discussed in \cite{Hohenegger:2015btj}, together with this new dual gauge theory associated with the same web gives us a "triality" of quiver gauge theories.  

The Calabi-Yau threefolds $X_{N,M}$ are resolution of $\mathbb{Z}_{N}\times \mathbb{Z}_{M}$ orbifold of $X_{1,1}$. Therefore, another very interesting and natural question is whether similar duality related by flop transitions can be obtained for different types of orbifolds where $\mathbb{Z}_N\times \mathbb{Z}_M$ is replaced by $\Gamma_1\times \Gamma_2$, with $\Gamma_1, \Gamma_2$ some other discrete subgroups of $SU(2)$.

\section*{Acknowledgement}
We are grateful to Dongsu Bak, Kang-Sin Choi, Taro Kimura and Washington Taylor for useful discussions. 
A.I. would like to acknowledge the ``2017 Simons Summer Workshop on Mathematics and Physics" for hospitality during this work. 
A.I. was supported in part by the Higher Education Commission grant HEC-20-2518.

\appendix
\section{Notation and Useful Identities}\label{App:Identities}
In this appendix, we first introduce some of our notation and provide some useful computational identities. We start by introducing some notation concerning integer partitions. Given an integer partition $\lambda$, we denote its transpose by $\lambda^t$. We define
\begin{align}
|\lambda|= \sum_{i=1}^{l(\lambda)}\lambda_i \,,  \quad ||\lambda||^2=\sum_{i=1}^{l(\lambda)}\lambda_i^2 \,,\quad  ||\lambda^t||^2= \sum_{i=1}^{l(\lambda^t)} (\lambda_i^t)^2 \label{parsum}
\end{align}
where $l(\lambda)$ is the length of the partition (\emph{i.e.} the number of non-zero terms). We also introduce the following functions that are indexed by two integer partitions $\mu$ and $\nu$:
\begin{align}
&\mathcal{J}_{\mu\nu}(x;t,q)=\prod_{k=1}^\infty J_{\mu\nu}(Q_\rho^{k-1}x;t,q)\,,\label{JfncPartBuilding}
\end{align}
where 
{\allowdisplaybreaks
\begin{align}
J_{\mu\nu}(x;t,q)=&\prod_{(i,j)\in\mu}\left(1-x\,t^{\nu^t_j-i+\frac{1}{2}}q^{\mu_i-j+\frac{1}{2}}\right)\nonumber\\
\times&\prod_{(i,j)\in\nu}\left(1-x\,t^{-\mu^t_j+i-\frac{1}{2}}q^{-\nu_i+j-\frac{1}{2}}\right)\,.\nonumber
\end{align}}
We further define
\begin{align}
\vartheta_{\mu\nu}(x;\rho)&=\prod_{(i,j)\in\mu}\vartheta\left(x^{-1}q^{-\nu_j^t+i-\frac{1}{2}}t^{-\mu_i+j-\frac{1}{2}};\rho\right)\,\nonumber\\
&\times\prod_{(i,j)\in\nu}\vartheta\left(x^{-1}q^{\mu_j^t-i+\frac{1}{2}}t^{\nu_i-j+\frac{1}{2}};\rho\right)\,,\label{DefTheta1}
\end{align}
where (with $x=e^{2\pi iz}$ and $Q_\rho=e^{2\pi i\rho}$)
\bea
\vartheta(x;\rho)&=&(x^{\frac{1}{2}}-x^{-\frac{1}{2}})\,\prod_{k=1}^\infty (1-x\,Q^k_{\rho})(1-x^{-1}Q_{\rho}^k)\nonumber\\
&=&\frac{i\,Q_{\rho}^{-\frac{1}{8}}\theta_1(\rho;z)}{\prod_{k=1}^\infty(1-Q_{\rho}^k)}\,.\label{DefTheta2}
\eea
Pairs of $\mathcal{J}_{\mu\nu}$-functions can be combined into $\vartheta_{\mu\nu}$-functions in Eq.~(\ref{DefTheta1}) by utilizing the following identities
\begin{align}
&\mathcal{J}_{\mu \nu}(x;q,t)\mathcal{J}_{\nu \mu}(Q_\rho x^{-1};q,t) \nonumber\\
&\hspace{0.2cm}=x^{\frac{|\mu|+|\nu|}{2}}q^{\frac{||\nu^t||^2-||\mu^t||^2}{4}}t^{\frac{||\mu||^2-||\nu||^2}{4}}  \,\vartheta_{\mu \nu}(x;\rho) \,,
\label{App:CurlJ1}
\end{align}
as well as
\begin{align}
&\frac{(-1)^{|\mu|}t^{|\frac{||\mu||^2}{2}}q^{\frac{||\mu^t||^2}{2}}\tilde{Z}_{\mu}(q,t)\tilde{Z}_{\mu^t}(t,q)}{\mathcal{J}_{\mu \mu}(Q_\rho \sqrt{\frac{t}{q}};q,t)\mathcal{J}_{\mu \mu}(Q_\rho \sqrt{\frac{q}{t}};q,t)} = \frac{1}{\vartheta_{\mu \mu}(\sqrt{\frac{q}{t}};\rho)}\nonumber\\
& = \frac{1}{\vartheta_{\mu \mu}(\sqrt{\frac{t}{q}};\rho)}\,. \label{App:CurlJ2}
\end{align}

In performing the calculations in Section~\ref{Sect:DirectComputation32} and Section~\ref{sec:strip}, we utilized a number of computational identities. Firstly, we found the following two identities helpful for performing sums of skew Schur functions (see \cite{Macdo} (page 93))
{\allowdisplaybreaks
\begin{align}
\sum_{\eta} s_{\eta^t /\mu}(\textbf{x}) s_{\eta/\nu}(\textbf{y})&= \prod_{i,j=1}^{\infty} (1+x_iy_j) \sum_{\tau} s_{\nu^t/\tau}(\textbf{x}) s_{\mu^t / \tau^t}(\textbf{y})  \nonumber \\
\sum_{\eta} s_{\eta /\mu}(\textbf{x}) s_{\eta/\nu}(\textbf{y})&= \prod_{i,j=1}^{\infty} (1-x_iy_j)^{-1} \sum_{\tau} s_{\nu^t/\tau}(\textbf{x}) s_{\mu / \tau}(\textbf{y}) \label{Schur1}
\end{align}}
Secondly, we also used the following identities between products over integer partitions and infinite products \cite{Macdo}
\begin{align}
\prod_{i,j=1}^{\infty} \frac{1-Qq^{-\mu_j^t+i-1}t^{-\nu_i+j}}{1-Qq^{i-1}t^j} &= \prod_{(i,j) \in \nu}(1-Qq^{-\mu_j^t+i-1}t^{-\nu_i+j}) \nonumber \\
&\times \prod_{(i,j)\in \mu} (1-Qq^{\nu_j^t-i}t^{\mu_i-j+1}) \nonumber \\
\prod_{i,j=1}^{\infty} \frac{1-Qq^{-\mu_j^t+i-1}t^{-\mu_i+j}}{1-Qq^{i-1}t^j} &= \prod_{(i,j) \in \mu}(1-Qq^{-\mu_j^t+i-1}t^{-\mu_i+j}) \nonumber \\
&\times (1-Qq^{\mu_j^t-i}t^{\mu_i-j+1}) \label{InfProd1}
\end{align}
Finally, we also recall the following identity \cite{Taki},
\begin{equation}
\sum_{(i,j) \in \nu} \mu_j^t = \sum_{(i,j)\in \mu} \nu_j^t\,. \label{TranSum}
\end{equation}
\section{Diagonal Partition Function $\pf{3}{2}$}\label{App:32ExpComputation}
In this appendix, we present details of the calculation of the building block $W$ for the computation of the diagonal expansion of $\pf{3}{2}$, which is introduced in Eq.~(\ref{Wblock}). Using the definition of the refined topological vertex
\begin{widetext}
\vskip-0.5cm
\begin{align}
C_{\lambda\mu\nu}(t,q)=q^{\tfrac{||\mu||^2}{2}}t^{-\tfrac{||\mu^t||^2}{2}}q^{\tfrac{||\nu||^2}{2}}\,\tilde{Z}_\nu(t,q)\,\sum_\eta\left(\frac{q}{t}\right)^{\frac{|\eta|+|\lambda|-|\mu|}{2}}\,s_{\lambda^t/\eta}(t^{-\rho} q^{-\nu})\,s_{\mu/\eta}(q^{-\rho}t^{-\nu^t})\, \label{DefRefinedVertex}
\end{align}
\end{widetext}
where $s_{\mu/\nu}$ are skew Schur functions and 
\begin{align}
\tilde{Z}_\nu(t,q)=\prod_{(i,j)\in\nu}\left(1-t^{\nu_j^t-i+1}q^{\nu_i-j}\right)^{-1} \, , 
\end{align}
the expression (\ref{Wblock}) can be written in the form (\ref{W}).

In what follows, we denote the whole set of variables as bold expressions, \emph{i.e.} $\mathbf{x}=\{x_i\}_{i=1,\ldots,6}$ and, for simplicity, we shall not consider the prefactor $\hat{Z}$ in (\ref{W}): indeed, the summation over the skew Schur function in (\ref{W}) can be written in the form
\begin{widetext}
\vskip-0.5cm
\begin{align}
G(\bold{x},\bold{y},\bold{w},\bold{z})= \sum_{\substack{\{\mu\} \{\nu\} \\ \{\eta\} \{\tilde{\eta}\}}} \prod_{i=1}^6 (-Q_{h_i})^{|\mu_i|} (-Q_{v_i})^{|\nu_i|} s_{\mu_i/ \tilde{\eta}_{i+3}}(x_{i+3})\,s_{\mu_i^t/ \eta_{i+5}}(y_{i+5}) s_{\nu_i^t/ \eta_i}(w_i) s_{\nu_i/ \tilde{\eta}_{i+3}}(z_{i+3})\,.\label{Sum}
\end{align}
\end{widetext}
We can perform the sum over skew Schur functions using a method similar to those in \cite{Haghighat:2013gba}: repeatedly using the identities (\ref{Schur1}) for summing skew Schur functions, we can write:
\begin{widetext}
\vskip-0.5cm
{\allowdisplaybreaks
\begin{align}
G(\bold{x},\bold{y},\bold{w},\bold{z})= P \times \sum_{\substack{\{\mu\} \{\nu\} \\ \{\eta\} \{\tilde{\eta}\}}} \prod_{i=1}^6 &(-Q_{h_i})^{|\mu_i|} (-Q_{v_i})^{|\nu_i|} s_{\mu_{i-1}/ \tilde{\eta}_{i+2}}(Q_{h_i}Q_{v_{i-1}}x_{i+3}) s_{\mu_{i+1}^t/ \eta_{i}}(\tilde{Q}_iy_{i+5})\nonumber \\
 &\times  \, s_{\nu_{i-1}^t/ \eta_{i-1}}(\tilde{Q}_iw_i) s_{\nu_{i+1}/ \tilde{\eta}_{i+2}}(Q_{h_{i+1}}Q_{v_i}z_{i+3})\,,\label{G4Resum}
\end{align}}
where we introduced the notation $\tilde{Q}_i=Q_{h_i} Q_{v_i}$ and
\begin{align}
P= \prod_{i=1}^6 \prod_{r,s=1}^{\infty} &\frac{(1-Q_{h_i}  x_{i+3,r}y_{i+5,s})(1-Q_{v_i} w_{i,r}z_{i+3,s})}{(1-Q_{h_i}Q_{v_{i-1}}  x_{i+3,r}z_{i+2,s})(1-\tilde{Q}_i y_{i+5,r}w_{i,s})}   \, \frac{(1-Q_{h_i}\tilde{Q}_{i-1} x_{i+3,r}y_{i+4,s})(1-Q_{v_i} \tilde{Q}_{i+1} w_{i+1,r}z_{i+3,s})}{(1-Q_{h_i}\tilde{Q}_{i-1}Q_{v_{i-2}}  x_{i+3,r}z_{i+1,s})(1-\tilde{Q}_i \tilde{Q}_{i+1}  y_{i+5,r}w_{i+1,s})}\,. \nonumber
\end{align}
\end{widetext}
Thus, in (\ref{G4Resum}), we find an expression similar to the expression (\ref{Sum}) except for the difference that the partitions have been shifted, \emph{e.g.} the Schur functions for $i=1$ have been replaced in the following fashion
\begin{widetext}
\vskip-0.5cm
\begin{align}
s_{\mu_{1}/ \tilde{\eta}_{4}}(x_{4}) s_{\mu_{1}^t/ \eta_{6}}(y_{6}) s_{\nu_{1}^t/ \eta_{1}}(w_1) s_{\nu_{1}/ \tilde{\eta}_{4}}(z_{4})
\to s_{\mu_{6}/ \tilde{\eta}_{3}}(Q_{h_1}Q_{v_6}x_{4}) s_{\mu_{2}^t/ \eta_{1}}(\tilde{Q}_1y_{6}) s_{\nu_{6}^t/ \eta_{6}}(\tilde{Q}_1w_1) s_{\nu_{2}/ \tilde{\eta}_{5}}(Q_{h_2}Q_{v_1}z_{4})\,.\nonumber
\end{align}
\end{widetext}
By repeating this procedure multiple times, we again obtain the quantity $G$ defined in (\ref{Sum}), up to a prefactor  $P_1$ (The precise form of $P_1$ turns out not to be important) and a shift of all arguments, as
\begin{align}
G(\bold{x},\bold{y},\bold{w},\bold{z})= P_1 \times G(\bar{Q}\bold{x},\bar{Q}\bold{y},\bar{Q}\bold{w},\bar{Q}\bold{z})\,.\label{RecursionRelation}
\end{align}
where we abbreviated $\bar{Q}=\tilde{Q}_1\tilde{Q}_2\tilde{Q}_3\tilde{Q}_4\tilde{Q}_5\tilde{Q}_6$. Repeating further, the relation (\ref{RecursionRelation}) leads to the following recursion relation for all integers $n$ 
\begin{align}
&G(\bold{x},\bold{y},\bold{w},\bold{z})=P_n \times G(\bar{Q}^n\bold{x},\bar{Q}^n\bold{y},\bar{Q}^n\bold{w},\bar{Q}^n\bold{z})\,.\label{Recursion6}
\end{align}
Using the fact that $\lim_{n\to \infty}\bar{Q}^n = 0$, we have the relation
\begin{widetext}
\vskip-0.5cm
{\allowdisplaybreaks
\begin{align}
\lim_{n \to \infty}& G(\bar{Q}^n\bold{x},\bar{Q}^n\bold{y},\bar{Q}^n\bold{w},\bar{Q}^n\bold{z})  \nonumber \\
&= \lim_{n \to \infty} \sum_{\substack{\{\mu\} \{\nu\} \\ \{\eta\} \{\tilde{\eta}\}}} \prod_i^6 (-Q_{h_i})^{|\mu_i|} (-Q_{v_i})^{|\nu_i|} s_{\mu_i/ \tilde{\eta}_{i+3}}((\bar{Q}^2)^nx_{i+3}) s_{\mu_i^t/ \eta_{i+5}}(y_{i+5}) s_{\nu_i^t/ \eta_i}((\bar{Q}^2)^nw_i) s_{\nu_i/ \tilde{\eta}_{i+3}}(z_{i+3}) \nonumber \\
&=  \sum_{\{\eta\} \{\tilde{\eta}\}} \prod_i^6 (-Q_{h_i})^{|\tilde{\eta}_{i+3}|} (-Q_{v_i})^{|\eta_i|} s_{\tilde{\eta}_{i+3}^t/\eta_{i+5}}(y_{i+5}) s_{\eta_i^t/\tilde{\eta}_{i+3}}(z_{i+3})\,,\label{ImposeConditionsOnEtas}
\end{align}}
\end{widetext}
as the only non-vanishing terms in this limit correspond to $\mu_i = \tilde{\eta}_{i+3}$ and $\nu_i^t= \eta_i$. The skew Schur functions are non-zero when $\eta_{i+2} \subset \tilde{\eta}_i^t$ and $\tilde{\eta}_{i+3} \subset \eta_i^t$, such that we have
{\allowdisplaybreaks
\bea\nonumber
&&\eta_3 \subset \tilde{\eta}_1^t \,, \eta_4 \subset \tilde{\eta}_2^t \,,\eta_5 \subset \tilde{\eta}_3^t \,,\eta_6 \subset \tilde{\eta}_4^t \,,\eta_1 \subset \tilde{\eta}_5^t \,,
\eta_2 \subset \tilde{\eta}_6^t\\\nonumber
&&\tilde{\eta}_4 \subset \eta_1^t\, ,\tilde{\eta}_5 \subset \eta_2^t\,, \tilde{\eta}_6 \subset \eta_3^t\,, \tilde{\eta}_1 \subset \eta_4^t\,, \tilde{\eta}_2 \subset \eta_5^t\,, \tilde{\eta}_3 \subset \eta_6^t\,.
\eea}
\noindent
By considering the transpose of the conditions in the second line conditions, we get
\bea\nonumber
&&\eta_3 \subset \tilde{\eta}_1^t \subset \eta_4 \subset \tilde{\eta}_2^t \subset \eta_5 \subset \tilde{\eta}_3^t \subset \eta_6 \subset \tilde{\eta}_4^t \subset \eta_1 \subset\\\nonumber
&& \tilde{\eta}_5^t \subset \eta_2 \subset \tilde{\eta}_6^t \subset \eta_3\,,
\eea
which implies that the summation in (\ref{ImposeConditionsOnEtas}) only receives contributions for
\bea\nonumber
&&\eta_1=\eta_2=\eta_3=\eta_4=\eta_5=\eta_6=\tilde{\eta}_1^t=\tilde{\eta}_2^t=\tilde{\eta}_3^t=\\\nonumber
&&\tilde{\eta}_4^t=\tilde{\eta}_5^t=\tilde{\eta}_6^t\,,
\eea
such that (\ref{ImposeConditionsOnEtas}) is reduced to
\begin{align}
\lim_{n \to \infty} G(\bar{Q}^n\bold{x},\bar{Q}^n\bold{y},\bar{Q}^n\bold{w},\bar{Q}^n\bold{z})=\sum_{\eta} \bar{Q}^{|\eta|} =\prod_{k=1}^{\infty} (1-\bar{Q}^k)^{-1}\,. \nonumber
\end{align}
We thus obtain
\begin{equation}
G(\bold{x},\bold{y},\bold{w},\bold{z})= P_{\infty} \cdot \prod_{k=1}^{\infty} (1-\bar{Q}^{k})^{-1} \, , 
\end{equation}
where the multiplicative prefactor can be worked out to be
\begin{widetext}
\vskip-0.5cm
{\allowdisplaybreaks
\begin{align}
&P_{\infty}=  \prod_{i=1}^6 \prod_{r,s=1}^{\infty}  \prod_{k=1}^{\infty}  \frac{(1-Q_{h_i} \bar{Q}^{k-1} x_{i+3,r}y_{i+5,s})(1-Q_{v_i} \bar{Q}^{k-1}w_{i,r}z_{i+3,s})}{(1-Q_{h_i}Q_{v_{i-1}} \bar{Q}^{k-1} x_{i+3,r}z_{i+2,s})(1-\tilde{Q}_i  \bar{Q}^{k-1}y_{i+5,r}w_{i,s})} \nonumber \\
&\times  \frac{(1-Q_{h_i}\tilde{Q}_{i-1}\bar{Q}^{k-1} x_{i+3,r}y_{i+4,s})(1-Q_{v_i} \tilde{Q}_{i+1} \bar{Q}^{k-1} w_{i+1,r}z_{i+3,s})}{(1-Q_{h_i}\tilde{Q}_{i-1}Q_{v_{i-2}} \bar{Q}^{k-1} x_{i+3,r}z_{i+1,s})(1-\tilde{Q}_i \tilde{Q}_{i+1} \bar{Q}^{k-1} y_{i+5,r}w_{i+1,s})} \nonumber \\
&\times \frac{(1-Q_{h_i} \tilde{Q}_{i-1}\tilde{Q}_{i-2}\bar{Q}^{k-1} x_{i+3,r}y_{i+3,s})(1-Q_{v_i} \tilde{Q}_{i+1} \tilde{Q}_{i+2} \bar{Q}^{k-1} w_{i+2,r}z_{i+3,s})}{(1-Q_{h_i}\tilde{Q}_{i-1}\tilde{Q}_{i-2}Q_{v_{i-3}} \bar{Q}^{k-1} x_{i+3,r}z_{i,s})(1-\tilde{Q}_{i}\tilde{Q}_{i+1}\tilde{Q}_{i+2} \bar{Q}^{k-1} y_{i+5,r}w_{i+2,s})} \nonumber \\
&\times \frac{(1-Q_{h_i} \tilde{Q}_{i-1}\tilde{Q}_{i-2}\tilde{Q}_{i-3}\bar{Q}^{k-1} x_{i+3,r}y_{i+2,s})(1-Q_{v_i} \tilde{Q}_{i+1} \tilde{Q}_{i+2} \tilde{Q}_{i+3} \bar{Q}^{k-1} w_{i+3,r}z_{i+3,s})}{(1-Q_{h_i}\tilde{Q}_{i-1}\tilde{Q}_{i-2}\tilde{Q}_{i-3}Q_{v_{i-4}} \bar{Q}^{k-1} x_{i+3,r}z_{i+5,s})(1-\tilde{Q}_{i}\tilde{Q}_{i+1}\tilde{Q}_{i+2} \tilde{Q}_{i+3} \bar{Q}^{k+5} y_{i+5,r}w_{i+3,s})} \nonumber \\
&\times \frac{(1-Q_{h_i} \tilde{Q}_{i-1}\tilde{Q}_{i+4}\tilde{Q}_{i-3}\tilde{Q}_{i+2}\bar{Q}^{k-1} x_{i+3,r}y_{i+1,s})(1-Q_{v_i} \tilde{Q}_{i+1} \tilde{Q}_{i+2} \tilde{Q}_{i+3} \tilde{Q}_{i+4}\bar{Q}^{k-1} w_{i+4,r}z_{i+3,s})}{(1-Q_{h_i}\tilde{Q}_{i-1}\tilde{Q}_{i+4}\tilde{Q}_{i-3}\tilde{Q}_{i-4}Q_{v_{i-5}} \bar{Q}^{k-1} x_{i+3,r}z_{i-2,s})(1-\tilde{Q}_{i}\tilde{Q}_{i+1}\tilde{Q}_{i+2} \tilde{Q}_{i+3} \tilde{Q}_{i+4}\bar{Q}^{k-1} y_{i+5,r}w_{i+4,s})} \nonumber \\
&\times \frac{(1-Q_{h_i} \tilde{Q}_{i-1}\tilde{Q}_{i-2}\tilde{Q}_{i-3}\tilde{Q}_{i-4}\tilde{Q}_{i-5}\bar{Q}^{k-1} x_{i+3,r}y_{i,s})(1-Q_{v_i} \tilde{Q}_{i+1} \tilde{Q}_{i+2} \tilde{Q}_{i+3} \tilde{Q}_{i+4} \tilde{Q}_{i+5}\bar{Q}^{k-1} w_{i+5,r}z_{i+3,s})}{(1-\bar{Q}^{k} x_{i+3,r}z_{i+3,s})(1-\bar{Q}^{k} y_{i+5,r}w_{i+5,s})} \,.
\end{align}
}
\end{widetext}
Finally, using the consistency conditions (\ref{InitC1})-(\ref{InitC6}), this expression can be simplified to Eq.(\ref{blockW}).

\section{Flop Transforms for Twisted Diagrams}
\label{StripFlop}
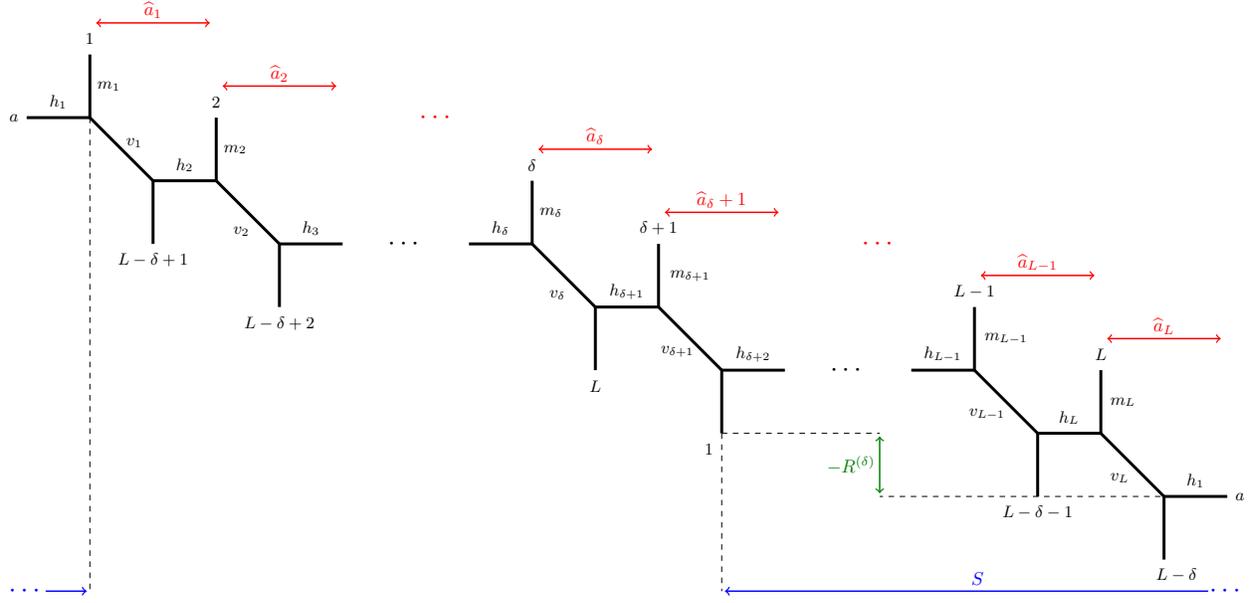
\begin{figure*}
\scalebox{0.70}{\parbox{23.5cm}{\begin{tikzpicture}[scale = 1.2]
\draw[ultra thick] (-1,0) -- (0,0);
\draw[ultra thick] (0,1) -- (0,0);
\draw[ultra thick] (0,0) -- (1,-1);
\draw[ultra thick] (1,-1) -- (2,-1);
\draw[ultra thick] (1,-1) -- (1,-2);
\draw[ultra thick] (2,-1) -- (2,0);
\draw[ultra thick] (2,-1) -- (3,-2);
\draw[ultra thick] (3,-2) -- (4,-2);
\draw[ultra thick] (3,-2) -- (3,-3);
\node at (5,-2) {\Large $\cdots$};
\draw[ultra thick] (6,-2) -- (7,-2) -- (8,-3) -- (9,-3) -- (10,-4) -- (11,-4);
\draw[ultra thick] (7,-2) -- (7,-1);
\draw[ultra thick] (8,-3) -- (8,-4);
\draw[ultra thick] (9,-3) -- (9,-2);
\draw[ultra thick] (10,-4) -- (10,-5);
\node at (12,-4) {\Large $\cdots$};
\draw[ultra thick] (13,-4) -- (14,-4) -- (15,-5) -- (16,-5) -- (17,-6) -- (18,-6);
\draw[ultra thick] (14,-4) -- (14,-3);
\draw[ultra thick] (15,-5) -- (15,-6);
\draw[ultra thick] (16,-5) -- (16,-4);
\draw[ultra thick] (17,-6) -- (17,-7);
\node at (-0.5,0.25) {{\small $h_1$}};
\node at (1.5,-0.75) {{\small $h_2$}};
\node at (3.5,-1.75) {{\small $h_3$}};
\node at (6.5,-1.75) {{\small $h_\delta$}};
\node at (8.5,-2.75) {{\small $h_{\delta+1}$}};
\node at (10.5,-3.75) {{\small $h_{\delta+2}$}};
\node at (13.5,-3.75) {{\small $h_{L-1}$}};
\node at (15.5,-4.75) {{\small $h_{L}$}};
\node at (17.5,-5.75) {{\small $h_{1}$}};
\node at (0.3,0.5) {{\small $m_1$}};
\node at (2.3,-0.5) {{\small $m_2$}};
\node at (7.3,-1.5) {{\small $m_\delta$}};
\node at (9.5,-2.5) {{\small $m_{\delta+1}$}};
\node at (14.5,-3.5) {{\small $m_{L-1}$}};
\node at (16.35,-4.5) {{\small $m_{L}$}};
\node at (0.7,-0.4) {{\small $v_1$}};
\node at (2.4,-1.8) {{\small $v_2$}};
\node at (7.4,-2.8) {{\small $v_\delta$}};
\node at (9.3,-3.7) {{\small $v_{\delta+1}$}};
\node at (14.2,-4.7) {{\small $v_{L-1}$}};
\node at (16.3,-5.7) {{\small $v_{L}$}};
\node at (-1.2,0) {{\small \bf $a$}};
\node at (18.2,-6) {{\small \bf $a$}};
\node at (0,1.25) {{\small \bf $1$}};
\node at (2,0.25) {{\small \bf $2$}};
\node at (7,-0.75) {{\small \bf $\delta$}};
\node at (9,-1.75) {{\small \bf $\delta+1$}};
\node at (14,-2.75) {{\small \bf $L-1$}};
\node at (16,-3.75) {{\small \bf $L$}};
\node at (1,-2.25) {{\small \bf $L-\delta+1$}};
\node at (3,-3.25) {{\small \bf $L-\delta+2$}};
\node at (8,-4.25) {{\small \bf $L$}};
\node at (9.8,-5.25) {{\small \bf $1$}};
\node at (15,-6.25) {{\small \bf $L-\delta-1$}};
\node at (17.2,-7.25) {{\small \bf $L-\delta$}};
%
\draw[thick,red,<->] (0.1,1.5) -- (1.9,1.5);
\node[red] at (1,1.7) {$\widehat{a}_1$};
\draw[thick,red,<->] (2.1,0.5) -- (3.9,0.5);
\node[red] at (3,0.7) {$\widehat{a}_2$};
\node[red] at (5.5,0) {\Large $\cdots$};
\draw[thick,red,<->] (7.1,-0.5) -- (8.9,-0.5);
\node[red] at (8,-0.3) {$\widehat{a}_\delta$};
\draw[thick,red,<->] (9.1,-1.5) -- (10.9,-1.5);
\node[red] at (10,-1.3) {$\widehat{a}_\delta+1$};
\node[red] at (12.5,-2) {\Large $\cdots$};
\draw[thick,red,<->] (14.1,-2.5) -- (15.9,-2.5);
\node[red] at (15,-2.3) {$\widehat{a}_{L-1}$};
\draw[thick,red,<->] (16.1,-3.5) -- (17.9,-3.5);
\node[red] at (17,-3.3) {$\widehat{a}_{L}$};
\draw[dashed] (10,-5) -- (12.5,-5);
\draw[dashed] (18,-6) -- (12.5,-6);
\draw[green!50!black,<->,thick] (12.5,-5.05) -- (12.5,-5.95);
\node[green!50!black] at (12,-5.5) {{ $-R^{(\delta)}$}};
\draw[dashed] (10,-5) -- (10,-7.5);
\draw[thick,<-,blue] (10.05,-7.5) -- (17.7,-7.5);
\node[blue] at (18,-7.5) {\Large $\cdots$};
\draw[dashed] (0,0) -- (0,-7.5);
\draw[thick,->,blue] (-0.7,-7.5) -- (-0.05,-7.5);
\node[blue] at (-1,-7.5) {\Large $\cdots$};
\node[blue] at (14,-7.3) {{ $S$}};
\end{tikzpicture}
}}
\caption{\sl Twisted web diagram with shift $\delta$. This diagram is obtained from \figref{Fig:StripLdelta} through an $SL(2,\mathbb{Z})$ transformation.}
\label{StripDelta}
\end{figure*}

In this appendix, we recapitulate a series of flop and symmetry transforms of a twisted web diagram of length $L$ with generic shift $\delta$ (as shown in \figref{Fig:StripLdelta} and equivalently in \figref{StripDelta}, along with a labelling of all relevant parameters and integer partitions) that relate it to a twisted web diagram of the same length but with shift $\delta+1$. The duality between the two twisted web diagrams was first discussed in \cite{Hohenegger:2016yuv}. It can be applied iteratively to obtain a web diagram with shift $\delta=0$, which was used in \cite{Hohenegger:2016yuv} to argue for a duality between $X_{N,M}$ and $X_{N',M'}$ for $NM=N'M'$ and $\text{gcd}(N,M)=k=\text{gcd}(N',M')$. Here, we are primarily interested in the case $\text{gcd}(N,M)=1$.

We start out by performing an $SL(2,\mathbb{Z})$ transformation of the twisted web diagram in \figref{Fig:StripLdelta}, whose result is shown in \figref{StripDelta}.
This does not change the K\"ahler parameters, as indicated in \figref{StripDelta}. 

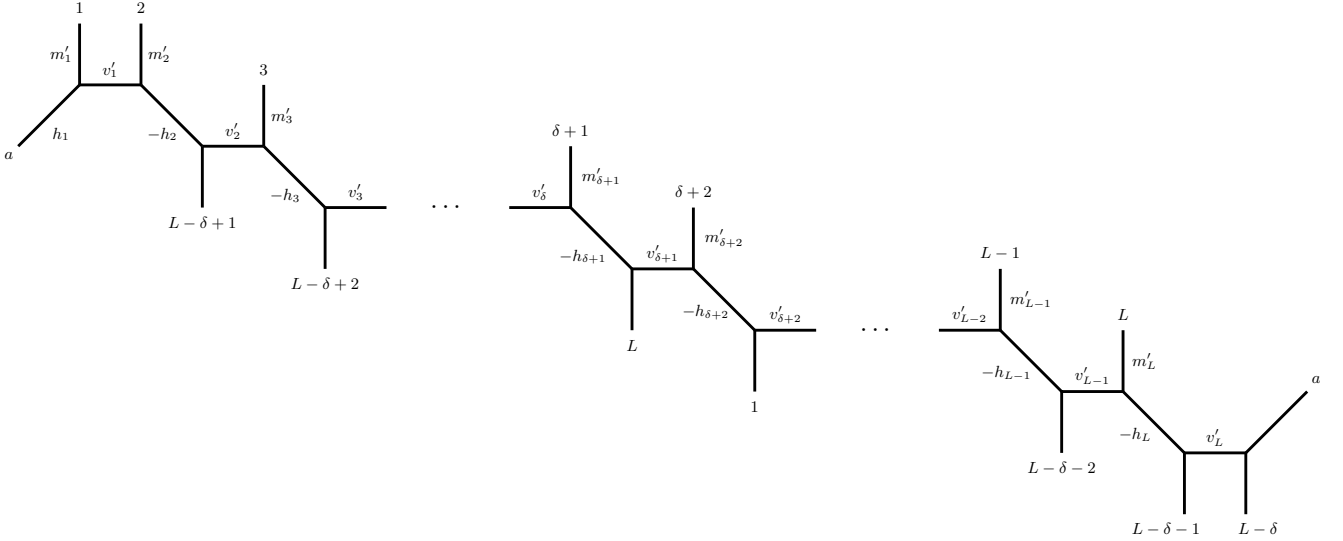
\begin{figure*}
\scalebox{0.68}{\parbox{25.8cm}{\begin{tikzpicture}[scale = 1.2]
\draw[ultra thick] (-2,-1) -- (-1,0) -- (0,0);
\draw[ultra thick] (-1,0) -- (-1,1);
\draw[ultra thick] (0,1) -- (0,0);
\draw[ultra thick] (0,0) -- (1,-1);
\draw[ultra thick] (1,-1) -- (2,-1);
\draw[ultra thick] (1,-1) -- (1,-2);
\draw[ultra thick] (2,-1) -- (2,0);
\draw[ultra thick] (2,-1) -- (3,-2);
\draw[ultra thick] (3,-2) -- (4,-2);
\draw[ultra thick] (3,-2) -- (3,-3);
\node at (5,-2) {\Large $\cdots$};
\draw[ultra thick] (6,-2) -- (7,-2) -- (8,-3) -- (9,-3) -- (10,-4) -- (11,-4);
\draw[ultra thick] (7,-2) -- (7,-1);
\draw[ultra thick] (8,-3) -- (8,-4);
\draw[ultra thick] (9,-3) -- (9,-2);
\draw[ultra thick] (10,-4) -- (10,-5);
\node at (12,-4) {\Large $\cdots$};
\draw[ultra thick] (13,-4) -- (14,-4) -- (15,-5) -- (16,-5) -- (17,-6) -- (18,-6) -- (19,-5);
\draw[ultra thick] (14,-4) -- (14,-3);
\draw[ultra thick] (15,-5) -- (15,-6);
\draw[ultra thick] (16,-5) -- (16,-4);
\draw[ultra thick] (17,-6) -- (17,-7);
\draw[ultra thick] (18,-6) -- (18,-7);
\node at (-0.5,0.25) {{\small $v'_1$}};
\node at (1.5,-0.75) {{\small $v'_2$}};
\node at (3.5,-1.75) {{\small $v'_3$}};
\node at (6.5,-1.75) {{\small $v'_\delta$}};
\node at (8.5,-2.75) {{\small $v'_{\delta+1}$}};
\node at (10.5,-3.75) {{\small $v'_{\delta+2}$}};
\node at (13.5,-3.75) {{\small $v'_{L-2}$}};
\node at (15.5,-4.75) {{\small $v'_{L-1}$}};
\node at (17.5,-5.75) {{\small $v'_{L}$}};
\node at (-1.3,0.5) {{\small $m'_1$}};
\node at (0.3,0.5) {{\small $m'_2$}};
\node at (2.3,-0.5) {{\small $m'_3$}};
\node at (7.5,-1.5) {{\small $m'_{\delta+1}$}};
\node at (9.5,-2.5) {{\small $m'_{\delta+2}$}};
\node at (14.5,-3.5) {{\small $m'_{L-1}$}};
\node at (16.35,-4.5) {{\small $m'_{L}$}};
\node at (-1.3,-0.8) {{\small $h_1$}};
\node at (0.35,-0.8) {{\small $-h_2$}};
\node at (2.35,-1.8) {{\small $-h_3$}};
\node at (7.2,-2.8) {{\small $-h_{\delta+1}$}};
\node at (9.2,-3.7) {{\small $-h_{\delta+2}$}};
\node at (14.1,-4.7) {{\small $-h_{L-1}$}};
\node at (16.2,-5.7) {{\small $-h_{L}$}};
\node at (-2.15,-1.15) {{\small \bf $a$}};
\node at (19.15,-4.8) {{\small \bf $a$}};
\node at (-1,1.25) {{\small \bf $1$}};
\node at (0,1.25) {{\small \bf $2$}};
\node at (2,0.25) {{\small \bf $3$}};
\node at (7,-0.75) {{\small \bf $\delta+1$}};
\node at (9,-1.75) {{\small \bf $\delta+2$}};
\node at (14,-2.75) {{\small \bf $L-1$}};
\node at (16,-3.75) {{\small \bf $L$}};
\node at (1,-2.25) {{\small \bf $L-\delta+1$}};
\node at (3,-3.25) {{\small \bf $L-\delta+2$}};
\node at (8,-4.25) {{\small \bf $L$}};
\node at (10,-5.25) {{\small \bf $1$}};
\node at (15,-6.25) {{\small \bf $L-\delta-2$}};
\node at (16.7,-7.25) {{\small \bf $L-\delta-1$}};
\node at (18.2,-7.25) {{\small \bf $L-\delta$}};
\end{tikzpicture}}}
\caption{\sl Web diagram obtained by flopping $h_2,\ldots,h_L$ in \figref{StripDelta}.}
\label{StripDeltaFlop}
\end{figure*}

As the next step, we perform a flop transformation on the intervals $\{h_2,\ldots,h_L\}$. After suitable cutting and re-gluing, the twisted web can be presented in the form shown in \figref{StripDeltaFlop}. Notice that not only $h_i\longrightarrow-h_i$ for $i=2,\ldots,L$, but also the remaining parameters have changed according to
{\allowdisplaybreaks
\begin{align}
&m'_1=m_1+h_{\delta+2}\,,&&v'_1=v_1+h_{2}\,,\nonumber\\
&m'_2=m_2+h_2+h_{\delta+3}\,,&&v'_2=v_2+h_2+h_{3}\,,\nonumber\\
&\ldots &&\ldots\nonumber\\
&m'_{L-1}=m_{L-1}+h_{L-1}+h_{\delta}\,, &&v'_{L-1}=v_{L-1}+h_{L-1}+h_L \,,\nonumber\\
&m'_L=m_L+h_L+h_{\delta+1}\,,&&v'_L=v_L+h_L\,,
\end{align}}
which reflect how the various intervals are connected to the ones that are being flopped.

As the final step, performing a flop transformation of $h_1$, we get the web shown in \figref{StripDeltaFlop2}, which is a twisted web diagram with shift $\delta+1$. Concerning the parameters that are associated with the individual line segments in the web, we have
\begin{widetext}
\begin{align}
&v_1''=v_1+h_1+h_2\,,&&v''_L=v_L+h_1+h_L\,,&&m''_1=m_1+h_1+h_{\delta+2}\,,&&m''_{L-\delta}=m_{L-\delta}+h_1+h_{L-\delta}\,,
\end{align}
\end{widetext}
and
\begin{align}
&v''_i=v'_i\,,&&h''_i=h'_i\,,&&m''_i=m'_i\,,&&\forall i\notin\{1,L\}\,.
\end{align}
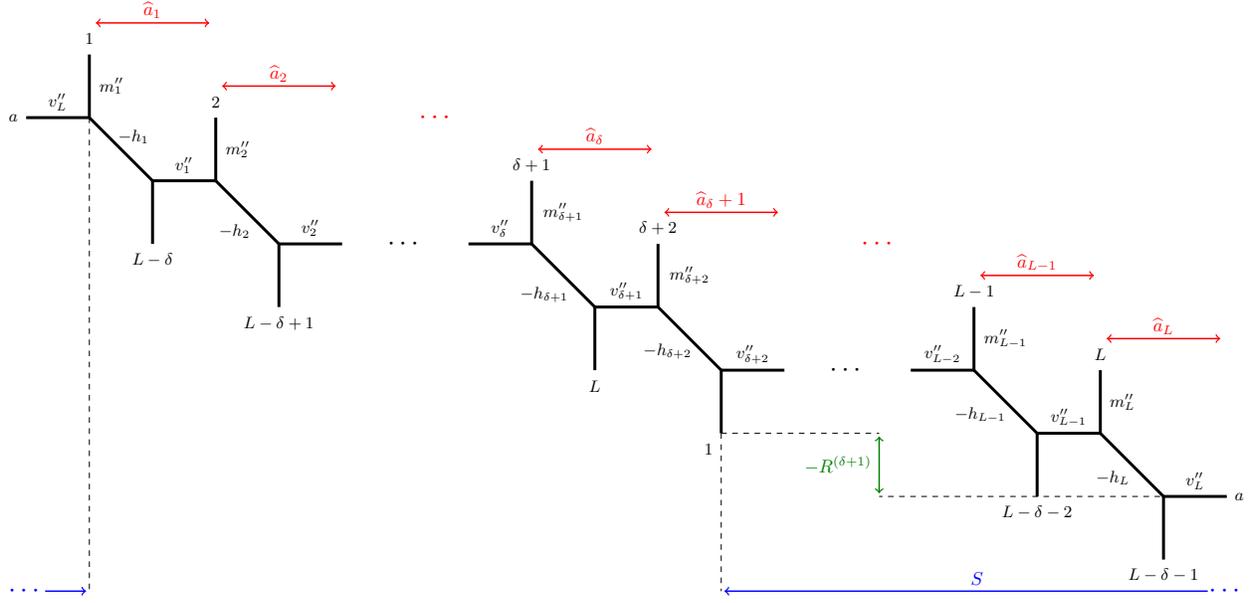
\begin{figure*}
\scalebox{0.70}{\parbox{24cm}{\begin{tikzpicture}[scale = 1.2]
\draw[ultra thick] (-1,0) -- (0,0);
\draw[ultra thick] (0,1) -- (0,0);
\draw[ultra thick] (0,0) -- (1,-1);
\draw[ultra thick] (1,-1) -- (2,-1);
\draw[ultra thick] (1,-1) -- (1,-2);
\draw[ultra thick] (2,-1) -- (2,0);
\draw[ultra thick] (2,-1) -- (3,-2);
\draw[ultra thick] (3,-2) -- (4,-2);
\draw[ultra thick] (3,-2) -- (3,-3);
\node at (5,-2) {\Large $\cdots$};
\draw[ultra thick] (6,-2) -- (7,-2) -- (8,-3) -- (9,-3) -- (10,-4) -- (11,-4);
\draw[ultra thick] (7,-2) -- (7,-1);
\draw[ultra thick] (8,-3) -- (8,-4);
\draw[ultra thick] (9,-3) -- (9,-2);
\draw[ultra thick] (10,-4) -- (10,-5);
\node at (12,-4) {\Large $\cdots$};
\draw[ultra thick] (13,-4) -- (14,-4) -- (15,-5) -- (16,-5) -- (17,-6) -- (18,-6);
\draw[ultra thick] (14,-4) -- (14,-3);
\draw[ultra thick] (15,-5) -- (15,-6);
\draw[ultra thick] (16,-5) -- (16,-4);
\draw[ultra thick] (17,-6) -- (17,-7);
\node at (-0.5,0.25) {{\small $v''_L$}};
\node at (1.5,-0.75) {{\small $v''_1$}};
\node at (3.5,-1.75) {{\small $v''_2$}};
\node at (6.5,-1.75) {{\small $v''_{\delta}$}};
\node at (8.5,-2.75) {{\small $v''_{\delta+1}$}};
\node at (10.5,-3.75) {{\small $v''_{\delta+2}$}};
\node at (13.5,-3.75) {{\small $v''_{L-2}$}};
\node at (15.5,-4.75) {{\small $v''_{L-1}$}};
\node at (17.5,-5.75) {{\small $v''_{L}$}};
\node at (0.35,0.5) {{\small $m''_1$}};
\node at (2.35,-0.5) {{\small $m''_2$}};
\node at (7.5,-1.5) {{\small $m''_{\delta+1}$}};
\node at (9.5,-2.5) {{\small $m''_{\delta+2}$}};
\node at (14.5,-3.5) {{\small $m''_{L-1}$}};
\node at (16.35,-4.5) {{\small $m''_{L}$}};
\node at (0.7,-0.3) {{\small $-h_1$}};
\node at (2.3,-1.8) {{\small $-h_2$}};
\node at (7.2,-2.8) {{\small $-h_{\delta+1}$}};
\node at (9.15,-3.7) {{\small $-h_{\delta+2}$}};
\node at (14.1,-4.7) {{\small $-h_{L-1}$}};
\node at (16.2,-5.7) {{\small $-h_{L}$}};
\node at (-1.2,0) {{\small \bf $a$}};
\node at (18.2,-6) {{\small \bf $a$}};
\node at (0,1.25) {{\small \bf $1$}};
\node at (2,0.25) {{\small \bf $2$}};
\node at (7,-0.75) {{\small \bf $\delta+1$}};
\node at (9,-1.75) {{\small \bf $\delta+2$}};
\node at (14,-2.75) {{\small \bf $L-1$}};
\node at (16,-3.75) {{\small \bf $L$}};
\node at (1,-2.25) {{\small \bf $L-\delta$}};
\node at (3,-3.25) {{\small \bf $L-\delta+1$}};
\node at (8,-4.25) {{\small \bf $L$}};
\node at (9.8,-5.25) {{\small \bf $1$}};
\node at (15,-6.25) {{\small \bf $L-\delta-2$}};
\node at (17,-7.25) {{\small \bf $L-\delta-1$}};
%
\draw[thick,red,<->] (0.1,1.5) -- (1.9,1.5);
\node[red] at (1,1.7) {$\widehat{a}_1$};
\draw[thick,red,<->] (2.1,0.5) -- (3.9,0.5);
\node[red] at (3,0.7) {$\widehat{a}_2$};
\node[red] at (5.5,0) {\Large $\cdots$};
\draw[thick,red,<->] (7.1,-0.5) -- (8.9,-0.5);
\node[red] at (8,-0.3) {$\widehat{a}_\delta$};
\draw[thick,red,<->] (9.1,-1.5) -- (10.9,-1.5);
\node[red] at (10,-1.3) {$\widehat{a}_\delta+1$};
\node[red] at (12.5,-2) {\Large $\cdots$};
\draw[thick,red,<->] (14.1,-2.5) -- (15.9,-2.5);
\node[red] at (15,-2.3) {$\widehat{a}_{L-1}$};
\draw[thick,red,<->] (16.1,-3.5) -- (17.9,-3.5);
\node[red] at (17,-3.3) {$\widehat{a}_{L}$};
\draw[dashed] (10,-5) -- (12.5,-5);
\draw[dashed] (18,-6) -- (12.5,-6);
\draw[green!50!black,<->,thick] (12.5,-5.05) -- (12.5,-5.95);
\node[green!50!black] at (11.8,-5.5) {{ $-R^{(\delta+1)}$}};
\draw[dashed] (10,-5) -- (10,-7.5);
\draw[thick,<-,blue] (10.05,-7.5) -- (17.7,-7.5);
\node[blue] at (18,-7.5) {\Large $\cdots$};
\draw[dashed] (0,0) -- (0,-7.5);
\draw[thick,->,blue] (-0.7,-7.5) -- (-0.05,-7.5);
\node[blue] at (-1,-7.5) {\Large $\cdots$};
\node[blue] at (14,-7.3) {{ $S$}};
\end{tikzpicture}}}
\caption{\sl Twisted web diagram with shift $\delta+1$ obtained from \figref{StripDeltaFlop} through flop of the curve $h_1$.}
\label{StripDeltaFlop2}
\end{figure*}
We can summarise this in the form of the duality map
\begin{align}
&v_i\longrightarrow -h_i\,, \nonumber \\
&h_i\longrightarrow v_{i-1}+h_i+h_{i-1}\,,\nonumber\\
&m_i\longrightarrow m_i+h_i+h_{i+\delta+1}\,.\label{DualityMapAbstract}
\end{align}
Moreover, the basis parameters $\widehat{a}_{1,\ldots,L}$ and $S$ remain the same as in \figref{StripDelta} (\emph{i.e.} they are invariant under the duality map), while the parameters $R^{(\delta)}$ and $R^{(\delta+1)}$ are related by
\begin{align}
R^{(\delta+1)}-R^{(\delta)}=S\,.
\end{align}

\end{document}